\documentstyle[preprint,aps]{revtex}
\begin{document}
\draft

\title
{The Classical Relativistic Quark Model in the Rest-Frame Wigner-Covariant
Coulomb Gauge.}

\author{David Alba}

\address
{Dipartimento di Fisica\\
Universita' di Firenze\\
L.go E.Fermi 2 (Arcetri)\\
50125 Firenze, Italy}

\author{and}

\author{Luca Lusanna}

\address
{Sezione INFN di Firenze\\
L.go E.Fermi 2 (Arcetri)\\
50125 Firenze, Italy\\
E-mail LUSANNA@FI.INFN.IT}

\maketitle

\begin{abstract}

The system of N scalar particles with Grassmann-valued color charges plus the 
color SU(3) Yang-Mills field is reformulated on spacelike hypersurfaces. The
Dirac observables are found and the physical invariant mass of the system in
the Wigner-covariant rest-frame instant form of dynamics (covariant Coulomb 
gauge) is given. From the reduced Hamilton equations we extract the second order
equations of motion both for the reduced transverse color field and the 
particles. Then, we study this relativistic scalar quark model, deduced from
the classical QCD Lagrangian and with the color field present, in the N=2 
(meson) case. A special form of the requirement of having only color singlets, 
suited for a field-independent quark model, produces a ``pseudoclassical 
asymptotic freedom" and a regularization of the quark self-energy.

\vskip 1truecm
\noindent April 1997
\vskip 1truecm
\noindent This work has been partially supported by the network ``Constrained 
Dynamical Systems" of the E.U. Programme ``Human Capital and Mobility".

\end{abstract}
\vfill\eject

\section
{Introduction}

The Dirac observables for Yang-Mills theory with Grassmann-valued fermions
\cite{lusa}, Abelian and non-Abelian S(2) Higgs models \cite{lv1,lv2} and for
the SU(3)xSU(2)xU(1) standard model of elementary particles\cite{lv3} and the
corresponding physical Hamiltonians have been found in a noncovariant way 
equivalent to a generalized Coulomb gauge, following the scheme used by Dirac
\cite{dira} to do the canonical reduction of the electromagnetic field with
charged fermions. See Refs.\cite{re} for reviews of the method and of the 
program, which aims to get a unified description of the standard model and of
tetrad gravity in terms of Dirac's observables.

Then, the problem of how to covariantize these results was started. In Ref.
\cite{lus1}, the system of N scalar particles with Grassmann electric charges
plus the electromagnetic field was described by defining it on arbitrary
spacelike hypersurfaces, which give a covariant 3+1 decomposition of Minkowski
spacetime $M^4$, following Refs.\cite{dira1,karpacz} (see also Ref.\cite{kuchar}
, where a theoretical study of this problem is done in curved spacetimes).
The new configuration variables are the points $z^{\mu}(\tau ,\vec \sigma )$ of
the spacelike hypersurface $\Sigma_{\tau}$ [the only ones carrying Lorentz
indices] and a set of Lorentz invariant variables containing a 3-vector ${\vec 
\eta}_i(\tau )$ for each particle [$x^{\mu}_i(\tau )=z^{\mu}(\tau ,{\vec \eta}
_i(\tau ))$] and the electromagnetic gauge potentials 
$A_{\check A}(\tau ,\vec \sigma )={{\partial z^{\mu}(\tau ,\vec \sigma )}\over
{\partial \sigma^{\check A}}} A_{\mu}(z(\tau ,\vec \sigma ))$ [$\sigma^{\check
A}=(\tau ,\vec \sigma )$], which know implicitely the embedding of
$\Sigma_{\tau}$ into $M^4$. One has to choose the sign of the energy of each
particle, because there are not mass-shell constraints (like $p_i^2-m^2_i\approx
0$) among the constraints of this formulation, due to the fact that one has only
3 degrees of freedom for particle since the intersection of a timelike
trajectory and of the spacelike hypersurface $\Sigma_{\tau}$, with Lorentz
scalar `time' parameter $\tau$ (labelling the leaves of the foliation of $M^4$
with the $\Sigma_{\tau}$ all diffeomorphic to a given $\Sigma$), is determined
by 3 numbers: $\vec \sigma ={\vec \eta}_i(\tau )$.

Besides a Lorentz scalar form of the electromagnetic first class constraints,
one has 4 further first class constraints ${\cal H}_{\mu}(\tau ,\vec \sigma )
\approx 0$ implying the independence of the description from the choice of the 
spacelike hypersufaces foliating $M^4$. Being in special relativity, it is 
convenient to restrict ourselves to arbitrary spacelike hyperplanes $z^{\mu}
(\tau ,\vec \sigma )=x^{\mu}_s(\tau )+b^{\mu}_{\check r}(\tau ) \sigma^{\check 
r}$. Since they are described by only 10 variables [an origin $x^{\mu}_s(\tau 
)$ and 3 orthogonal spacelike unit vectors generating the fixed constant 
timelike unit normal to the hyperplane], we remain only with 10 first class
constraints determining the 10 variables conjugate to the hyperplane [they are
a 4-momentum $p^{\mu}_s$ and the 6 independent degrees of freedom hidden in a
spin tensor $S^{\mu\nu}_s$] in terms of the variables of the particles and of
the electromagnetic field. We can make the canonical reduction of the
electromagnetic field variables to transverse gauge potentials and electric
fields at the hypersurface as well as at the hyperplane level.

If we now restrict ourselves to timelike ($p^2_s > 0$) 4-momenta [the set of 
particles plus electromagnetic field configurations with $p^{\mu}_s$ not
timelike is of zero measure in the space of all configurations], we can 
restrict the description to the so-called Wigner hyperplanes orthogonal to 
$p^{\mu}_s$ itself. To get this result, we must boost at rest all the 
variables with Lorentz indices by using the standard Wigner boost $L^{\mu}{}
_{\nu}(p_s,{\buildrel \circ \over p}_s)$ for timelike Poincar\'e orbits, and
then add the gauge-fixings $b^{\mu}_{\check r}(\tau )-L^{\mu}{}_{\check r}(p_s,
{\buildrel \circ \over p}_s)\approx 0$. Since these gauge-fixings depend on 
$p^{\mu}_s$, the final canonical variables, apart $p^{\mu}_s$ itself, are of 3
types: i) there is a non-covariant center-of-mass variable ${\tilde x}^{\mu}
(\tau )$ [the classical basis of the Newton-Wigner position operator]; ii) all
the 3-vector variables become Wigner spin 1 3-vectors [boosts in $M^4$ induce
Wigner rotations on them]; iii) all the other variables are Lorentz scalars 
[in the case under consideration they are absent after the canonical reduction
to the transverse electromagnetic degrees of freedom]. Only the 4 first class
constraints determining $p^{\mu}_s$ are left. One obtains in this way a new kind
of instant form of the dynamics (see Ref.\cite{dira2}), the Euclidean covariant 
1-time rest-frame instant form. It is the special relativistic generalization of
the nonrelativistic separation of the center of mass from the relative motion
[$H={{ {\vec P}^2}\over {2M}}+H_{rel}$]. The role of the center of mass is 
taken by the Wigner hyperplane, identified by the point ${\tilde x}^{\mu}(\tau 
)$ and by its normal $p^{\mu}_s$. 
The 4 first class constraints can be put in the 
following form: i) the vanishing of the total (Wigner spin 1) 3-momentum of the
particles plus electromagnetic field $\vec p[system]\approx 0$ , saying that 
the Wigner hyperplane $\Sigma_W(\tau )$ is the intrinsic rest frame
[instead, ${\vec p}_s$ is left arbitrary, since it reflects the orientation of
the Wigner hyperplane with respect to arbitrary reference frames in Minkowski
spacetime]; 
ii) $\pm \sqrt{p^2_s}-M[system]\approx 0$, saying that the
invariant mass M of the system replaces the nonrelativistic  Hamiltonian
$H_{rel}$ for the relative degrees of freedom, after the addition of the
gauge-fixing $T_s-\tau \approx 0$ [identifying the time parameter $\tau$ with 
the Lorentz scalar time of the center of mass in the rest frame; M generates the
evolution in this time]. When one is able, as in the case of N free particles,
to find the (Wigner spin 1) 3-vector $\vec \eta (\tau )$ conjugate to
$\vec p[system]$($\approx 0$), the gauge-fixing $\vec \eta \approx 0$ eliminates
the gauge variables describing the 3-dimensional 
intrinsic center of mass inside the
Wigner hyperplane [$\vec \eta \approx 0$ forces it to coincide with $x^{\mu}_s
(\tau )=z^{\mu}(\tau ,\vec \sigma =\vec \eta =0)$ and breaks the translation
invariance $\vec \sigma \mapsto \vec \sigma +\vec a$], so that we remain only 
with Newtonian-like degrees of freedom with rotational covariance: i) a 
3-coordinate (not Lorentz covariant) ${\vec z}_s=\sqrt{p_s^2}({\vec {\tilde x}}
_s-{{{\vec p}_s}\over {p_s^o}}{\tilde x}^o)$ and its conjugate momentum 
${\vec k}_s={\vec p}_s/\sqrt{p^2_s}$ 
for the absolute center of mass in Minkowski spacetime; ii) a set of relative
conjugate pairs of variables with Wigner covariance inside the Wigner hyperplane
. As noted in Ref.\cite{lus1}, the noncovariance of the center of mass of
extended relativistic systems defines a classical intrinsic unit of lenght [the
M$\o$ller radius $\rho =\sqrt{-W^2}/cP^2=|{\check {\vec S}}|/c\sqrt{P^2}$
determined by the Poincar\'e Casimirs] to be used as a ultraviolet cutoff in 
the spirit of Dirac and Yukawa in future attempts of quantization. Let us
remark that this ultraviolet cutoff exists also in asymptotically flat general
relativity, taking into account the asymptotic Poincar\'e charges.

In this paper we will extend these results to N scalar particles with Grassmann
SU(3) color charges plus the SU(3) Yang-Mills field. The resulting
covariantization provides the tools to covariantize the bosonic part of the
SU(3)xSU(2)xU(1) model. To complete its covariantization, the description of 
Dirac and chiral fields on spacelike hypersurfaces in Minkowski spacetime is
needed (this problem is under investigation\cite{dep}).

The final result will be an expression for the physical invariant mass of the
system scalar quarks plus the color SU(3) Yang-Mills field in terms of gauge
invariant quantities (covariant generalized Coulomb gauge in the rest-frame 
instant form of dynamics). This pseudoclassical expression is the starting point
for defining a relativistic quark model derived from classical QCD. Till now, 
only the nonrelativistic quantum quark model is available \cite{q1} and there 
is no satisfactory derivation of it from QCD, notwithstanding its 
phenomenological relevance. Instead, we have here the full model with 
relativistic scalar quarks and classical color field and the classical reduced
Yang-Mills equations with transverse sources analogue of the equations
$\Box {\vec A}_{\perp}={\vec j}_{\perp}$ of the electromagnetic case in the
rest frame\cite{alba} : if one could guess a
reasonable solution of the equations for the color field and put it into the
invariant mass, one would obtain an effective invariant mass for a true 
relativistic quark model, even still with scalar quarks. Since this is not yet
possible, we shall limit ourselves to study some properties of pseudoclassical
mesons (N=2). In particular, we will show that the requirement of having only
color singlets, realized in a way suited to define a quark model without
color field, immediately produces a kind of ``pseudoclassical asymptotic
freedom" besides regularizing the quark self-energy. Also some comments on the
difficult problem of confinement are done.

In Section II we give some definitions and some results on spacelike
hypersurfaces. In Section III the Lagrangian of the system on spacelike 
hypersurfaces is given and the Hamiltonian formalism is developed till the 
reduction to the Wigner hyperplane. In Section IV the Dirac observables of the 
system are found and the final form of the physical invariant mass is given. In
Section V the reduced Hamilton equations and then the associated Euler-Lagrange 
equations in the rest frame are given. In Section VI the relativistic quark 
model in the case N=2 (mesons) is defined and it is shown that there is 
pseudoclassical asymptotic freedom.

In the final Section there are some conclusions and some comments on 
confinement and the open problems.

\vfill\eject

\section
{Dynamics on spacelike hypersurfaces}

In this Section we will introduce the background material from Ref.\cite{lus1} 
needed in the description of physical systems on spacelike hypersurfaces, 
integrating it with the definitions of non-Abelian SU(3) Yang-Mills fields
\cite{lusa}.

Let $\lbrace \Sigma_{\tau}\rbrace$ be a one-parameter family of spacelike
hypersurfaces foliating Minkowski spacetime $M^4$ and giving a 3+1 decomposition
of it. At fixed $\tau$, let 
$z^{\mu}(\tau ,\vec \sigma )$ be the coordinates of the points on $\Sigma
_{\tau}$ in $M^4$, $\lbrace \vec \sigma \rbrace$ a system of coordinates on
$\Sigma_{\tau}$. If $\sigma^{\check A}=(\sigma^{\tau}=\tau ;\vec \sigma 
=\lbrace \sigma^{\check r}\rbrace)$ [the notation ${\check A}=(\tau ,
{\check r})$ with ${\check r}=1,2,3$ will be used; note that ${\check A}=
\tau$ and ${\check A}={\check r}=1,2,3$ are Lorentz-scalar indices] and 
$\partial_{\check A}=\partial /\partial \sigma^{\check A}$, 
one can define the vierbeins

\begin{equation}
z^{\mu}_{\check A}(\tau ,\vec \sigma )=\partial_{\check A}z^{\mu}(\tau ,\vec 
\sigma ),\quad\quad
\partial_{\check B}z^{\mu}_{\check A}-\partial_{\check A}z^{\mu}_{\check B}=0,
\label {a1}
\end{equation}

\noindent so that the metric on $\Sigma_{\tau}$ is

\begin{eqnarray}
&&g_{{\check A}{\check B}}(\tau ,\vec \sigma )=z^{\mu}_{\check A}(\tau ,\vec 
\sigma )\eta_{\mu\nu}z^{\nu}_{\check B}(\tau ,\vec \sigma ),\quad\quad 
g_{\tau\tau}(\tau ,\vec \sigma ) > 0,\nonumber \\
&&g(\tau ,\vec \sigma )=-det\, ||\, g_{{\check A}{\check B}}(\tau ,\vec 
\sigma )\, || ={(det\, ||\, z^{\mu}_{\check A}(\tau ,\vec \sigma )\, ||)}^2,
\nonumber \\
&&\gamma (\tau ,\vec \sigma )=-det\, ||\, g_{{\check r}{\check s}}(\tau ,\vec 
\sigma )\, ||.
\label{a2}
\end{eqnarray}

If $\gamma^{{\check r}{\check s}}(\tau ,\vec \sigma )$ is the inverse of the 
3-metric $g_{{\check r}{\check s}}(\tau ,\vec \sigma )$ [$\gamma^{{\check r}
{\check u}}(\tau ,\vec \sigma )g_{{\check u}{\check s}}(\tau ,\vec 
\sigma )=\delta^{\check r}_{\check s}$], the inverse $g^{{\check A}{\check B}}
(\tau ,\vec \sigma )$ of $g_{{\check A}{\check B}}(\tau ,\vec \sigma )$ 
[$g^{{\check A}{\check C}}(\tau ,\vec \sigma )g_{{\check c}{\check b}}(\tau ,
\vec \sigma )=\delta^{\check A}_{\check B}$] is given by

\begin{eqnarray}
&&g^{\tau\tau}(\tau ,\vec \sigma )={{\gamma (\tau ,\vec \sigma )}\over
{g(\tau ,\vec \sigma )}},\nonumber \\
&&g^{\tau {\check r}}(\tau ,\vec \sigma )=-[{{\gamma}\over g} g_{\tau {\check 
u}}\gamma^{{\check u}{\check r}}](\tau ,\vec \sigma ),\nonumber \\
&&g^{{\check r}{\check s}}(\tau ,\vec \sigma )=\gamma^{{\check r}{\check s}}
(\tau ,\vec \sigma )+[{{\gamma}\over g}g_{\tau {\check u}}g_{\tau {\check v}}
\gamma^{{\check u}{\check r}}\gamma^{{\check v}{\check s}}](\tau ,\vec \sigma ),
\label{a3}
\end{eqnarray}

\noindent so that $1=g^{\tau {\check C}}(\tau ,\vec \sigma )g_{{\check C}\tau}
(\tau ,\vec \sigma )$ is equivalent to

\begin{equation}
{{g(\tau ,\vec \sigma )}\over {\gamma (\tau ,\vec \sigma )}}=g_{\tau\tau}
(\tau ,\vec \sigma )-\gamma^{{\check r}{\check s}}(\tau ,\vec \sigma )
g_{\tau {\check r}}(\tau ,\vec \sigma )g_{\tau {\check s}}(\tau ,\vec \sigma ).
\label{a4}
\end{equation}

We have

\begin{equation}
z^{\mu}_{\tau}(\tau ,\vec \sigma )=(\sqrt{ {g\over {\gamma}} }l^{\mu}+
g_{\tau {\check r}}\gamma^{{\check r}{\check s}}z^{\mu}_{\check s})(\tau ,
\vec \sigma ),
\label{a5}
\end{equation}

\noindent and

\begin{eqnarray}
\eta^{\mu\nu}&=&z^{\mu}_{\check A}(\tau ,\vec \sigma )g^{{\check A}{\check B}}
(\tau ,\vec \sigma )z^{\nu}_{\check B}(\tau ,\vec \sigma )=\nonumber \\
&=&(l^{\mu}l^{\nu}+z^{\mu}_{\check r}\gamma^{{\check r}{\check s}}
z^{\nu}_{\check s})(\tau ,\vec \sigma ),
\label{a6}
\end{eqnarray}

\noindent where

\begin{eqnarray}
l^{\mu}(\tau ,\vec \sigma )&=&({1\over {\sqrt{\gamma}} }\epsilon^{\mu}{}_{\alpha
\beta\gamma}z^{\alpha}_{\check 1}z^{\beta}_{\check 2}z^{\gamma}_{\check 3})
(\tau ,\vec \sigma ),\nonumber \\
&&l^2(\tau ,\vec \sigma )=1,\quad\quad l_{\mu}(\tau ,\vec \sigma )z^{\mu}
_{\check r}(\tau ,\vec \sigma )=0,
\label{a7}
\end{eqnarray}

\noindent is the unit (future pointing) normal to $\Sigma (\tau )$ at
$z^{\mu}(\tau ,\vec \sigma )$.

For the volume element in Minkowski spacetime we have

\begin{eqnarray}
d^4z&=&z^{\mu}_{\tau}(\tau ,\vec \sigma )d\tau d^3\Sigma_{\mu}=d\tau [z^{\mu}
_{\tau}(\tau ,\vec \sigma )l_{\mu}(\tau ,\vec \sigma )]\sqrt{\gamma
(\tau ,\vec \sigma )}d^3\sigma=\nonumber \\
&=&\sqrt{g(\tau ,\vec \sigma )} d\tau d^3\sigma.
\label{a8}
\end{eqnarray}

Let us remark that according to the geometrical approach of 
Ref.\cite{kuchar},one 
can use Eq.(\ref{a5}) in the form $z^{\mu}_{\tau}(\tau ,\vec \sigma )=N(\tau ,
\vec \sigma )l^{\mu}(\tau ,\vec \sigma )+N^{\check r}(\tau ,\vec \sigma )
z^{\mu}_{\check r}(\tau ,\vec \sigma )$, where $N=\sqrt{g/\gamma}=\sqrt{g
_{\tau\tau}-\gamma^{{\check r}{\check s}}g_{\tau{\check r}}g_{\tau{\check s}}}$ 
and $N^{\check r}=g_{\tau \check s}\gamma^{\check s\check r}$ are the 
standard lapse and shift functions, so that $g_{\tau \tau}=N^2+
g_{\check r\check s}N^{\check r}N^{\check s}, g_{\tau \check r}=
g_{\check r\check s}N^{\check s},
g^{\tau \tau}=N^{-2}, g^{\tau \check r}=-N^{\check r}/N^2, g^{\check r\check
s}=\gamma^{\check r\check s}+{{N^{\check r}N^{\check s}}\over {N^2}}$,
${{\partial}\over {\partial z^{\mu}_{\tau}}}=l_{\mu}\, {{\partial}\over
{\partial N}}+z_{{\check s}\mu}\gamma^{{\check s}{\check r}} {{\partial}\over
{\partial N^{\check r}}}$, $d^4z=N\sqrt{\gamma}d\tau d^3\sigma$.

The rest frame form of a timelike fourvector $p^{\mu}$ is $\stackrel
{\circ}{p}{}^{\mu}=\eta \sqrt{p^2} (1;\vec 0)= \eta^{\mu o}\eta \sqrt{p^2}$,
$\stackrel{\circ}{p}{}^2=p^2$, where $\eta =sign\, p^o$.
The standard Wigner boost transforming $\stackrel{\circ}{p}{}^{\mu}$ into
$p^{\mu}$ is

\begin{eqnarray}
L^{\mu}{}_{\nu}(p,\stackrel{\circ}{p})&=&\epsilon^{\mu}_{\nu}(u(p))=
\nonumber \\
&=&\eta^{\mu}_{\nu}+2{ {p^{\mu}{\stackrel{\circ}{p}}_{\nu}}\over {p^2}}-
{ {(p^{\mu}+{\stackrel{\circ}{p}}^{\mu})(p_{\nu}+{\stackrel{\circ}{p}}_{\nu})}
\over {p\cdot \stackrel{\circ}{p} +p^2} }=\nonumber \\
&=&\eta^{\mu}_{\nu}+2u^{\mu}(p)u_{\nu}(\stackrel{\circ}{p})-{ {(u^{\mu}(p)+
u^{\mu}(\stackrel{\circ}{p}))(u_{\nu}(p)+u_{\nu}(\stackrel{\circ}{p}))}
\over {1+u^o(p)} },\nonumber \\
&&{} \nonumber \\
\nu =0 &&\epsilon^{\mu}_o(u(p))=u^{\mu}(p)=p^{\mu}/\eta \sqrt{p^2}, \nonumber \\
\nu =r &&\epsilon^{\mu}_r(u(p))=(-u_r(p); \delta^i_r-{ {u^i(p)u_r(p)}\over
{1+u^o(p)} }).
\label{a9}
\end{eqnarray}

The inverse of $L^{\mu}{}_{\nu}(p,\stackrel{\circ}{p})$ is $L^{\mu}{}_{\nu}
(\stackrel{\circ}{p},p)$, the standard boost to the rest frame, defined by

\begin{equation}
L^{\mu}{}_{\nu}(\stackrel{\circ}{p},p)=L_{\nu}{}^{\mu}(p,\stackrel{\circ}{p})=
L^{\mu}{}_{\nu}(p,\stackrel{\circ}{p}){|}_{\vec p\rightarrow -\vec p}.
\label{a10}
\end{equation}

Therefore, we can define the following vierbeins [the $\epsilon^{\mu}_r(u(p))$'s
are also called polarization vectors; the indices r, s will be used for A=1,2,3
and $\bar o$ for A=0]

\begin{eqnarray}
&&\epsilon^{\mu}_A(u(p))=L^{\mu}{}_A(p,\stackrel{\circ}{p}),\nonumber \\
&&\epsilon^A_{\mu}(u(p))=L^A{}_{\mu}(\stackrel{\circ}{p},p)=\eta^{AB}\eta
_{\mu\nu}\epsilon^{\nu}_B(u(p)),\nonumber \\
&&{} \nonumber \\
&&\epsilon^{\bar o}_{\mu}(u(p))=\eta_{\mu\nu}\epsilon^{\nu}_o(u(p))=u_{\mu}(p),
\nonumber \\
&&\epsilon^r_{\mu}(u(p))=-\delta^{rs}\eta_{\mu\nu}\epsilon^{\nu}_r(u(p))=
(\delta^{rs}u_s(p);\delta^r_j-\delta^{rs}\delta_{jh}{{u^h(p)u_s(p)}\over
{1+u^o(p)} }),\nonumber \\
&&\epsilon^A_o(u(p))=u_A(p),
\label{a11}
\end{eqnarray}

\noindent which satisfy

\begin{eqnarray}
&&\epsilon^A_{\mu}(u(p))\epsilon^{\nu}_A(u(p))=\eta^{\mu}_{\nu},\nonumber \\
&&\epsilon^A_{\mu}(u(p))\epsilon^{\mu}_B(u(p))=\eta^A_B,\nonumber \\
&&\eta^{\mu\nu}=\epsilon^{\mu}_A(u(p))\eta^{AB}\epsilon^{\nu}_B(u(p))=u^{\mu}
(p)u^{\nu}(p)-\sum_{r=1}^3\epsilon^{\mu}_r(u(p))\epsilon^{\nu}_r(u(p)),
\nonumber \\
&&\eta_{AB}=\epsilon^{\mu}_A(u(p))\eta_{\mu\nu}\epsilon^{\nu}_B(u(p)),\nonumber 
\\
&&p_{\alpha}{{\partial}\over {\partial p_{\alpha}} }\epsilon^{\mu}_A(u(p))=
p_{\alpha}{{\partial}\over {\partial p_{\alpha}} }\epsilon^A_{\mu}(u(p))
=0.
\label{a12}
\end{eqnarray}

The Wigner rotation corresponding to the Lorentz transformation $\Lambda$ is

\begin{eqnarray}
R^{\mu}{}_{\nu}(\Lambda ,p)&=&{[L(\stackrel{\circ}{p},p)\Lambda^{-1}L(\Lambda
p,\stackrel{\circ}{p})]}^{\mu}{}_{\nu}=\left(
\begin{array}{cc}
1 & 0 \\
0 & R^i{}_j(\Lambda ,p)
\end{array}  
\right) ,\nonumber \\
{} && {}\nonumber \\
R^i{}_j(\Lambda ,p)&=&{(\Lambda^{-1})}^i{}_j-{ {(\Lambda^{-1})^i{}_op_{\beta}
(\Lambda^{-1})^{\beta}{}_j}\over {p_{\rho}(\Lambda^{-1})^{\rho}{}_o+\eta 
\sqrt{p^2}} }-\nonumber \\
&-&{{p^i}\over {p^o+\eta \sqrt{p^2}} }[(\Lambda^{-1})^o{}_j- { {((\Lambda^{-1})^o
{}_o-1)p_{\beta}(\Lambda^{-1})^{\beta}{}_j}\over {p_{\rho}(\Lambda^{-1})^{\rho}
{}_o+\eta \sqrt{p^2}} }].
\label{a13}
\end{eqnarray}

The polarization vectors transform under the 
Poincar\'e transformations $(a,\Lambda )$ in the following way

\begin{equation}
\epsilon^{\mu}_r(u(\Lambda p))=(R^{-1})_r{}^s\, \Lambda^{\mu}{}_{\nu}\, 
\epsilon^{\nu}_s(u(p)).
\label{a14}
\end{equation}

On the hypersurface $\Sigma_{\tau}$, we describe the color SU(3) potential
and field strength with Lorentz-scalar variables $A_{a\check A}(\tau ,\vec 
\sigma )$ and $F_{a{\check A}{\check B}}(\tau ,\vec \sigma )$ respectively: 
they contain the embedding $\Sigma (\tau ) \rightarrow M^4$ and are defined by

\begin{eqnarray}
&&A_{a\check A}(\tau ,\vec \sigma )=z^{\mu}_{\check A}(\tau ,\vec \sigma )
A_{a\mu}(z(\tau ,\vec \sigma )),\nonumber \\
&&F_{a{\check A}{\check B}}(\tau ,\vec \sigma )={\partial}_{\check A}A_{a\check
B}(\tau ,\vec \sigma )-{\partial}_{\check B}A_{a\check A}(\tau ,\vec \sigma )+
c_{abc}A_{b\check A}(\tau ,\vec \sigma )A_{b\check B}(\tau ,\vec \sigma )=
\nonumber \\
&&=z^{\mu}_{\check A}(\tau ,\vec \sigma )z^{\nu}_{\check B}(\tau ,\vec \sigma )
F_{a\mu\nu}(z(\tau ,\vec \sigma ))=z^{\mu}_{\check A}(\tau ,\vec \sigma )
z^{\nu}_{\check B}(\tau ,\vec \sigma )[\partial_{\mu}A_{a\nu}(z(\tau ,\vec 
\sigma ))-\partial_{\nu}A_{a\mu}(z(\tau ,\vec \sigma ))+\nonumber \\
&&+c_{abc}A_{b\mu}
(z(\tau ,\vec \sigma ))A_{c\nu}(z(\tau ,\vec \sigma ))].
\label{a15}
\end{eqnarray}

We could have written $A_{\mu}=z_{\mu}^{\check A}A_{\check A}=l_{\mu}\, A_l+z
_{\mu}^{\check r}A_{\check r}$ [$z_{\mu}^{\check A}$ are the inverse vierbeins],
so to get

\begin{equation}
A_{\tau }(\tau ,\vec \sigma )=
N(\tau ,\vec \sigma )A_l(\tau ,\vec \sigma )+N^{\check r}
(\tau ,\vec \sigma )A_{\check r}(\tau ,\vec \sigma ),
\label{a16}
\end{equation}

\noindent and we could have used $A_l
(\tau ,\vec \sigma )$ as the genuine field configuration variable independent
from the motion of the embedded hypersurface, as suggested in Ref.\cite{kuchar}.
However, this more geometric formulation is equivalent to the simpler one of
Ref.\cite{lus1} [see its Appendix C] for spin 1 fields, so that in this paper we
shall go on to use $A_{\tau}$ rather than $A_l$.

The generators ${\hat T}^a$
of the Lie algebra su(3) of color [${\hat A}_{\mu}=
A_{a\mu}{\hat T}^a$] in the 8-dimensional adjoint representation of SU(3) 
and those $T^a$ in the 3-dimensional fundamental one 
are [$c_{abc}$ are the SU(3) totally antisymmetric structure
constants]

\begin{eqnarray}
&&{\hat T}^a=-{\hat T}^{a\dagger},\quad\quad ({\hat T}^a)_{bc}=c_{abc},
\quad\quad [{\hat T}^a,{\hat T}^b]=c_{abc}{\hat  T}^c,\nonumber \\
&&T^a=-T^{a\dagger},\quad\quad [T^a,T^b]=c_{abc}T^c,\quad\quad T^a=-{i\over 2}
\lambda_a,
\label{a17}
\end{eqnarray}

\noindent where the $\lambda_a$'s era the $3\times 3$ Gell-Mann matrices.

The covariant derivative associated with $A_{a\mu}$ is 

\begin{equation}
({\hat D}^{(A)}_{\mu})_{ac}=\delta_{ac}\partial_{\mu}+c_{abc}A_{b\mu}=
(\partial_{\mu}-{\hat A}_{\mu})_{ac}
\label{a18}
\end{equation}

\noindent and the gauge transformations are defined as [${\hat F}_{\mu\nu}=
F_{a\mu\nu}{\hat T}^a$]

\begin{eqnarray}
{\hat A}_{\mu}(x)&\mapsto& {\hat A}^U_{\mu}(x)=U^{-1}(x){\hat A}_{\mu}(x)U
(x)+U^{-1}(x)\partial_{\mu}U(x)=\nonumber \\
&=&{\hat A}_{\mu}(x)+U^{-1}(x)\, (\partial_{\mu}U(x)+[{\hat A}_{\mu}(x),U
(x)]),\nonumber \\
{\hat F}_{\mu\nu}(x)&\mapsto& {\hat F}^U_{\mu\nu}(x)=U^{-1}(x){\hat F}
_{\mu\nu}(x)U(x)=
{\hat F}_{\mu\nu}(x)+U^{-1}(x)\, [{\hat F}_{\mu\nu}(x),U(x)].
\label{a19}
\end{eqnarray}

\noindent Here $U$ is the realization in the
adjoint representation of the SU(3) gauge transformations.

The scalar particles with Minkowski coordinates $x^{\mu}_i(\tau )$, i=1,..,N,
are identified on the spacelike hypersurface $\Sigma 
_{\tau}$ by 3 numbers ${\vec \eta}_i(\tau )$, i=1,..,N, by the equation $x
^{\mu}_i(\tau )= z^{\mu}(\tau ,{\vec \eta}_i(\tau ))$ [so that ${\dot x}^{\mu}
_i(\tau )=z^{\mu}_{\tau}(\tau ,{\vec \eta}_i(\tau ))+z^{\mu}_{\check r}(\tau ,
{\vec \eta}_i(\tau )){\dot \eta}_i^{\check r}(\tau )$ and $\eta_i={\dot x}^o_i
(\tau )$]. As shown in Ref.\cite{lus1}
this implies that the mass shell constraint $p^2_i-m^2_i=0$ has been solved
and a choice of the sign of the energy, $\eta_i=sign\, p^o_i$, has been done.

In this paper we consider the case of N relativistic scalar 
particles  with the color SU(3) charge of each particle 
described in a pseudoclassical way\cite{lus2} [see also 
Refs.\cite{casalb} for
pseudoclassicla mechanics] by means of 3 pairs of
complex conjugate Grassmann variables $\theta_{i\alpha}(\tau ), 
\theta^{*}_{i\alpha}(\tau )$, $\alpha =1,2,3$,
which belong to the fundamental representation of SU(3). They satisfy

\begin{eqnarray}
&&\theta_{i\alpha}\theta_{i\beta}+\theta_{i\beta}\theta_{i\alpha}=0,
\nonumber \\
&&\theta^{*}_{i\alpha}\theta^{*}_{i\beta}+\theta^{*}_{i\beta}\theta^{*}
_{i\alpha}=0,\nonumber \\
&&\theta_{i\alpha}\theta^{*}_{i\beta}+\theta^{*}_{i\beta}\theta_{i\alpha}=0,
\label{a20}
\end{eqnarray}

\noindent and the Grassmann variables of different particles are assumed to
commute

\begin{eqnarray}
&&\theta_{i\alpha}\theta_{j\beta}=\theta_{j\beta}\theta_{i\alpha},\quad\quad
i\not=j\nonumber \\
&&\theta^{*}_{i\alpha}\theta^{*}_{j\beta}=\theta^{*}_{j\beta}\theta^{*}
_{i\alpha},\nonumber \\
&&\theta_{i\alpha}\theta^{*}_{j\beta}=\theta^{*}_{j\beta}\theta_{i\alpha}.
\label{a21}
\end{eqnarray}

The color charges of the particles are 
$Q_{ia}(\tau )=i\sum^3_{\alpha ,\beta =1}\theta^{*}_{i\alpha}(\tau ) 
(T^a)_{\alpha\beta}\theta_{i\beta}(\tau )$ [$Q^{*}_{ia}=Q_{ia}$ since 
$T^{a\dagger}=-T^a$].

At the quantum level \cite{lus2} (see also Ref.\cite{ht}) the Grassmann
variables $\theta^{*}_{i\alpha}$, $\theta_{i\alpha}$ go into Fermi
oscillators $b^{\dagger}_{i\alpha}$, $b_{i\alpha}$ satisfying the
anticommutation relations $[b_{i\alpha},b^{\dagger}_{j\beta}]_{+}=\delta
_{ij}\delta_{\alpha\beta}$, $[b_{i\alpha},b_{j\beta}]_{+}=[b^{\dagger}
_{i\alpha},b^{\dagger}_{j\beta}]_{+}=0$. For each particle there is a 
8-dimensional Hilbert space of charge states [the charge operator is ${\hat Q}
_{ia}=i\sum_{\alpha\beta}b^{\dagger}_{i\alpha}(T^a)_{\alpha\beta}b_{i\beta}$] 
with basis $|\, 0_i\, >$, $b^{\dagger}_{i\alpha} |\, 0_i\, >$, $b^{\dagger}
_{i\alpha}b^{\dagger}_{i\beta} |\, 0_i\, >$, $b^{\dagger}_{i\alpha}b^{\dagger}
_{i\beta}b^{\dagger}_{i\gamma} |\, 0_i\, >$ : the states with k=0,1,2,3, 
oscillators transform like a completely antisymmetric representation of
dimension $\left( \begin{array}{c} 3 \\ k \end{array} \right)$ of SU(3).
Therefore, the space of charge states for each particle transforms like the 
reducible representation $1\oplus 3\oplus 3^{*}\oplus 1$ of SU(3) of dimension
$\sum_{k=0}^3 \left( \begin{array}{c} 3 \\ k \end{array} \right) =2^3$.

To select the triplet (quark) or antitriplet (antiquark) representation, we
shall add to the pseudoclassical theory the constraint

\begin{equation}
N_i=\sum_{\alpha}\theta_{i\alpha}^{*}\theta_{i\alpha}\approx 0.
\label{a22}
\end{equation}

\noindent As shown in Ref.\cite{lus3}, after quantization $N_i$ is replaced by
${\hat N}_i[A_i]=\sum_{\alpha=0}^3 b^{\dagger}_{i\alpha}b_{i\alpha}-A_i=
{\hat n}_i-A_i$, where ${\hat n}_i$ is the occupation number selecting the
$\left( \begin{array}{c} 3 \\ k \end{array} \right)$-dimensional
representation of SU(3) and $A_i$ is an arbitrary c-number present due to
ordering problems. Therefore, with the prescription $A_i=1\, or\, 2$, the
constraint $N_i\approx 0$ becomes  the quantum constraint ${\hat N}_i[1]
|\, 0_i\, > =0$ or ${\hat N}_i[2] |\, 0_i\, > =0$, selecting the triplet or
the antitriplet representation respectively for particle `i'.

With the constraint $N_i\approx 0$ for each i one has $Q_{ia}Q_{ib}Q_{ic}
\equiv 0$ because it is proportional to $N^3_i$ [$\equiv 0$ in the Dirac strong
sense]. Moreover, since one has $\sum_aQ^2_{ia}=-\theta^{*}_{i\alpha}\theta
_{i\beta}\theta^{*}_{i\gamma}\theta_{i\delta} \sum_a(T^a)_{\alpha\beta}
(T^a)_{\gamma\delta}$ with $\sum_a (T^a)_{\alpha\beta}(T^a)_{\gamma\delta}=
{1\over 6}\delta_{\alpha\beta}\delta_{\gamma\delta}-{1\over 2}\delta_{\alpha
\delta}\delta_{\beta\gamma}$ [valid in the fundamental representation of
SU(3)], one gets

\begin{equation}
\sum_a Q^2_{ia}=-{2\over 3} N^2_i \equiv 0.
\label{a23}
\end{equation}

\vfill\eject

\section
{The Lagrangian for colored particles plus Yang-Mills field}

The system of N colored scalar particles plus the SU(3) Yang-Mills field
is described by the action

\begin{eqnarray}
S&=& \int d\tau d^3\sigma \, {\cal L}(\tau ,\vec
\sigma )=\int d\tau L(\tau ),\nonumber \\
L(\tau )&=&\int d^3\sigma {\cal L}(\tau ,\vec \sigma ),\nonumber \\
{\cal L}(\tau ,\vec \sigma )&=&{i\over 2}\sum_{i=1}^N\delta^3(\vec \sigma 
-{\vec \eta}_i(\tau ))\sum^3_{\alpha =1}[\theta^{*}_{i\alpha}(\tau ){\dot 
\theta}_{i\alpha}(\tau )-{\dot \theta}^{*}_{i\alpha}(\tau )\theta_{i\alpha}
(\tau )+\lambda_i(\tau )\sum^3_{\alpha =1}\theta^{*}_{i\alpha}(\tau )\theta
_{i\alpha}(\tau )]-\nonumber \\
&&-\sum_{i=1}^N\delta^3(\vec \sigma -{\vec \eta}_i
(\tau ))[\eta_im_i\sqrt{ g_{\tau\tau}(\tau ,\vec \sigma )+2g_{\tau {\check r}}
(\tau ,\vec \sigma ){\dot \eta}^{\check r}_i(\tau )+g_{{\check r}{\check s}}
(\tau ,\vec \sigma ){\dot \eta}_i^{\check r}(\tau ){\dot \eta}_i^{\check s}
(\tau )  }-\nonumber \\
&&-\sum_aQ_{ia}(A_{a\tau }(\tau ,\vec \sigma )+
A_{a\check r}(\tau ,\vec \sigma ){\dot \eta}^{\check r}_i(\tau ))]-\nonumber \\
&&-{1\over {4g_s^2}}\, \sqrt {g(\tau ,\vec \sigma )}
g^{{\check A}{\check C}}(\tau ,\vec \sigma )g^{{\check B}{\check D}}
(\tau ,\vec \sigma )\sum_aF_{a{\check A}{\check B}}(\tau ,\vec \sigma )
F_{a{\check C}{\check D}}(\tau ,\vec \sigma ),
\label{b1}
\end{eqnarray}

\noindent where the configuration variables are $z^{\mu}(\tau ,\vec \sigma )$
$A_{a\check A}(\tau ,\vec \sigma )$, ${\vec \eta}_i(\tau )$, $\theta_{i\alpha}
(\tau )$ and $\theta^{*}_{i\alpha}(\tau )$, i=1,..,N. The particles have 
Grassmann-valued charges $Q_{ai}(\tau )=i \sum_{\alpha ,\beta =1}^3\theta^{*}
_{i\alpha}(\tau ) (T^a)_{\alpha\beta}\theta_{i\beta}(\tau )$. 

We have

$-{1\over 4}\sqrt{g} g^{\check A\check C}g^{\check B\check D}
\sum_aF_{a\check A
\check B}F_{a\check C\check D}=\hfill \break
=-{1\over 4}\sqrt{g}\sum_a [2(g^{\tau\tau}
g^{{\check r}{\check s}}-g^{\tau{\check r}}g^{\tau{\check s}}) 
F_{a\tau{\check r}}F_{a\tau{\check s}} +4g^{{\check r}{\check s}}
g^{\tau {\check u}}
F_{a\tau {\check r}}F_{a{\check s}{\check u}}+g^{{\check r}{\check u}}
g^{{\check s}{\check v}}F_{a{\check r}{\check s}}F_{a{\check u}{\check v}}]=
\hfill \break
=-\sqrt{\gamma}\sum_a[{1\over 2}\sqrt{ {{\gamma}\over
g} } F_{a\tau \check r}\gamma^{\check r\check s}F_{a\tau \check s}-
\sqrt{{{\gamma}
\over g}} g_{\tau \check v}\gamma^{\check v\check r}F_{a\check r\check s}
\gamma^{\check s\check u}F_{a\tau \check u}+{1\over 4}\sqrt{ {g\over {\gamma}}}
\gamma^{\check r\check s}F_{a\check r\check u}F_{a\check s\check v}(\gamma
^{\check
u\check v}+2{{\gamma}\over g}g_{\tau \check m}\gamma^{\check m\check u}
g_{\tau \check n}\gamma^{\check n\check v})]=-{{\sqrt{\gamma}}\over {2N}}
(F_{\tau \check r}-N^{\check u}F_{\check u\check r})\gamma^{\check r\check s}
(F_{\tau \check s}-N^{\check v}F_{\check v\check s})-{{N\sqrt{\gamma}}\over 4}
\gamma^{\check r\check s}\gamma^{\check u\check v}F_{\check r\check u}
F_{\check s\check v}.$

The action is invariant under separate
$\tau$- and $\vec \sigma$-reparametrizations, since $A_{a\tau}(\tau ,\vec \sigma
)$ transforms as a $\tau$-derivative; moreover, it is invariant under the odd
phase transformations $\delta \theta_{i\alpha}\mapsto \alpha_a(T^a)
_{\alpha\beta} \theta_{i\beta}$.

The canonical momenta are [$E_{a\check r}=F_{a{\check r}\tau}$ and $B
_{a\check r}={1\over 2}\epsilon_{{\check r}{\check s}{\check t}}F_{a{\check s}
{\check t}}$ 
($\epsilon_{{\check r}{\check s}{\check t}}=\epsilon^{{\check r}{\check s}
{\check t}}$) are the electric and magnetic fields respectively; for
$g_{\check A\check B}\rightarrow \eta_{\check A\check B}$ one gets 
$\pi_a^{\check r}=-E_{a\check r}=E_a^{\check r}$]

\begin{eqnarray}
\rho_{\mu}(\tau ,\vec \sigma )&=&-{ {\partial {\cal L}(\tau ,\vec \sigma )}
\over {\partial z^{\mu}_{\tau}(\tau ,\vec \sigma )} }=\sum_{i=1}^N\delta^3
(\vec \sigma -{\vec \eta}_i(\tau ))\eta_im_i\nonumber \\
&&{ {z_{\tau\mu}(\tau ,\vec \sigma )+z_{{\check r}\mu}(\tau ,\vec \sigma )
{\dot \eta}_i^{\check r}(\tau )}\over {\sqrt{g_{\tau\tau}(\tau ,\vec \sigma )+
2g_{\tau {\check r}}(\tau ,\vec \sigma ){\dot \eta}_i^{\check r}(\tau )+
g_{{\check r}{\check s}}(\tau ,\vec \sigma ){\dot \eta}_i^{\check r}(\tau ){\dot
\eta}_i^{\check s}(\tau ) }} }+\nonumber \\
&&+{ {\sqrt {g(\tau ,\vec \sigma )}}\over 4}[(g^{\tau \tau}z_{\tau \mu}+
g^{\tau {\check r}}z_{\check r\mu})(\tau ,\vec \sigma )g^{{\check A}{\check C}}
(\tau ,\vec \sigma )g^{{\check B}{\check D}}(\tau ,\vec \sigma )
\sum_aF_{a{\check A}
{\check B}}(\tau ,\vec \sigma )F_{a{\check C}{\check D}}(\tau ,\vec \sigma )
-\nonumber \\
&-&2[z_{\tau \mu}(\tau ,\vec \sigma )(g^{\check A\tau}g^{\tau \check C}
g^{{\check B}{\check D}}+g^{{\check A}{\check C}}g^{\check B\tau}g^{\tau 
\check D})(\tau ,\vec \sigma )+\nonumber \\
&+&z_{\check r\mu}(\tau ,\vec \sigma )
(g^{{\check A}{\check r}}g^{\tau {\check C}}+g^{{\check A}\tau}g^{{\check r}
{\check C}})(\tau ,\vec \sigma )g^{{\check B}{\check D}}
(\tau ,\vec \sigma )]\sum_aF_{a{\check A}{\check B}}(\tau ,\vec \sigma )
F_{a{\check C}{\check D}}(\tau ,\vec \sigma )]=
\nonumber \\
&&=[(\rho_{\nu}l^{\nu})l_{\mu}+(\rho_{\nu}z^{\nu}_{\check r})\gamma^{{\check r}
{\check s}}z_{{\check s}\mu}](\tau ,\vec \sigma ),\nonumber \\
&&{}\nonumber \\
\pi_a^{\tau}(\tau ,\vec \sigma )&=&{ {\partial L}\over {\partial \partial_{\tau}
A_{a\tau}(\tau ,\vec \sigma )} }=0,\nonumber \\
\pi_a^{\check r}(\tau ,\vec \sigma )&=&{ {\partial L}\over {\partial \partial
_{\tau}A_{a\check r}(\tau ,\vec \sigma )} }=-
g_s^{-2}{ {\gamma (\tau ,\vec \sigma )}
\over {\sqrt {g(\tau ,\vec \sigma )}} }\gamma^{{\check r}{\check s}}(\tau ,
\vec \sigma )(F_{a\tau {\check s}}+g_{\tau {\check v}}\gamma^{{\check v}
{\check u}}F_{a{\check u}{\check s}})(\tau ,\vec \sigma )=\nonumber \\
&&=g_s^{-2}
{ {\gamma (\tau ,\vec \sigma )}\over {\sqrt {g(\tau ,\vec \sigma )}} }
\gamma^{{\check r}{\check s}}(\tau ,\vec \sigma )(E_{a\check s}(\tau ,\vec 
\sigma )+g_{\tau {\check v}}(\tau ,\vec \sigma )\gamma^{{\check v}{\check u}}
(\tau ,\vec \sigma )\epsilon_{{\check u}{\check s}{\check t}} B_{a\check t}
(\tau ,\vec \sigma )),\nonumber \\
&&{}\nonumber \\
\kappa_{i{\check r}}(\tau )&=&-{ {\partial L(\tau )}\over {\partial {\dot
\eta}_i^{\check r}(\tau )} }=\nonumber \\
&=&\eta_im_i{ {g_{\tau {\check r}}(\tau ,{\vec \eta}_i(\tau ))+g_{{\check r}
{\check s}}(\tau ,{\vec \eta}_i(\tau )){\dot \eta}_i^{\check s}(\tau )}\over
{ \sqrt{g_{\tau\tau}(\tau ,{\vec \eta}_i(\tau ))+
2g_{\tau {\check r}}(\tau ,{\vec \eta}_i(\tau )){\dot \eta}_i^{\check r}(\tau )+
g_{{\check r}{\check s}}(\tau ,{\vec \eta}_i(\tau )){\dot \eta}_i^{\check r}
(\tau ){\dot \eta}_i^{\check s}(\tau ) }} }-\nonumber \\
&-&\sum_aQ_{ia}(\tau )A_{a\check r}(\tau ,{\vec 
\eta}_i(\tau )),\nonumber \\
&&{}\nonumber \\
\pi_{\theta \,i\alpha}(\tau )&=&{{\partial L(\tau )}\over {\partial {\dot 
\theta}_{i\alpha}(\tau )}}=-{i\over 2}\theta^{*}_{i\alpha}(\tau )\nonumber \\
\pi_{\theta^{*} \, i\alpha}(\tau )&=&{{\partial L(\tau )}\over {\partial {\dot 
\theta}^{*}_{i\alpha}(\tau )}}=-{i\over 2}\theta_{i\alpha}(\tau ),
\label{b2}
\end{eqnarray}

\noindent and the following Poisson brackets are assumed

\begin{eqnarray}
&&\lbrace z^{\mu}(\tau ,\vec \sigma ),\rho_{\nu}(\tau ,{\vec \sigma}^{'}\rbrace
=-\eta^{\mu}_{\nu}\delta^3(\vec \sigma -{\vec \sigma}^{'}),\nonumber \\
&&\lbrace A_{a\check A}(\tau ,\vec \sigma ),\pi_b^{\check B}(\tau ,\vec 
\sigma^{'} )\rbrace =\eta^{\check B}_{\check A}\delta_{ab}
\delta^3(\vec \sigma -\vec \sigma^{'}),\nonumber \\
&&\lbrace \eta^{\check r}_i(\tau ),\kappa_{j{\check s}}(\tau )\rbrace =-
\delta_{ij}\delta^{\check r}_{\check s},\nonumber \\
&&\lbrace \theta_{i\alpha}(\tau ),\pi_{\theta \, j\beta}(\tau )\rbrace =-\delta
_{ij}\delta_{\alpha\beta},
\nonumber \\
&&\lbrace \theta^{*}_{i\alpha}(\tau ),\pi_{\theta^{*} \, j\beta}(\tau )\rbrace 
=-\delta_{ij}\delta_{\alpha\beta}.
\label{b3}
\end{eqnarray}

The Grassmann momenta give rise to the second class constraints $\pi_{\theta
\, i\alpha}+{i\over 2}\theta^{*}_{i\alpha}\approx 0$, $\pi_{\theta^{*}\, 
i\alpha}+{i\over 2}\theta_{i\alpha}\approx 0$ [$\lbrace \pi_{\theta \, i\alpha}
+{i\over 2}\theta^{*}_{i\alpha},\pi_{\theta^{*}\, j\beta}+{i\over 2}\theta
_{j\beta}\rbrace =-i\delta_{ij}\delta_{\alpha\beta}$]; $\pi
_{\theta \, i\alpha}$ and $\pi_{\theta^{*}\, i\alpha}$ are then eliminated 
with the help of Dirac brackets

\begin{eqnarray}
\lbrace A,B\rbrace {}^{*}&=&\lbrace A,B\rbrace -i\sum_{\alpha =1}^3[\lbrace 
A,\pi_{\theta \, i\alpha}+{i\over 2}\theta^{*}_{i\alpha}\rbrace \lbrace \pi
_{\theta^{*}\, i\alpha}+{i\over 2}\theta_{i\alpha},B\rbrace  +\nonumber \\
&+&\lbrace A,\pi
_{\theta^{*}\, i\alpha}+{i\over 2}\theta_{i\alpha} \rbrace \lbrace \pi
_{\theta \, i\alpha}+{i\over 2}\theta^{*}_{i\alpha},B\rbrace ]
\label{b4}
\end{eqnarray}

\noindent so that the remaining Grassmann variables have the fundamental
Dirac brackets [which we will still denote $\lbrace .,.\rbrace$ for the sake of
simplicity]

\begin{eqnarray}
&&\lbrace \theta_{i\alpha}(\tau ),\theta_{j\beta}(\tau )\rbrace = \lbrace 
\theta_{i\alpha}^{*}(\tau ),\theta_{j\beta}^{*}(\tau )\rbrace =0,\nonumber \\
&&\lbrace \theta_{i\alpha}(\tau ),\theta_{j\beta}^{*}(\tau )\rbrace =-i\delta
_{ij}\delta_{\alpha\beta}.
\label{b5}
\end{eqnarray}

These equations imply $\lbrace Q_{ia},Q_{jb}\rbrace =\delta_{ij}
c_{abc}Q_{ic}$ and that the gauge transformations
$\delta \theta_{i\alpha}=(\alpha_a T^a)_{\alpha\beta}\theta_{i\beta}$, under
which the action is invariant, are generated by the $Q_{ia}$'s.

By varying the Lagrange multipliers $\lambda_i(\tau )$,
we get the Grassmann constraints mentioned in Section II

\begin{equation}
N_i(\tau )=\sum_{\alpha =1}^3 \theta^{*}_{i\alpha}(\tau )\theta_{i\alpha}(\tau )
\approx 0.
\label{b6}
\end{equation}

\noindent We could also treat the multipliers $\lambda_i(\tau )$ as 
configuration variables: we would get the first class constraints $\pi
_{\lambda_i}(\tau )\approx 0$ and we could get free of the $\lambda_i$'s by 
adding the gauge-fixings $\lambda_i(\tau )\approx 0$ and going to the Dirac
brackets for the resulting 2N second class constraints.

>From the expression of the momenta
we obtain the four primary constraints

\begin{eqnarray}
&{\cal H}_{\mu}&(\tau ,\vec \sigma )= \rho_{\mu}(\tau ,\vec \sigma )-
l_{\mu}(\tau ,\vec \sigma )[T_{\tau\tau}(\tau ,\vec \sigma )+
\nonumber \\
&+&\sum_{i=1}^N\delta^3(\vec \sigma -{\vec \eta}_i(\tau ))\times
\nonumber \\
&\eta_i&\sqrt{ m^2_i-\gamma^{{\check r}{\check s}}(\tau ,\vec \sigma )
[\kappa_{i{\check r}}(\tau )+\sum_aQ_{ia}(\tau )A_{a\check r}
(\tau ,\vec \sigma )][\kappa_{i{\check s}}(\tau ) +\sum_bQ_{ib}
(\tau )A_{b\check s}(\tau ,\vec \sigma )]  }\, \, ]-\nonumber \\
&-&z_{{\check r}\mu}(\tau ,\vec \sigma )\gamma^{{\check r}{\check s}}
(\tau ,\vec \sigma )\lbrace
-T_{\tau \check s}(\tau ,\vec \sigma )+\sum_{i=1}^N\delta^3(\vec \sigma -
{\vec \eta}_i(\tau ))[\kappa_{i{\check s}}+\sum_aQ_{ia}
(\tau )A_{a\check s}(\tau ,\vec \sigma )] \rbrace \approx 0,
\label{b7}
\end{eqnarray}

\noindent where

\begin{eqnarray}
T_{\tau \tau}(\tau ,\vec \sigma )
&=&-{1\over 2}\sum_a
({g_s^2\over {\sqrt {\gamma}} }\pi_a^{\check r}g_{{\check r}{\check s}}
\pi_a^{\check s}-{ {\sqrt {\gamma}}\over {2g_s^2}}\gamma^{{\check r}{\check s}}
\gamma^{{\check u}{\check v}}
F_{a{\check r}{\check u}}F_{a{\check s}{\check v}})
(\tau ,\vec \sigma ),\nonumber \\
T_{\tau {\check s}}(\tau ,\vec \sigma )&=&-
\sum_aF_{a{\check s}{\check t}}(\tau ,\vec 
\sigma )\pi_a^{\check t}(\tau ,\vec \sigma )=-\epsilon_{{\check s}{\check t}
{\check u}}\sum_a\pi_a^{\check t}(\tau ,\vec \sigma )B_{a\check u}(\tau ,\vec 
\sigma )=\nonumber \\
&=&\sum_a{[{\vec \pi}_a (\tau ,\vec \sigma )\times 
{\vec B}_a(\tau ,\vec \sigma )]}_{\check s},
\label{b8}
\end{eqnarray}

\noindent are the energy density and the Poynting vector respectively. We use
the notation $\sum_a({\vec \pi}_a \times {\vec B}_a)_{\check s}=({\vec E}_a
\times {\vec B}_a)_{\check s}$ because it is consistent
with  $\epsilon_{{\check s}{\check t}{\check u}}\sum_a\pi_a^{\check t}
B_{a\check u}$ in the flat metric limit 
$g_{{\check A}{\check B}}\rightarrow \eta _{{\check A}
{\check B}}$; in this limit $T_{\tau\tau}\rightarrow {1\over 2}\sum_a({\vec E}
_a^2+{\vec B}_a^2)$.

Since the canonical Hamiltonian is (we assume boundary conditions for the
electromagnetic potential such that all the surface terms can be neglected;
see Ref.\cite{lusa}) 

\begin{eqnarray}
H_c&=&-\sum_{i=1}^N\kappa_{i{\check r}}(\tau ){\dot \eta}_i^{\check r}(\tau )+
\int d^3\sigma [\sum_a\pi_a^{\check A}(\tau ,\vec \sigma )\partial_{\tau}
A_{a\check A}
(\tau ,\vec \sigma )-\rho_{\mu}(\tau ,\vec \sigma )z^{\mu}_{\tau}(\tau ,\vec
\sigma )-{\cal L}(\tau ,\vec \sigma )]=\nonumber \\
&=&\int d^3\sigma \sum_a[\partial_{\check r}(\pi_a^{\check r}(\tau ,\vec 
\sigma )A_{a\tau}(\tau ,\vec \sigma ))-A_{a\tau}(\tau ,\vec \sigma )\Gamma_a
(\tau ,\vec \sigma )]=\nonumber \\
&=&-\int d^3\sigma \sum_aA_{a\tau}(\tau ,\vec \sigma )
\Gamma_a (\tau ,\vec \sigma ),
\label{b9}
\end{eqnarray}

\noindent with [$\partial_{\check r}={{\partial}\over {\partial \sigma
^{\check r}}}=-\partial^{\check r}$, $\vec \partial =\{ \partial^{\check r}\}$,
$\triangle -{\vec \partial}^2$, ${\hat {\vec D}}^{(A)}_{ab}=\{ \delta_{ab}
\partial^{\check r}+c_{acb}A^{\check r}_c\}$ ]

\begin{eqnarray}
\Gamma_a(\tau ,\vec \sigma )&=&\partial_{\check r}\pi_a^{\check r}
(\tau ,\vec \sigma )+c_{abc}
A_{b\check r}(\tau ,\vec \sigma )\pi_c^{\check r}(\tau ,\vec \sigma )
+\sum_{i=1}^NQ_{ia}(\tau )\delta^3(\vec \sigma -{\vec \eta}_i(\tau ))=
\nonumber \\
&=&-{\hat {\vec D}}^{(A)}_{ab}(\tau ,\vec \sigma )\cdot {\vec \pi}_b(\tau ,\vec 
\sigma )+\sum_{i=1}^NQ_{ia}(\tau )\delta^3(\vec \sigma -{\vec \eta}_i(\tau )),
\label{b10}
\end{eqnarray}

\noindent we have the Dirac Hamiltonian ($\lambda^{\mu}(\tau ,\vec \sigma )$, 
$\lambda_{\tau}(\tau ,\vec \sigma )$  and $\mu_i(\tau )$
are Dirac's multipliers)

\begin{eqnarray}
H_D&=&\int d^3\sigma [\lambda^{\mu}(\tau ,\vec \sigma ){\cal H}_{\mu}(\tau ,
\vec  \sigma )+\sum_a
\lambda_{a\tau}(\tau ,\vec \sigma )\pi_a^{\tau}(\tau ,\vec \sigma )-\sum_a
A_{a\tau}(\tau ,\vec \sigma )\Gamma_a (\tau ,\vec \sigma )]+\nonumber \\
&+&\sum_{i=1}^N\mu_i(\tau )N_i(\tau ).
\label{b11}
\end{eqnarray}

The Lorentz scalar constraints $\pi_a^{\tau}(\tau ,\vec \sigma )\approx 0$ are
generated by the electromagnetic gauge invariance of the action; their time 
constancy will produce the only secondary constraints (Gauss laws) 

\begin{equation}
\Gamma_a (\tau ,\vec \sigma )\approx 0.
\label{b12}
\end{equation}

The six constraints ${\cal H}_{\mu}(\tau ,\vec \sigma )\approx 0$, $\pi_a^{\tau}
(\tau ,\vec \sigma )\approx 0$, $\Gamma_a (\tau ,\vec \sigma )\approx 0$ are
first class with the only non vanishing Poisson brackets

\begin{eqnarray}
\lbrace {\cal H}_{\mu}(\tau ,\vec \sigma )&,&{\cal H}_{\nu}(\tau ,{\vec 
\sigma}^{'} )\rbrace =\nonumber \\
&=&\lbrace [l_{\mu}(\tau ,\vec \sigma )z_{{\check r}\nu}(\tau ,\vec \sigma )
-l_{\nu}(\tau ,\vec \sigma )z_{{\check r}\mu}(\tau ,\vec \sigma )]
{ {\pi^{\check r}(\tau ,\vec 
\sigma )}\over {\sqrt{\gamma (\tau ,\vec \sigma )}} }-\nonumber \\
&-&z_{\check u\mu}(\tau ,\vec \sigma )\gamma^{{\check u}{\check r}}(\tau ,\vec 
\sigma )
\sum_aF_{a{\check r}{\check s}}(\tau ,\vec \sigma )\gamma^{{\check s}{\check 
v}}(\tau ,\vec \sigma )z_{\check v\nu}(\tau ,\vec \sigma )\rbrace
\Gamma_a (\tau ,\vec \sigma )\delta^3(\vec \sigma -{\vec \sigma}^{'})\approx 0.
\label{b13}
\end{eqnarray}

\noindent Moreover, since $\lbrace Q_{ia},N_j\rbrace =0$, also the constraints
$N_i\approx 0$ are first class and, being constants of the motion for each i,
they do not generate secondaries.

Let us remark that the simplicity of Eqs.(\ref{b13}) is due to the 
use of Cartesian coordinates: had we used the constraints ${\cal H}_l(\tau ,
\vec \sigma )=l^{\mu}(\tau ,\vec \sigma ){\cal H}_{\mu}(\tau ,\vec \sigma )$, 
${\cal H}_{\check r}(\tau ,\vec \sigma )=z^{\mu}_{\check r}(\tau ,\vec \sigma )
{\cal H}_{\mu}(\tau ,\vec \sigma )$ (i.e. nonholonomic coordinates), so that 
their associated Dirac multipliers $\lambda_l(\tau ,\vec \sigma )$, $\lambda
_{\check r}(\tau ,\vec \sigma )$ would have been
the lapse and shift functions of general relativity, one would have obtained
the universal algebra of Ref.\cite{dira1}.

The ten conserved Poincar\'e generators are

\begin{eqnarray}
P^{\mu}=
p_s^{\mu}&=&\int d^3\sigma \, \rho^{\mu}(\tau ,\vec \sigma ),\nonumber \\
J^{\mu\nu}=
J_s^{\mu\nu}&=&\int d^3\sigma \, (z^{\mu}(\tau ,\vec \sigma )\rho^{\nu}
(\tau ,\vec \sigma )-z^{\nu}(\tau ,\vec \sigma )\rho^{\mu}(\tau ,\vec \sigma )),
\label{b14}
\end{eqnarray}

\noindent so that the total momentum is built starting from the existing
energy momentum densities on the hypersurface

\begin{eqnarray}
&\int& d^3\sigma {\cal H}^{\mu}(\tau ,\vec \sigma )=p^{\mu}_s-
\int d^3\sigma l_{\mu}(\tau ,\vec \sigma )[T_{\tau\tau}(\tau ,\vec \sigma )+
\nonumber \\
&+&\sum_{i=1}^N\delta^3(\vec \sigma -{\vec \eta}_i(\tau ))\times
\nonumber \\
&\eta_i&\sqrt{ m^2_i-\gamma^{{\check r}{\check s}}(\tau ,\vec \sigma )
[\kappa_{i{\check r}}(\tau )+\sum_aQ_{ia}(\tau )A_{a\check r}
(\tau ,\vec \sigma )][\kappa_{i{\check s}}(\tau )+\sum_bQ_{ib}
(\tau )A_{b\check s}(\tau ,\vec \sigma )]  }]-\nonumber \\
&-&\int d^3\sigma
z_{{\check r}\mu}(\tau ,\vec \sigma )\gamma^{{\check r}{\check s}}
(\tau ,\vec \sigma )\lbrace
T_{\tau \check s}(\tau ,\vec \sigma )+\nonumber \\
&+&\sum_{i=1}^N\delta^3(\vec \sigma -
{\vec \eta}_i(\tau ))[\kappa_{i{\check s}}(\tau )+\sum_aQ_{ia}
(\tau )A_{a\check s}(\tau ,\vec \sigma )] \rbrace \approx 0.
\label{b15}
\end{eqnarray}

We add the gauge-fixings to arbitrary hyperplanes as in Ref.\cite{lus1}

\begin{equation}
\zeta^{\mu}(\tau ,\vec \sigma )=z^{\mu}(\tau ,\vec \sigma )-x_s^{\mu}(\tau )-
b^{\mu}_{\check r}(\tau )\sigma^{\check r}\approx 0,
\label{b16}
\end{equation}

\noindent where $b^{\mu}_{\check r}(\tau )$, ${\check r}=1,2,3$, are three
orthonormal vectors such that the constant (future pointing) normal to the
hyperplane is

\begin{equation}
l^{\mu}(\tau ,\vec \sigma )\approx l^{\mu}=b^{\mu}_{\tau}=\epsilon^{\mu}
{}_{\alpha\beta\gamma}b^{\alpha}_{\check 1}(\tau )b^{\beta}_{\check 2}(\tau )
b^{\gamma}_{\check 3}(\tau ).
\label{b17}
\end{equation}

Therefore, we get

\begin{eqnarray}
&&z^{\mu}_{\check r}(\tau ,\vec \sigma )\approx b^{\mu}_{\check r}(\tau ),
\nonumber \\
&&z^{\mu}_{\tau}(\tau ,\vec \sigma )\approx {\dot x}^{\mu}_s(\tau )+{\dot
b}^{\mu}_{\check r}(\tau )\sigma^{\check r},\nonumber \\
&&g_{{\check r}{\check s}}(\tau ,\vec \sigma )\approx -\delta_{{\check
r}{\check s}},\quad\quad \gamma^{{\check r}{\check s}}(\tau ,\vec \sigma )
\approx -\delta^{{\check r}{\check s}},\quad\quad \gamma (\tau ,\vec \sigma )
\approx 1.
\label{b18}
\end{eqnarray}

By introducing the Dirac brackets for the resulting second class constraints
[now we have $\lbrace \eta^{\check r}_i(\tau ),\kappa_j^{\check s}(\tau )
\rbrace =\delta_{ij}\delta^{\check r\check s}$]

\begin{equation}
\lbrace A,B\rbrace {}^{*}=\lbrace A,B\rbrace -\int d^3\sigma [\lbrace A,\zeta
^{\mu}(\tau ,\vec \sigma )\rbrace \lbrace {\cal H}_{\mu}(\tau ,\vec \sigma ),
B\rbrace -\lbrace A,{\cal H}_{\mu}(\tau ,\vec \sigma )\rbrace \lbrace \zeta
^{\mu}(\tau ,\vec \sigma ),B\rbrace ],
\label{b19}
\end{equation}

\noindent we find $\lbrace x_s^{\mu}(\tau ),p^{\nu}_s(\tau )\rbrace {}^{*}=
-\eta^{\mu\nu}$ [with the assumumption 
$\lbrace b^{\mu}_{\check r}(\tau ), p^{\nu}_s\rbrace =0$].

The ten degrees of freedom describing the hyperplane are $x^{\mu}_s(\tau )$
with conjugate momentum $p^{\mu}_s$ and six variables $\phi_{\lambda}(\tau )$,
$\lambda =1,..,6$, which parametrize the orthonormal tetrad $b^{\mu}_{\check A}
(\tau )$, with their conjugate momenta $T_{\lambda}(\tau )$.

The preservation of the gauge-fixings $\zeta^{\mu}(\tau ,\vec \sigma )
\approx 0$ in time implies

\begin{equation}
{d\over {d\tau}}\zeta^{\mu}(\tau ,\vec \sigma )=\lbrace \zeta^{\mu}(\tau ,\vec 
\sigma ),H_D\rbrace =-\lambda^{\mu}(\tau ,\vec \sigma )-{\dot x}^{\mu}_s(\tau
)-{\dot b}^{\mu}_{\check r}(\tau )\sigma^{\check r}\approx 0,
\label{b20}
\end{equation}

\noindent so that one has [using ${\dot b}^{\mu}_{\tau}=0$ and ${\dot b}
^{\mu}_{\check r}(\tau )b^{\nu}_{\check r}(\tau )=-b^{\mu}_{\check r}(\tau )
{\dot b}^{\nu}_{\check r}(\tau )$]

\begin{eqnarray}
\lambda^{\mu}(\tau ,\vec \sigma )&\approx&{\tilde \lambda}^{\mu}(\tau )+
{\tilde \lambda}^{\mu}{}_{\nu}(\tau )b^{\nu}_{\check r}(\tau )
\sigma^{\check r},\nonumber \\
&&{\tilde \lambda}^{\mu}(\tau )=-{\dot x}^{\mu}_s(\tau ),\nonumber \\
&&{\tilde \lambda}^{\mu\nu}(\tau )=-{\tilde \lambda}^{\nu\mu}(\tau )={1\over 2}
[{\dot b}^{\mu}_{\check r}(\tau )b^{\nu}_{\check r}(\tau )-b^{\mu}_{\check r}
(\tau ){\dot b}^{\nu}_{\check r}(\tau )].
\label{b21}
\end{eqnarray}

Thus, the Dirac Hamiltonian becomes

\begin{equation}
H_D={\tilde \lambda}^{\mu}(\tau ){\tilde {\cal H}}_{\mu}(\tau )-{1\over 2}
{\tilde \lambda}^{\mu\nu}(\tau ){\tilde {\cal H}}_{\mu\nu}(\tau )+\sum_{i=1}^N
\mu_i(\tau )N_i(\tau ),
\label{b22}
\end{equation}

\noindent and only the following  first class constraints are left
[now we remain with the variables $x^{\mu}_s, p^{\mu}_s,b^{\mu}_{\check
A}, S_s^{\mu\nu}, A_{\check A}, \pi^{\check A}, {\vec \eta}_i, {\vec \kappa}_i,
\theta_{i\alpha}, \theta^{*}_{i\alpha}$]

\begin{eqnarray}
{\tilde {\cal H}}^{\mu}(\tau )
&=&\int d^3\sigma {\cal H}^{\mu}(\tau ,\vec \sigma )=
p^{\mu}_s-l_{\mu}\lbrace {1\over 2}\int d^3\sigma \sum_a
[g_s^2{\vec \pi}_a^2(\tau ,\vec
\sigma )+g_s^{-2}{\vec B}_a^2(\tau ,\vec \sigma )]+\nonumber \\
&+&\sum_{i=1}^N\eta_i\sqrt{ m^2_i+
{[{\vec \kappa}^2_i(\tau )+\sum_aQ_{ia}(\tau ){\vec A}_a
(\tau ,{\vec \eta}_i(\tau ))]}^2 }\rbrace -\nonumber \\
&-&b_{\check r}^{\mu}(\tau )\lbrace \int d^3\sigma \sum_a
{[{\vec \pi}_a (\tau ,\vec 
\sigma )\times {\vec B}_a(\tau ,\vec \sigma )]}_{\check r}+\sum_{i=1}^N
[\kappa_{i{\check r}}(\tau )+\sum_aQ_{ia}
(\tau )A_{a\check r}(\tau ,{\vec \eta}_i(\tau ))] \rbrace 
\approx 0,\nonumber \\
{\tilde {\cal H}}^{\mu\nu}(\tau )&=& b^{\mu}_{\check r}(\tau )\int d^3\sigma
\sigma^{\check r}\, {\cal H}^{\nu}(\tau ,\vec \sigma )-b^{\nu}_{\check r}(\tau )
\int d^3\sigma \sigma^{\check r}\, {\cal H}^{\mu}(\tau ,\vec \sigma )=
\nonumber \\
&=&S_s^{\mu\nu}(\tau )-[b^{\mu}_{\check r}(\tau )b^{\nu}_{\tau}-b^{\nu}_{\check
r}(\tau )b^{\mu}_{\tau}]\, [{1\over 2}\int d^3\sigma \sigma^{\check r}\, \sum_a
[g^2_s{\vec \pi}_a^2(\tau ,\vec \sigma )+g^{-2}_s{\vec B}_a^2(\tau ,\vec 
\sigma )]+\nonumber \\
&+&\sum_{i=1}^N\eta_i^{\check r}(\tau )\eta_i
\sqrt{m^2_i+[{\vec \kappa}_i(\tau )+\sum_aQ_{ia}(\tau )
{\vec A}_a(\tau ,{\vec \eta}_i(\tau ))]{}^2 }]+\nonumber \\
&+&[b^{\mu}_{\check r}(\tau )b^{\nu}_{\check s}(\tau )-b^{\nu}_{\check r}(\tau )
b^{\mu}_{\check s}(\tau )]\, [\int d^3\sigma \sigma^{\check r}\, \sum_a
{[{\vec \pi}_a (\tau ,\vec \sigma )\times {\vec B}_a(\tau ,\vec \sigma )]}
_{\check s}+\nonumber \\
&+&\sum_{i=1}^N\eta_i^{\check r}(\tau )[\kappa_i^{\check s}(\tau )+\sum_a
Q_{ia}(\tau )A_a^{\check s}(\tau ,{\vec \eta}_i(\tau ))]\,\, ]
\approx 0,\nonumber \\
\pi_a^{\tau}(\tau ,\vec \sigma )&\approx& 0,\nonumber \\
\Gamma_a (\tau ,\vec \sigma )&\approx& 0,\nonumber \\
N_i&\approx& 0.
\label{b23}
\end{eqnarray}

Here $S^{\mu\nu}_s$ is the spin part of the Lorentz generators

\begin{eqnarray}
J^{\mu\nu}_s&=&x^{\mu}_sp^{\nu}_s-x^{\nu}_sp^{\mu}_s+S^{\mu\nu}_s,\nonumber \\
&&S^{\mu\nu}_s=b^{\mu}_{\check r}(\tau )\int d^3\sigma \sigma^{\check r}
\rho^{\nu}(\tau ,\vec \sigma )-b^{\nu}_{\check r}(\tau )\int d^3\sigma
\sigma^{\check r}\rho^{\mu}(\tau ,\vec \sigma ).
\label{b24}
\end{eqnarray}

As shown in Ref.\cite{lus1} instead of finding
$\phi_{\lambda}(\tau ), T_{\lambda}(\tau )$, one can use the redundant
variables $b^{\mu}_{\check A}(\tau ), S_s^{\mu\nu}(\tau )$, with the
following Dirac brackets assuring the validity of the orthonormality
condition $\eta^{\mu\nu}-b^{\mu}_{\check A}\eta^{{\check A}{\check b}}b^{\nu}
_{\check B}=0$ [$C^{\mu\nu\alpha\beta}_{\gamma\delta}=\eta^{\nu}_{\gamma}
\eta^{\alpha}_{\delta}\eta^{\mu\beta}+\eta^{\mu}_{\gamma}\eta^{\beta}_{\delta}
\eta^{\nu\alpha}-\eta^{\nu}_{\gamma}\eta^{\beta}_{\delta}\eta^{\mu\alpha}-
\eta^{\mu}_{\gamma}\eta^{\alpha}_{\delta}\eta^{\nu\beta}$
are the structure constants of the Lorentz group]

\begin{eqnarray}
&&\lbrace S_s^{\mu\nu},b^{\rho}_{\check A}\rbrace {}^{*}=\eta^{\rho\nu}
b^{\mu}_{\check A}-\eta^{\rho\mu}b^{\nu}_{\check A}\nonumber \\
&&\lbrace S^{\mu\nu}_s,S_s^{\alpha\beta}\rbrace {}^{*}=C^{\mu\nu\alpha\beta}
_{\gamma\delta}S_s^{\gamma\delta},
\label{b25}
\end{eqnarray}

The Dirac brackets of the left constraints are

\begin{eqnarray}
&&\lbrace {\tilde {\cal H}}^{\mu}(\tau ),{\tilde {\cal H}}^{\nu}
(\tau )\rbrace {}^{*}=\sum_a
\int d^3\sigma \lbrace [b^{\mu}_{\tau}b^{\nu}_{\check r}
(\tau )-b^{\nu}_{\tau}b^{\mu}_{\check r}(\tau )] g^2_s \pi_{a\check r}(\tau ,
\vec \sigma )-\nonumber \\
&&-b^{\mu}_{\check r}(\tau )F_{a{\check r}{\check s}}(\tau ,\vec \sigma )
b^{\nu}_{\check s}(\tau )\rbrace \Gamma_a (\tau ,\vec \sigma ),\nonumber \\
&&\lbrace {\tilde {\cal H}}^{\mu}(\tau ),{\tilde {\cal H}}^{\alpha
\beta}(\tau )\rbrace {}^{*}=-\sum_a\int d^3\sigma \, \sigma^{\check t}\, \lbrace
[(b^{\alpha}_{\check r}(\tau )b^{\beta}_{\check t}(\tau )-b^{\alpha}_{\check t}
(\tau )b^{\beta}_{\check r}(\tau ))b^{\mu}_{\tau}-\nonumber \\
&-&(b^{\alpha}_{\tau}b^{\beta}_{\check t}(\tau )-b^{\beta}_{\tau}b^{\alpha}
_{\check t}(\tau ))b^{\mu}_{\check r}(\tau )]
g^2_s\pi^{a\check r}(\tau ,\vec \sigma )+\nonumber \\
&&+b^{\mu}_{\check r}(\tau )F_{a{\check r}{\check s}}(\tau ,\vec \sigma )
[b^{\alpha}_{\check t}(\tau )b^{\beta}_{\check s}(\tau )-b^{\beta}_{\check t}
(\tau )b^{\alpha}_{\check s}(\tau )]\rbrace
\Gamma_a (\tau ,\vec \sigma ),\nonumber \\
&&\lbrace {\tilde {\cal H}}^{\mu\nu}(\tau ),{\tilde {\cal H}}
^{\alpha\beta}(\tau )\rbrace {}^{*}=C^{\mu\nu\alpha\beta}
_{\gamma\delta}{\tilde {\cal H}}^{\gamma\delta}(\tau )+\sum_a\int d^3\sigma \,
\sigma^{\check u}\, \sigma^{\check v}\nonumber \\
&&\lbrace b^{\mu}_{\check u}(\tau )b^{\alpha}_{\check v}(\tau )([b^{\nu}
_{\tau}b^{\beta}_{\check r}(\tau )-b^{\beta}_{\tau}b^{\nu}_{\check r}(\tau )]
g^2_s\pi_{a\check r}(\tau ,\vec \sigma )-
b^{\nu}_{\check r}(\tau )F_{a{\check r}{\check
s}}(\tau ,\vec \sigma )b^{\beta}_{\check s}(\tau ))-\nonumber \\
&&-b^{\mu}_{\check u}(\tau )b^{\beta}_{\check v}(\tau )([b^{\nu}
_{\tau}b^{\alpha}_{\check r}(\tau )-b^{\alpha}_{\tau}b^{\nu}_{\check r}(\tau )]
g^2_s\pi_{a\check r}(\tau ,\vec \sigma )-
b^{\nu}_{\check r}(\tau )F_{a{\check r}{\check
s}}(\tau ,\vec \sigma )b^{\alpha}_{\check s}(\tau ))-\nonumber \\
&&-b^{\nu}_{\check u}(\tau )b^{\alpha}_{\check v}(\tau )([b^{\mu}
_{\tau}b^{\beta}_{\check r}(\tau )-b^{\beta}_{\tau}b^{\mu}_{\check r}(\tau )]
g^2_s\pi_{a\check r}(\tau ,\vec \sigma )-
b^{\mu}_{\check r}(\tau )F_{a{\check r}{\check
s}}(\tau ,\vec \sigma )b^{\beta}_{\check s}(\tau ))+\nonumber \\
&&+b^{\nu}_{\check u}(\tau )b^{\beta}_{\check v}(\tau )([b^{\mu}
_{\tau}b^{\alpha}_{\check r}(\tau )-b^{\alpha}_{\tau}b^{\mu}_{\check r}(\tau )]
g^2_s\pi_{a\check r}(\tau ,\vec \sigma )-
b^{\mu}_{\check r}(\tau )F_{a{\check r}{\check
s}}(\tau ,\vec \sigma )b^{\alpha}_{\check s}(\tau ))\rbrace
\Gamma_a (\tau ,\vec \sigma ),
\label{b26}
\end{eqnarray}

Let us now restrict ourselves to configurations with $p_s^2 > 0$ and let us
use the Wigner boost $L^{\mu}{}_{\nu}(\stackrel{\circ}{p_s},
p_s)$ to boost to rest the variables $b^{\mu}_{\check A}$, $S_s^{\mu\nu}$ of 
the following non-Darboux basis 

${}$

$x^{\mu}_s, p^{\mu}_s, b^{\mu}_{\check A}, S_s^{\mu\nu}, \eta_i^{\check r},
\kappa_i^{\check r}$ 

${}$

\noindent of the Dirac brackets. The following new non-Darboux basis is obtained
[${\tilde x}^{\mu}_s$ is no more a 4-vector]

\begin{eqnarray}
&&{\tilde x}^{\mu}_s=x_s^{\mu}+{1\over 2}\, \epsilon^A_{\nu}(u(p_s))\eta_{AB}
{ {\partial \epsilon^B_{\rho}(u(p_s))}\over {\partial p_{s\mu}} }\, 
S^{\rho\nu}_s=\nonumber \\
&&=x_s^{\mu}-{ 1\over {\eta \sqrt{p_s^2}(p_s^o+\eta \sqrt{p_s^2})} }\, [p_{s\nu}
S^{\nu\mu}_s+\eta \sqrt{p_s^2} (S_s^{o\mu}-S_s^{o\nu}{ {p_{s\nu}p_s^{\mu}}\over 
{p_s^2} })]=\nonumber \\
&&=x_s^{\mu}-{1\over {\eta_s \sqrt{p_s^2}} }[\eta^{\mu}_A({\bar S}_s^{\bar oA}
-{ {{\bar S}_s^{Ar}p_s^r}\over {p_s^o+\eta_s \sqrt{p_s^2}} })+{ {p_s^{\mu}+2
\eta_s \sqrt{p_s^2}\eta^{\mu o}}\over {\eta_s \sqrt{p_s^2}(p_s^o+\eta_s 
\sqrt{p_s^2})} }{\bar S}_s^{\bar or}p_s^r],\nonumber \\
&&{}\nonumber \\
&&p^{\mu}_s=p_s^{\mu},\nonumber \\
&&{}\nonumber \\
&&\eta_i^{\check r}=\eta_i^{\check r},\nonumber \\
&&\kappa_i^{\check r}=\kappa_i^{\check r},\nonumber \\
&&{}\nonumber \\
&&b^A_{\check r}=\epsilon^A_{\mu}(u(p_s))b^{\mu}_{\check r},\nonumber \\
&&{\tilde S}_s^{\mu\nu}=S^{\mu\nu}_s-{1\over 2}\epsilon^A_{\rho}(u(p_s))\eta
_{AB}({ {\partial \epsilon^B_{\sigma}(u(p_s))}\over {\partial p_{s\mu}} }\, 
p^{\nu}_s-{ {\partial \epsilon^B_{\sigma}(u(p_s))}\over {\partial p_{s\nu}} }\,
 p_s^{\mu})S^{\rho\sigma}_s=\nonumber \\
&&=S^{\mu\nu}_s+{1\over {\eta \sqrt{p_s^2}(p_s^o+\eta \sqrt{p_s^2})} } 
[p_{s\beta}(S^{\beta\mu}_sp_s^{\nu}-S_s^{\beta\nu}p_s^{\mu})+\eta \sqrt{p_s^2}
(S_s^{o\mu}p_s^{\nu}-S^{o\nu}_sp_s^{\mu})],\nonumber \\
&&{}\nonumber \\
&&J^{\mu\nu}_s={\tilde x}_s^{\mu}p_s^{\nu}-{\tilde x}^{\nu}_sp_s^{\mu}+{\tilde
S}_s^{\mu\nu}={\tilde L}^{\mu\nu}_s+{\tilde S}^{\mu\nu}_s.
\label{b27}
\end{eqnarray}

We have 

\begin{eqnarray}
&&\lbrace {\tilde x}^{\mu}_s,p^{\nu}_s\rbrace {}^{*}=-\eta^{\mu\nu},\nonumber \\
&&\lbrace {\tilde S}^{oi}_s,b^r_{\check A}\rbrace {}^{*}={ {\delta^{is}(p^r_s
b^s_{\check A}-p^s_sb^r_{\check A})}\over {p^o_s+\eta_s\sqrt{p^2_s}} },
\nonumber \\
&&\lbrace {\tilde S}_s^{ij},b^r_{\check A}\rbrace {}^{*}=(\delta^{ir}\delta
^{js}-\delta^{is}\delta^{jr})b^s_{\check A},\nonumber \\
&&\lbrace {\tilde S}_s^{\mu\nu},{\tilde S}_s^{\alpha\beta}\rbrace {}^{*}=
C^{\mu\nu\alpha\beta}_{\gamma\delta}{\tilde S}_s^{\gamma\delta},
\label{b28}
\end{eqnarray}

As shown in Ref.\cite{lus1}, under Poincar\'e transformations $(a,\Lambda )$ 
we get

\begin{eqnarray}
p^{'\mu}&=&\Lambda^{\mu}{}_{\nu} p^{\nu},\nonumber \\
{\tilde x}^{'\mu}_s&=&\Lambda^{\mu}{}_{\nu}[{\tilde x}^{\nu}_s+{1\over 2}{\bar S}
_{s,rs}R^r{}_k(\Lambda ,p_s){{\partial}\over {\partial p_{s\nu}} }R^s{}_k(\Lambda ,
p_s)]+a^{\mu}=\nonumber \\
&=&\Lambda^{\mu}{}_{\nu}\{ {\tilde x}^{\nu}_s+{ {{\bar S}_{s,rs}}\over {\Lambda^o
{}_{\alpha}p^{\alpha}_s+\eta \sqrt{p^2_s}} }[\eta^{\nu}_r(\Lambda^o{}_s-
{ {(\Lambda^o{}_o-1)p_{s,s}}\over {p^o_s+\eta \sqrt{p^2_s}} })-\nonumber \\
&-&{ {(p^{\nu}_s+\eta^{\nu}_o\eta \sqrt{p^2_s})p_{s,r}\Lambda^o{}_s}\over 
{\eta \sqrt{p^2_s}(p^o_s+\eta \sqrt{p^2_s})} }] \} +a^{\mu}.
\label{b28a}
\end{eqnarray}

\noindent Therefore, ${\tilde x}^{\mu}_s$ is not a 4-vector: its infinitesimal
transformation properties under Lorentz transformations generated by $J^{\mu\nu}
_s$ are

\begin{eqnarray}
\lbrace {\tilde x}^{\mu}_s,J^{\alpha\beta}_s\rbrace &=&\eta^{\mu\alpha}{\tilde 
x}^{\beta}_s-\eta^{\mu\beta}{\tilde x}^{\alpha}_s+\lbrace {\tilde x}_s^{\mu},
{\tilde S}_s^{\alpha\beta}\rbrace ,\nonumber \\
&&\lbrace {\tilde x}^{\mu}_s,{\tilde S}^{oi}_s\rbrace =-{1\over {p^o_s+\eta 
\sqrt{p^2_s}} }[\eta^{\mu j}{\tilde S}^{ji}_s+{ {(p^{\mu}_s+\eta^{\mu o}\eta
\sqrt{p^2_s}){\tilde S}^{ik}_sp_s^k}\over {\eta \sqrt{p^2_s}(p^o_s+\eta 
\sqrt{p^2_s})} }],\nonumber \\
&&\lbrace {\tilde x}^{\mu}_s,{\tilde S}_s^{ij}\rbrace =0.
\label{b29b}
\end{eqnarray}

We can define [the
new variables are ${\tilde x}^{\mu}_s, p_s^{\mu}, b^A_{\check A}, {\tilde S}_s
^{\mu\nu}, A_{a\check A}, \pi_a^{\check A}, {\vec \eta}_i, {\vec \kappa}_i, 
\theta_{i\alpha}, \theta_{i\alpha}^{*}$]

\begin{eqnarray}
{\bar S}_s^{AB}&=&\epsilon^A_{\mu}(u(p_s))\epsilon^B_{\nu}(u(p_s))S_s^{\mu\nu}
\approx \nonumber \\
&\approx&[b^A_{\check r}(\tau )b^B_{\tau}-b^B_{\check r}(\tau )b^A_{\tau}]\,
[{1\over 2}\int d^3\sigma \sigma^{\check r}\, \sum_a
[g^2_s{\vec \pi}_a^2(\tau ,\vec \sigma )+
g^{-2}_s{\vec B}_a^2(\tau ,\vec \sigma )]+\nonumber \\
&+&\sum_{i=1}^N\eta_i^{\check r}(\tau )\eta_i
\sqrt{m^2_i+[{\vec \kappa}_i(\tau )+\sum_aQ_{ia}(\tau )
{\vec A}_a(\tau ,{\vec \eta}_i(\tau ))]{}^2 }]-\nonumber \\
&-&[b^A_{\check r}(\tau )b^B_{\check s}(\tau )-b^B_{\check r}(\tau )b^A
_{\check s}(\tau )]\, [\int d^3\sigma \sigma^{\check r}\, \sum_a
{[{\vec \pi}_a (\tau ,\vec \sigma )\times {\vec B}_a(\tau ,\vec \sigma )]}
_{\check s}+\nonumber \\
&+&\sum_{i=1}^N\eta_i^{\check r}(\tau )[\kappa_i^{\check s}(\tau )+\sum_a
Q_{ia}(\tau )A_a^{\check s}(\tau ,{\vec \eta}_i(\tau ))]\,\, ].
\label{b29}
\end{eqnarray}

Let us now add six more gauge-fixings by selecting the special family of
spacelike hyperplanes orthogonal to $p^{\mu}_s$ (this is possible for
$p^2_s > 0$), which can be called the `Wigner foliation' of Minkowski
spacetime. This can be done by requiring (only six conditions are
independent)

\begin{eqnarray}
T^{\mu}_{\check A}(\tau )&=&b^{\mu}_{\check A}(\tau )-\epsilon^{\mu}
_{A={\check A}}(u(p_s))\approx 0\nonumber \\
&&\Rightarrow \quad b^A_{\check A}(\tau )=\epsilon^A_{\mu}(u(p_s))b^{\mu}
_{\check A}(\tau )\approx \eta^A_{\check A}.
\label{b30}
\end{eqnarray}

Now the tetrad $b^{\mu}_{\check A}$ has become $\epsilon^{\mu}_A(u(p_s))$ and
the indices `${\check r}$' are forced to coincide with the Wigner spin-1 indices
`r', while $\bar o=\tau$ is a Lorentz-scalar index. 
The final Wigner-covariant variables are ${\tilde x}^{\mu}_s,
p^{\mu}_s, A_{a\tau}, \pi_a^{\tau}, A_{ar}, \pi_a^r, \eta^r_i, \kappa_{ir}, 
\theta_{i\alpha}, \theta^{*}_{i\alpha}$. One has

\begin{eqnarray}
{\bar S}_s^{AB}&\approx& (\eta^A_{\check r}\eta^B_{\tau}-\eta^B_{\check r}\eta^A
_{\tau})\, [{1\over 2}\int d^3\sigma \sigma^{\check r}\, \sum_a
[g^2_s{\vec \pi}_a^2 (\tau ,\vec \sigma )+
g^{-2}_s{\vec B}_a^2(\tau ,\vec \sigma )]+\nonumber \\
&+&\sum_{i=1}^N\eta_i^{\check r}(\tau )\eta_i
\sqrt{m^2_i+[{\vec \kappa}_i(\tau )+\sum_aQ_{ia}(\tau )
{\vec A}_a(\tau ,{\vec \eta}_i(\tau ))]{}^2 }]-\nonumber \\
&-&(\eta^A_{\check r}\eta^B_{\check s}-\eta^B_{\check r}\eta^A_{\check s})\,
[\int d^3\sigma \sigma^{\check r}\, \sum_a
{[{\vec \pi}_a (\tau ,\vec \sigma )\times {\vec B}_a(\tau ,\vec \sigma )]}
_{\check s}+\nonumber \\
&+&\sum_{i=1}^N\eta_i^{\check r}(\tau )[\kappa_i^{\check s}(\tau )+\sum_a
Q_{ia}(\tau )A_a^{\check s}(\tau ,{\vec \eta}_i(\tau ))]\,\, ],
\nonumber \\
{\bar S}_s^{rs}&\approx& 
\sum_{i=1}^N(\eta_i^r(\tau )[\kappa_i^s(\tau )+\sum_a
Q_{ia}(\tau )A_a^s(\tau ,{\vec \eta}_i(\tau ))]-
\nonumber \\
&-&\sum_{i=1}^N\eta_i^s(\tau )[\kappa_i^r(\tau )+\sum_a
Q_{ia}(\tau )A_a^r(\tau ,{\vec \eta}_i(\tau ))]\, )+
\nonumber \\
&+&\int d^3\sigma \, \sum_a(\sigma^r\,
{[{\vec \pi}_a(\tau ,\vec \sigma )\times {\vec B}_a(\tau ,\vec \sigma )]}^s-
\sigma^s\,
{[{\vec \pi}_a(\tau ,\vec \sigma )\times {\vec B}_a(\tau ,\vec \sigma )]}^r),
\nonumber \\
{\bar S}_s^{\bar or}&\approx& -{\bar S}_s^{r\bar o}=-
\sum_{i=1}^N\eta_i^r(\tau )\eta_i
\sqrt{m^2_i+[{\vec \kappa}_i(\tau )+\sum_aQ_{ia}(\tau )
{\vec A}_a(\tau ,{\vec \eta}_i(\tau ))]{}^2 }-\nonumber \\
&-&{1\over 2}\int d^3\sigma \sigma^r\, \sum_a
[g^2_s{\vec \pi}_a^2(\tau ,\vec \sigma )+g^{-2}_s{\vec B}_a^2(\tau ,\vec 
\sigma )],\nonumber \\
&&{}\nonumber \\
J_s^{ij}&\approx& {\tilde x}^i_sp_s^j-{\tilde x}_s^jp^i_s+\delta^{ir}\delta
^{js}{\bar S}_s^{rs},\nonumber \\
J_s^{oi}&\approx&{\tilde x}_s^op^i_s-{\tilde x}_s^ip^o_s-{ {\delta^{ir}{\bar
S}_s^{rs}p^s_s}\over {p^o_s+\eta_s\sqrt{p^2_s}} }.
\label{b31}
\end{eqnarray}

The time constancy of $T^{\mu}_{\check A}\approx 0$ with respect to the Dirac
Hamiltonian gives

\begin{eqnarray}
{d\over {d\tau}}[b^{\mu}_{\check r}(\tau )-\epsilon^{\mu}_r(u(p_s))]&=&
\lbrace b^{\mu}_{\check r}(\tau )-\epsilon^{\mu}_r(u(p_s)),H_D\rbrace {}^{*}=
\nonumber \\
&=&{1\over 2}{\tilde \lambda}^{\alpha\beta}(\tau )\lbrace b^{\mu}_{\check r}
(\tau ),S_{s\alpha\beta}(\tau )\rbrace {}^{*}={\tilde \lambda}^{\mu\alpha}
(\tau )b_{\check r\alpha}(\tau )\approx 0\nonumber \\
&\Rightarrow& {\tilde \lambda}^{\mu\nu}(\tau )\approx 0,
\label{b32}
\end{eqnarray}

\noindent so that the independent gauge-fixings contained in Eqs.(\ref{b30}) 
and 
the constraints ${\tilde {\cal H}}^{\mu\nu}(\tau )\approx 0$ form six pairs of
second class constraints.

Now we have [remember that ${\dot x}_s^{\mu}(\tau )=
-{\tilde \lambda}^{\mu}(\tau )$]

\begin{eqnarray}
&&l^{\mu}=b^{\mu}_{\tau}=u^{\mu}(p_s),\nonumber \\
&&z^{\mu}_{\tau}(\tau )={\dot x}^{\mu}_s(\tau )=\sqrt{g(\tau )}u^{\mu}(p_s)-
{\dot x}_{s\nu}(\tau )\epsilon^{\mu}_r(u(p_s))\epsilon^{\nu}_r(u(p_s)),
\nonumber \\
&&g(\tau )={[{\dot x}_{s\mu}(\tau )u^{\mu}(p_s))]}^2,\nonumber \\
&&g_{\tau\tau}={\dot x}^2_s,\quad\quad g^{\tau\tau}={1\over g},\quad\quad
g^{\tau{\check r}}={1\over g}{\dot x}_{s\mu}\delta^{{\check r}s}\epsilon^{\mu}
_s(u(p_s)),\nonumber \\
&&g_{\tau{\check r}}={\dot x}_{s\mu}\delta_{{\check r}s}\epsilon_s^{\mu}
(u(p_s)),\quad\quad g^{{\check r}{\check s}}=-\delta^{{\check r}{\check s}}+
{ {\delta^{{\check r}u}\delta^{{\check s}v}}\over {g(\tau )}}
{\dot x}_{s\mu}\epsilon^{\mu}_u(u(p_s))
{\dot x}_{s\nu}\epsilon^{\nu}_v(u(p_s)).
\label{b33}
\end{eqnarray}

On the hyperplane $\Sigma_{W\, \tau}$ all the degrees of freedom $z^{\mu}(\tau ,
\vec \sigma )$ are reduced to the four degrees of freedom ${\tilde x}^{\mu}_s
(\tau )$, which replace $x^{\mu}_s$. The Dirac
Hamiltonian is now $H_D={\tilde \lambda}^{\mu}(\tau ){\tilde {\cal H}}_{\mu}
(\tau )+\sum_{i=1}^N\mu_i(\tau )N_i(\tau )+\int d^3\sigma \sum_a[\lambda
_{a\tau}(\tau ,\vec \sigma )\pi^{\tau}_a(\tau ,\vec \sigma )+\lambda_a
(\tau ,\vec \sigma )\Gamma_a(\tau ,\vec \sigma )]$

To find the new Dirac brackets, one needs to evaluate the matrix of the
old Dirac brackets of the second class constraints (without extracting the 
independent ones)

\begin{eqnarray}
C=\left(
\begin{array}{cccc}
\lbrace {\tilde {\cal H}}^{\alpha\beta},{\tilde {\cal H}}^{\gamma\delta}
\rbrace {}^{*}\approx 0 & \lbrace {\tilde {\cal H}}^{\alpha\beta},T^{\sigma}
_{\check B}\rbrace {}^{*}=\\
{} & =\delta_{{\check B}B}[\eta^{\sigma\beta}\epsilon
^{\alpha}_B(u(p_s))-\eta^{\sigma\alpha}\epsilon_B^{\beta}(u(p_s))] \\
\lbrace T^{\rho}_{\check A},{\tilde {\cal H}}^{\gamma\delta}\rbrace {}^{*}
= & \lbrace T^{\rho}_{\check A},T^{\sigma}_{\check B}\rbrace {}^{*}=0\\
=\delta_{{\check A}A}[\eta^{\rho\gamma}\epsilon^{\delta}_A(u(p_s))-\eta
^{\rho\delta}\epsilon^{\gamma}_A(u(p_s))] & {}. 
\end{array} \right)
\label{b35}
\end{eqnarray}

Since the constraints are redundant, this matrix has the following left and
right null eigenvectors: $\left( \begin{array}{c}  a_{\alpha\beta}=a_{\beta
\alpha} \\ 0 \end{array} \right)$  [$a_{\alpha\beta}$ arbitrary], 
$\left( \begin{array}{c}   0 \\ \epsilon^B_{\sigma}(u(p_s)) 
\end{array} \right)$. Therefore,  one has to find 
a left and right quasi-inverse
$\bar C$, $\bar CC=C\bar C=D$, such that $\bar C$ and D have the same left and
right null eigenvectors. One finds

\begin{eqnarray}
\bar C&=&\left( \begin{array}{cc} 
0_{\gamma\delta\mu\nu} & {1\over 4}[\eta_{\gamma\tau}\epsilon^D_{\delta}(u(p_s))
-\eta_{\delta\tau}\epsilon^D_{\gamma}(u(p_s))] \\ {1\over 4}[\eta_{\sigma\nu}
\epsilon^B_{\mu}(u(p_s))-\eta_{\sigma\mu}\epsilon^B_{\nu}(u(p_s))] &
0^{BD}_{\sigma\tau} \end{array} \right) \nonumber \\
&&{}\nonumber \\
\bar CC=C\bar C=D&=&\left( \begin{array}{cc} 
{1\over 2}(\eta^{\alpha}_{\mu}\eta^{\beta}_{\nu}-\eta^{\alpha}_{\nu}\eta_{\mu}
^{\beta}) & 0^{\alpha\beta D}_{\tau} \\
0^{\rho}_{A\mu\nu} & {1\over 2}(\eta^{\rho}_{\tau}\eta^D_A-\epsilon^{D\rho}
(u(p_s))\epsilon_{A\tau}(u(p_s)) \end{array} \right)
\label{b36}
\end{eqnarray}

\noindent and the new Dirac brackets are

\begin{eqnarray}
\lbrace A,B\rbrace {}^{**}&=&\lbrace A,B\rbrace {}^{*}-{1\over 4}[\lbrace
A,{\tilde {\cal H}}^{\gamma\delta}\rbrace {}^{*}[\eta_{\gamma\tau}\epsilon^D
_{\delta}(u(p_s))-\eta_{\delta\tau}\epsilon^D_{\gamma}(u(p_s))]\lbrace
T^{\tau}_D,B\rbrace {}^{*}+\nonumber \\
&+&\lbrace A,T^{\sigma}_B\rbrace {}^{*}[\eta_{\sigma\nu}\epsilon^B_{\mu}(u(p
_s))-\eta_{\sigma\mu}\epsilon^B_{\nu}(u(p_s))]\lbrace {\tilde {\cal H}}^{\mu\nu}
,B\rbrace {}^{*}].
\label{b37}
\end{eqnarray}

\noindent While the check of $\lbrace {\tilde {\cal H}}^{\alpha\beta},B\rbrace
{}^{**}=0$ is immediate, we must use the relation $b_{{\check A}\mu}T^{\mu}
_D\epsilon^{D\rho}=-T^{\rho}_{\check A}$ [at this level we have $T^{\mu}_{\check
A}=T^{\mu}_A$] to check $\lbrace T^{\rho}_A,B\rbrace {}^{**}=0$.

Then, we find the following brackets for the remaining variables ${\tilde
x}^{\mu}_s, p_s^{\mu}, \eta^r_i, \kappa_i^r$ [the metric $\gamma^{rs}=-\delta
^{rs}$ will be used, so that $\lbrace \eta^r_i,.\rbrace =\partial /\partial 
\kappa_i^r=-\partial /\partial \kappa_{ir}$]

\begin{eqnarray}
&&\lbrace {\tilde x}^{\mu}_s,p^{\nu}_s\rbrace {}^{**}=-\eta^{\mu\nu},
\nonumber \\
&&\lbrace \eta_i^r,\kappa^s_j\rbrace {}^{**}=\delta_{ij}\delta^{rs}=-\delta_{ij}
\gamma^{rs},
\label{b38}
\end{eqnarray}

\noindent and the following form of the Poincar\'e generators [$\lbrace 
{\tilde L}^{\mu\nu}_s,{\tilde S}^{\mu\nu}_s\rbrace {}^{**}\not= 0$]

\begin{eqnarray}
p^{\mu}_s,&&{}\nonumber \\
J^{\mu\nu}_s&=&{\tilde x}^{\mu}_sp^{\nu}_s-{\tilde x}^{\nu}_sp_s^{\mu}+{\tilde
S}_s^{\mu\nu}={\tilde L}_s^{\mu\nu}+{\tilde S}_s^{\mu\nu},\nonumber \\
&&{\tilde S}_s^{oi}=-{ {\delta^{ir}{\bar S}_s^{rs}p_s^s}\over {p^o_s+\eta_s
\sqrt{p_s^2}} },\nonumber \\
&&{\tilde S}_s^{ij}=\delta^{ir}\delta^{js}{\bar S}_s^{rs}.
\label{b39}
\end{eqnarray}

\noindent Therefore, ${\tilde x}_s^{\mu}$ is not a fourvector and ${\vec \eta}
_i, {\vec \kappa}i$ transform as Wigner spin-1 3-vectors. Indeed, as shown in 
Ref.\cite{lus1}, under global Poincar\'e and infinitesimal Lorentz 
transformations one has

\begin{eqnarray}
\eta^{'}_{ir}&=&\eta_{is}R^s{}_r(\Lambda ,p),\nonumber \\
\kappa^{'}_{ir}&=&\kappa_{is}R^s{}_r(\Lambda ,p),\nonumber \\
&&{}\nonumber \\
\lbrace \eta^r_i,J_s^{oi}\rbrace &=&-{ {\delta^{is}(p_s^r\eta^s_i-p_s^s\eta^r_i
)}\over {p_s^o+\eta \sqrt{p^2_s}} },\nonumber \\
\lbrace \eta^r_i,J_s^{ij}\rbrace &=&\delta^{is}\delta^{jt}\lbrace \eta^r_i,{\bar
S}^{st}_s\rbrace =(\delta^{is}\delta^{jr}-\delta^{ir}\delta^{js})\eta^s_i,
\nonumber \\
\lbrace \kappa^r_i,J_s^{oi}\rbrace &=&-{ {\delta^{is}(p_s^r\kappa^s_i-p_s^s
\kappa^r_i)}\over {p_s^o+\eta \sqrt{p^2_s}} },\nonumber \\
\lbrace \kappa^r_i,J_s^{ij}\rbrace &=&(\delta^{is}\delta^{jr}-\delta^{ir}
\delta^{js})\kappa^s_i.
\label{b39a}
\end{eqnarray}

The only left first class constraints are

\begin{eqnarray}
{\tilde {\cal H}}^{\mu}(\tau )&=&p_s^{\mu}-u^{\mu}(u(p_s))\,
[{1\over 2}\int d^3\sigma \sum_a 
[g^2_s{\vec \pi}_a^2(\tau ,\vec \sigma )+
g^{-2}_s{\vec B}_a^2(\tau ,\vec \sigma )]+\nonumber \\
&+&\sum_{i=1}^N\eta_i
\sqrt{m^2_i+[{\vec \kappa}_i(\tau )+\sum_aQ_{ia}(\tau )
{\vec A}_a(\tau ,{\vec \eta}_i(\tau ))]{}^2 }]-\nonumber \\
&-&\epsilon^{\mu}_r(u(p_s))\, [\int d^3\sigma \sum_a 
{[{\vec \pi}_a(\tau ,\vec \sigma )\times {\vec B}_a(\tau ,\vec \sigma )]}^r+
\nonumber \\
&+&\sum_{i=1}^N[\kappa_i^r(\tau )+\sum_a
Q_{ia}(\tau )A_a^r(\tau ,{\vec \eta}_i(\tau ))]\,\, ]
\approx 0,\nonumber \\
\pi_a^{\tau}(\tau ,\vec \sigma )&\approx& 0,\nonumber \\
\Gamma_a (\tau ,\vec \sigma )&\approx& 0,\nonumber \\
N_i(\tau )&\approx& 0,\nonumber \\
&&{}\nonumber \\
&&\lbrace {\tilde {\cal H}}^{\mu},{\tilde {\cal H}}^{\nu}\rbrace {}^{**}=\sum_a
\int d^3\sigma \lbrace [\epsilon^{\mu}_{\tau}(u(p_s))\epsilon^{\nu}_r
(u(p_s))-\epsilon^{\nu}_{\tau}(u(p_s))\epsilon^{\mu}_r(u(p_s))] \pi_{ar}(\tau ,
\vec \sigma )+\nonumber \\
&&+\epsilon^{\mu}_r(u(p_s))F_{ars}(\tau ,\vec \sigma )
\epsilon^{\nu}_s(u(p_s))\rbrace \Gamma_a (\tau ,\vec \sigma ),
\label{b40}
\end{eqnarray}

\noindent or

\begin{eqnarray}
{\cal H}(\tau )&=&\eta_s\sqrt{p^2_s}-
[\sum_{i=1}^N\eta_i
\sqrt{m^2_i+[{\vec \kappa}_i(\tau )+\sum_aQ_{ia}(\tau )
{\vec A}_a(\tau ,{\vec \eta}_i(\tau ))]{}^2 }+\nonumber \\
&+&{1\over 2}\int d^3\sigma \sum_a 
[g^2_s{\vec \pi}_a^2(\tau ,\vec \sigma )+g^{-2}_s{\vec B}_a^2(\tau ,\vec 
\sigma ) ]\, ]\approx 0,\nonumber \\
{\vec {\cal H}}_p(\tau )&=&
\sum_{i=1}^N[{\vec \kappa}_i(\tau )+\sum_a
Q_{ia}(\tau ){\vec A}_a(\tau ,{\vec \eta}_i(\tau ))]+
\nonumber \\
&+&\int d^3\sigma \sum_a
{\vec \pi}_a (\tau ,\vec \sigma )\times {\vec B}_a(\tau ,\vec \sigma )
\approx 0,\nonumber \\
\pi_a^{\tau}(\tau ,\vec \sigma )&\approx& 0,\nonumber \\
\Gamma_a (\tau ,\vec \sigma )&\approx& 0,\nonumber \\
N_i(\tau )&\approx& 0.
\label{b41}
\end{eqnarray}

The first one gives the mass spectrum of the isolated system, while
the other three say that the total (Wigner spin 1)
3-momentum of the N particles on the
hyperplane $\Sigma_{W\, \tau}$ vanishes. The Dirac Hamiltonian is now $H_D=
\lambda (\tau ){\cal H}(\tau )-\vec \lambda (\tau )\cdot {\vec {\cal H}}_p
(\tau )+\sum_{i=1}^N\mu_i(\tau )N_i(\tau )+\int d^3\sigma \sum_a[\lambda
_{a\tau}(\tau ,\vec \sigma )\pi^{\tau}_a(\tau ,\vec \sigma )+\lambda_a
(\tau ,\vec \sigma )\Gamma_a(\tau ,\vec \sigma )]$ 
and we have ${\dot {\tilde x}}_s^{\mu}=\lbrace {\tilde x}_s^{\mu},
H_D\rbrace {}^{**}=-\lambda (\tau )u^{\mu}(p_s)$. Therefore, while the old
$x^{\mu}_s$ had a velocity ${\dot x}_s^{\mu}$ not parallel to the normal
$l^{\mu}=u^{\mu}(p_s)$ to the hyperplane as shown by Eqs.(\ref{b33}), the new
${\tilde x}_s^{\mu}$ has ${\dot {\tilde x}}^{\mu}_s \| l^{\mu}$ and
no classical zitterbewegung. Moreover, we have that $T_s=l\cdot {\tilde x}_s=
l\cdot x_s$ is the Lorentz-invariant rest frame time.

For $N\geq 2$, let us perform the following canonical transformation 
[$\bar a=1,..,N-1$]

\begin{equation}
\begin{minipage}[t]{2cm}
\begin{tabular}{|ll|} \hline
${\tilde x}_s^{\mu}$    &  $p_s^{\mu}$  \\ \hline
${\vec \eta}_i$  & ${\vec \kappa}_i$ \\ \hline
\end{tabular}
\end{minipage} \ {\longrightarrow \hspace{.2cm}} \
\begin{minipage}[t]{3cm}
\begin{tabular}{|ll|l|} \hline
{} & {} & $T_s$  \\
${\vec z}_s$ & ${\vec k}_s$ & { }  \\
{} & {} & $\epsilon_s$  \\ \hline
${\vec \rho}^{'}_{\bar a}$  &  ${\vec \pi}^{'}_{\bar a}$  & {}  \\ \hline
${\vec \eta}_{+}$  &  ${\vec \kappa}_{+}$  &  {}  \\ \hline
\end{tabular}
\end{minipage}
\label{b42}
\end{equation}

\begin{eqnarray}
&&T_s={ {p_s\cdot {\tilde x}_s}\over {\eta_s\sqrt{p^2_s}}}=
{ {p_s\cdot x_s}\over {\eta_s\sqrt{p^2_s}}},\nonumber \\
&&\epsilon_s=\eta_s\sqrt{p_s^2},\nonumber \\
&&{\vec z}_s=\eta_s\sqrt{p_s^2}({\vec {\tilde x}}_s-{ {{\vec p}_s}\over 
{p^o_s}}{\tilde x}^o_s),\nonumber \\
&&{\vec k}_s={{{\vec p}_s}\over {\eta_s\sqrt{p_s^2}}},\nonumber \\
&&{\vec \eta}_{+}={1\over N}\sum_{i=1}^N{\vec \eta}_i,\nonumber \\
&&{\vec \kappa}_{+}=\sum_{i=1}^N{\vec \kappa}_i,\nonumber \\
&&{\vec \rho}^{'}_{\bar a}=\sqrt{N}\sum_{i=1}^N{\hat \gamma}_{\bar ai}{\vec 
\eta}_i,\nonumber \\
&&{\vec \pi}^{'}_{\bar a}={1\over {\sqrt{N}}}\sum_{i=1}^N{\hat \gamma}_{\bar 
ai}{\vec \kappa}_i,\nonumber \\
&&{}\nonumber \\
&&\sum_{i=1}^N{\hat \gamma}_{\bar ai}=0,\quad\quad \sum_{i=1}^N{\hat \gamma}
_{\bar ai}{\hat \gamma}_{\bar bi}=\delta_{\bar a\bar b},\nonumber \\
&&\sum_{\bar a=1}^{N-1}{\hat \gamma}_{\bar ai}{\hat \gamma}_{\bar aj}=\delta
_{ij}-{1\over N},
\label{b43}
\end{eqnarray}

\noindent whose inverse is

\begin{eqnarray}
&&{\tilde x}^o_s=\sqrt{1+{\vec k}^2_s}(T_s+{ { {\vec k}_s\cdot {\vec z}_s}
\over {\epsilon_s} }),\nonumber \\
&&{\vec {\tilde x}}_s={ { {\vec z}_s}\over {\epsilon_s}}+(T_s+{ { {\vec k}_s
\cdot {\vec z}_s}\over {\epsilon_s} }){\vec k}_s,\nonumber \\
&&p^o_s=\epsilon_s\sqrt{1+{\vec k}^2_s},\nonumber \\
&&{\vec p}_s=\epsilon_s{\vec k}_s,\nonumber \\
&&{\vec \eta}_i={\vec \eta}_{+}+{1\over {\sqrt{N}}}\sum_{\bar a=1}^{N-1}{\hat 
\gamma}_{\bar ai}{\vec \rho}^{'}_{\bar a},\nonumber \\
&&{\vec \kappa}_i={1\over N}{\vec \kappa}_{+}+\sqrt{N}\sum_{\bar a=1}^{N-1}
{\hat \gamma}_{\bar ai}{\vec \pi}^{'}_{\bar a}.
\label{b44}
\end{eqnarray}

The new form of the constraints is

\begin{eqnarray}
&&{\cal H}(\tau )=\epsilon_s-\lbrace \sum_{i=1}^N\eta_i\times \nonumber \\
&& \sqrt{m^2_i+[{1\over N}{\vec \kappa}_{+}(\tau )+\sqrt{N}
\sum_{\bar a=1}^{N-1}{\hat \gamma}_{\bar ai}{\vec \pi}_{\bar a}(\tau )+
\sum_aQ_{ia}(\tau ){\vec A}_a(\tau ,{\vec \eta}_{+}(\tau )+{1\over {\sqrt{N}}}
\sum_{\bar a=1}^{N-1}{\hat \gamma}_{\bar ai}{\vec \rho}_{\bar a}(\tau ))]{}^2 
}+\nonumber \\
&+&{1\over 2}\int d^3\sigma \sum_a 
[g^2_s{\vec \pi}_a^2(\tau ,\vec \sigma )+
g^{-2}_s{\vec B}_a^2(\tau ,\vec \sigma ) ]\rbrace
=\epsilon_s-E_{(P+I)s}-E_{(F)s}\approx 0,\nonumber \\
&&{}\nonumber \\
&&{\vec {\cal H}}_p(\tau )={\vec \kappa}_{+}(\tau )+\sum_{i=1}^N
\sum_aQ_{ia}(\tau ){\vec A}_a(\tau ,{\vec \eta}_{+}(\tau )+{1\over {\sqrt{N}}}
\sum_{\bar a=1}^{N-1}{\hat \gamma}_{\bar ai}{\vec \rho}_{\bar a}(\tau ))+
\nonumber \\
&+& \int d^3\sigma \sum_a{\vec \pi}_a (\tau ,\vec \sigma )\times 
{\vec B}_a(\tau ,\vec 
\sigma )={\vec P}_{(P+I)s}+{\vec P}_{(F)s}\approx 0,\nonumber \\
&&\pi^{\tau}_a(\tau ,\vec \sigma )\approx 0,\nonumber \\
&&\Gamma_a (\tau ,\vec \sigma )\approx 0,\nonumber \\
&&N_i(\tau )\approx 0,
\label{b45}
\end{eqnarray}

\noindent where $E_{(F)s}={1\over 2}\int d^3\sigma \sum_a
[g^2_s{\vec \pi}_a^2(\tau ,\vec \sigma )+
g^{-2}_s{\vec B}_a^2(\tau ,\vec \sigma )]$ and ${\vec P}_{(F)s}=
\int d^3\sigma \sum_a{\vec \pi}_a (\tau ,\vec \sigma )\times 
{\vec B}_a(\tau ,\vec 
\sigma )$ are the rest-frame field energy and three-momentum respectively
[now we have ${\vec \pi}_a (\tau ,\vec \sigma )=
g^{-2}_s{\vec E}_a(\tau ,\vec \sigma )$],
while $E_{(P+I)s}$ and ${\vec P}_{(P+I)s}$ denote the particle+interaction
total rest-frame energy and three-momentum, before the decoupling from the
electromagnetic gauge degrees of freedom.

The final form of the rest-frame spin tensor is

\begin{eqnarray}
{\bar S}_s^{rs}&=&\sum_{i=1}^N\lbrace (\eta^r_{+}(\tau )+{1\over {\sqrt{N}}}
\sum_{\bar a=1}^{N-1} {\hat \gamma}_{\bar ai}\rho^r_{\bar a}(\tau ))({{\kappa
^s_{+}(\tau )}\over N}+\sqrt{N}\sum_{\bar b=1}^{N-1}{\hat \gamma}_{\bar bi}\pi
^s_{\bar b}(\tau )+\nonumber \\
&+&\sum_aQ_{ia}(\tau )
A_a^s(\tau ,{\vec \eta}_{+}(\tau )+{1\over {\sqrt{N}}}
\sum_{\bar c=1}^{N-1}{\hat \gamma}_{\bar ci}{\vec \rho}_{\bar c}(\tau )))-
\nonumber \\
&-&(\eta^s_{+}(\tau )+{1\over {\sqrt{N}}}
\sum_{\bar a=1}^{N-1} {\hat \gamma}_{\bar ai}\rho^s_{\bar a}(\tau ))({{\kappa
^r_{+}(\tau )}\over N}+\sqrt{N}\sum_{\bar b=1}^{N-1}{\hat \gamma}_{\bar bi}\pi
^r_{\bar b}(\tau )+\nonumber \\
&+&\sum_aQ_{ia}(\tau )
A_a^r(\tau ,{\vec \eta}_{+}(\tau )+{1\over {\sqrt{N}}}
\sum_{\bar c=1}^{N-1}{\hat \gamma}_{\bar ci}{\vec \rho}_{\bar c}(\tau )))
\rbrace +\nonumber \\
&+&\int d^3\sigma \, \sum_a (\sigma^r\, {[{\vec \pi}_a (\tau ,\vec \sigma )
\times {\vec B}_a(\tau ,\vec \sigma )]}^s-\sigma^s\,
{[{\vec \pi}_a (\tau ,\vec \sigma )\times {\vec B}_a(\tau ,\vec \sigma )]}^r)=
{\bar S}_{(P+I)s}^{rs}+{\bar S}_{(F)s}^{rs},\nonumber \\
&&{}\nonumber \\
{\bar S}_s^{\bar or}&=&-{\bar S}_s^{r\bar o}=-\sum_{i=1}^N(\eta^r_{+}(\tau )
+{1\over {\sqrt{N}}}\sum_{\bar a=1}^{N-1}{\hat \gamma}_{\bar ai}\rho^r_{\bar a}
(\tau ))\eta_i \times \nonumber \\
&&\sqrt{m^2_i+[{1\over N}{\vec \kappa}_{+}(\tau )+\sqrt{N}
\sum_{\bar a=1}^{N-1}{\hat \gamma}_{\bar ai}{\vec \pi}_{\bar a}(\tau )+
\sum_aQ_{ia}(\tau ){\vec A}_a(\tau ,{\vec \eta}_{+}(\tau )+{1\over {\sqrt{N}}}
\sum_{\bar a=1}^{N-1}{\hat \gamma}_{\bar ai}{\vec \rho}_{\bar a}(\tau ))]{}^2 
}\nonumber \\
&-&{1\over 2}\int d^3\sigma \sigma^r\, \sum_a
[g^2_s{\vec \pi}_a^2(\tau ,\vec \sigma )+
g^{-2}_s{\vec B}_a^2(\tau ,\vec \sigma )],
\label{b46}
\end{eqnarray}

\noindent while the Dirac Hamiltonian is

\begin{eqnarray}
H_D&=&\lambda (\tau ){\cal H}-\vec \lambda (\tau ){\vec {\cal H}}_p+\int d^3
\sigma \sum_a[\lambda_{a\tau}(\tau ,\vec \sigma )\pi_a^{\tau}(\tau ,\vec 
\sigma )-A_{a\tau}(\tau ,\vec \sigma )\Gamma_a (\tau ,\vec \sigma )]
+\nonumber \\
&+&\sum_{i=1}^N\mu_i(\tau )N_i(\tau ).
\label{b47}
\end{eqnarray}

On an arbitrary spacelike hypersurface or on
the Wigner hyperplane one has in the free case [but also in the interacting
one looking at the four constraints ${\tilde {\cal H}}^{\mu}(\tau )\approx 0$
of Eq.(\ref{b40}) and of Ref.\cite{lus1}]

$p^{\mu}_i{|}_{\Sigma_{\tau}}(\tau )=\eta_i\sqrt{m^2_i-\gamma
^{\check r\check s}(\tau ,\vec \sigma )\kappa_{i\check s}(\tau )\kappa
_{i\check s}(\tau )} l^{\mu}(\tau ,\vec \sigma )-z_{\check r}^{\mu}(\tau ,
\vec \sigma )\gamma^{\check r\check s}(\tau ,\vec \sigma )\kappa_{i\check s}
(\tau )$

$p^{\mu}_i{|}_{Wigner hyperplane}(\tau )=\eta_i\sqrt{m^2_i+{\vec \kappa}^2_i
(\tau )}u^{\mu}(p_s)+\epsilon^{\mu}_{\check r}(u(p_s)) \kappa_i^{\check r}
(\tau ),$

\noindent which are solutions of $p^2_i-m^2_i=0$.

\vfill\eject

\section
{The Dirac Observables}

In this Section we will make the canonical reduction with respect to the SU(3)
Yang-Mills gauge transformations. From now on we shall use the notation
$\lbrace .,.\rbrace$ for the Dirac brackets $\lbrace .,.\rbrace {}^{**}$ on the
Wigner hyperplane $\Sigma_{W\, \tau}$.

The  decompositions in the non-Abelian SU(3) gauge potential can  be
obtained from the second paper in Ref.\cite{lusa}.
For the vector potential we use Eqs.(4-13), (4-16), (4-26), (4-29), (4-30),
(4-31), (4-33), (5-21), (5-24), of that paper to get

\begin{eqnarray}
{\vec A}_a(\tau ,\vec \sigma )&=&A_{ab}(\eta^{(A)}(\tau ,\vec \sigma )) \vec 
\partial \eta^{(A)}_b(\tau ,\vec \sigma ) +(P\, e^{\Omega_s^{(\hat \gamma )}
(\eta^{(A)}(\tau ,\vec \sigma ))} )_{ab} {\check {\vec A}}_{b\perp}(\tau ,
\vec \sigma )=\nonumber \\
&&={\vec \Theta}_a(\eta^{(A)}(\tau ,\vec \sigma ), \vec \partial \eta^{(A)}
(\tau ,\vec \sigma ))+(P\, e^{\Omega_s^{(\hat \gamma )}
(\eta^{(A)}(\tau ,\vec \sigma ))} )_{ab} {\check {\vec A}}_{b\perp}(\tau ,
\vec \sigma ),\nonumber \\
&&\vec \partial \cdot {\check {\vec A}}_{a\perp}(\tau ,
\vec \sigma )=0\nonumber \\
&&{}\nonumber \\
&&{\hat T}^a A_{ab}(\eta^{(A)}(\tau ,\vec \sigma )) \vec \partial \eta^{(A)}
_b(\tau ,\vec \sigma )\cdot d\vec \sigma =H_b(\eta^{(A)}(\tau ,\vec \sigma )) 
\vec \partial \eta^{(A)}_b(\tau ,\vec \sigma )\cdot d\vec \sigma =\nonumber \\
&=&\Theta_a(\eta^{(A)}(\tau ,\vec \sigma ), \vec \partial \eta^{(A)}(\tau ,
\vec \sigma )) {\hat  T}^a= 
d_{(\hat \gamma )} \Omega_s^{(\hat \gamma )}(\eta^{(A)}(\tau ,\vec \sigma )),
\quad\quad \Theta_a={\vec \Theta}_a\cdot d\vec \sigma ,\nonumber \\
&&{}\nonumber \\
&&\Omega_s^{(\hat \gamma )}(\eta^{(A)}(\tau ,\vec \sigma ))=\Omega_{sa}
^{(\hat \gamma )}(\eta^{(A)}(\tau ,\vec \sigma )) {\hat T}^a =\nonumber \\
&=& {}_{(\hat 
\gamma )}\int_0^{\eta^{(A)}(\tau ,\vec \sigma ,s)} H_b(\eta^{(A)}(\tau ,\vec
\sigma ;s)) {\cal D}\eta_b^{(A)}(\tau ,\vec \sigma ;s).
\label{c1}
\end{eqnarray}

If $\eta_a$ are coordinates in a chart of the group manifold of SU(3), the
matrices $A_{ab}(\eta )$ satisfy the Maurer-Cartan equations, which can be
written in the zero curvature form ${{\partial H_a(\eta )}\over {\partial
\eta_b}}-{{\partial H_b(\eta )}\over {\partial \eta_a}}+[H_a(\eta ),H_b
(\eta )]=0$. We shall use only canonical coordinates of the first kind, defined
by $A_{ab}(\eta )\eta_b=\eta_a$ [ so that $A(\eta )={{e^{T\eta}-1}\over 
{T\eta}}$ with $(T\eta )_{ab}=({\hat T}^c)_{ab}\eta_c=c_{abc}\eta_c$]. If 
$\theta_a=A_{ab}(\eta ) d\eta_b$ are the left-invariant (or Maurer-Cartan) 
one-forms on SU(3), the abstract Maurer-Cartan equations are $d\theta_a=-
{1\over 2}c_{abc} \theta_b\wedge \theta_c$; then, by using the preferred line 
$\gamma_{\eta}(s)$ (s is the parameter along it) defining the canonical
coordinates of the first kind in a neighbourhood of the identity I of SU(3),
one can define $d_{(\gamma_{\eta})} \omega_a^{(\gamma_{\eta})}(\eta (s))=
\theta_a(\eta (s))$ [$d_{(\gamma_{\eta})}$ is the exterior derivative along
$\gamma_{\eta}$] with $\omega^{(\gamma_{\eta})}(\eta (s))=\omega_a^{(\gamma
_{\eta})}(\eta (s)) {\hat T}^a={}_{(\gamma_{\eta})}\int_0^{\eta (s)}{\hat T}
^a A_{ab}(\bar \eta ) d{\bar \eta}_b ={}_{(\gamma_{\eta})}\int_I^{\gamma_{eta}
(s)} \theta_a{|}_{\gamma_{\eta}} {\hat T}^a ={}_{(\gamma_{\eta})}\int_I
^{\gamma_{\eta}(s)} \omega_G{|}_{\gamma_{\eta}}$, where $\omega_G=\theta_a{\hat
T}^a$ is the canonical one-form on SU(3) in the adjoint representation.
In our case of a trivial principal SU(3)-bundle $P(\Sigma (\tau ),SU(3))$ over 
the spacelike hypersurface $\Sigma_{\tau}$ [diffeomorphic to $R^3$]
of Minkowski spacetime, it is shown in Ref.\cite{lusa}
that $\Theta_a(\eta^{(A)}(\tau ,\vec \sigma ), \vec \partial \eta^{(A)}(\tau ,
\vec \sigma ))$ and $\Omega_s^{(\hat \gamma )}(\eta^{(A)}(\tau ,\vec \sigma ))$ 
are just the extension of these SU(3) objects: in the second paper of 
Ref.\cite{lusa}, a connection-dependent coordinatization 
$(\tau ,\vec \sigma ; \eta^{(A)}(\tau ,\vec \sigma ))$ of the principal bundle 
is given with the SU(3) fibers parametrized with parallely transported (with 
respect to the given connection) canonical coordinates of the first kind from 
a reference fiber over an arbitrarily chosen origin in $R^3$. The functions 
$\eta^{(A)}_a(\tau ,\vec \sigma )$ and their gradients $\vec \partial \eta
^{(A)}_a(\tau ,\vec \sigma )$
vanish on the identity cross section $\sigma_I$ of the trivial principal
bundle. The path $\hat \gamma$ is a surface (in the total bundle space) of
preferred paths, associated with these generalized canonical coordinates of
the first kind, starting from the identity cross section $\sigma_I$ till
a cross section parametrized by the parameter s, in a tubolar neighbourhood of
$\sigma_I$. The operator $d_{(\hat \gamma )}$ is the exterior derivative on the 
principal SU(3)-bundle total space restricted to $\hat \gamma$; it can be
identified with the vertical derivative on the principal bundle and with the
Hamiltonian BRST operator. With these conventions, one has $\lbrace .,\Gamma
_a(\tau ,\vec \sigma )\rbrace \equiv \lbrace .,
-{\hat {\vec D}}_{ab}\cdot {\vec {\tilde \pi}}_b(\tau ,\vec \sigma )
\rbrace =- B_{ba}(\eta^{(A)}(\tau ,\vec \sigma )) {{\tilde \delta}\over 
{\delta \eta^{(A)}_b(\tau ,\vec \sigma )}}$ [$B(\eta )=A^{-1}(\eta )$] with the 
functional derivative to be interpreted as a directional derivative along 
the surface of paths $\hat \gamma$. The longitudinal gauge variables 
have a complicated formal implicit expression given in Eq.(4-49) of
the second paper of Ref.\cite{lusa} and satisfy $\lbrace \eta^{(A)}_a(\tau ,
\vec \sigma ),{\tilde \Gamma}_b(\tau ,{\vec \sigma}^{'})\rbrace =-\delta_{ab}
\delta^3(\vec \sigma -{\vec \sigma}^{'})$, where ${\tilde \Gamma}_a(\tau ,\vec
\sigma )=\Gamma_b(\tau ,\vec \sigma ) A_{ba}(\eta^{(A)}(\tau ,\vec \sigma ))$ 
are the Abelianized Gauss laws [$\lbrace {\tilde \Gamma}_a(\tau ,\vec \sigma
), {\tilde \Gamma}_b(\tau ,{\vec \sigma}^{'})\rbrace =0$]. In Eq.(\ref{c1}),
$A_{ab}(\eta^{(A)}(\tau ,\vec \sigma )) \vec \partial \eta^{(A)}_b(\tau ,\vec
\sigma )$ is the pure gauge part (saturated with $d\vec \sigma$ it is the BRST 
ghost) of the vector potential ${\vec A}_a(\tau ,\vec \sigma )$: the magnetic 
field ${\vec B}_a(\tau ,\vec \sigma )$ is generated only by 
the second term of Eq.(\ref{c1}). In this sense, $\eta_a^{(A)}(\tau ,\vec
\sigma )=0$ is the true generalized non-Abelian Coulomb gauge with all the same 
properties of the Abelian Coulomb gauge. In suitable weighted Sobolev spaces, 
as discussed in
Ref.\cite{lusa}, this gauge-fixing is well defined, since all the connections 
over the principal SU(3)-bundle are completely irreducible [their holonomy
bundles (i.e. the set of points of $P(R^3,SU(3))$ which can be joined by
horizontal curves) coincide with the principal bundle itself] and there is no
form of Gribov ambiguity (i.e. of stability subgroups of the group of gauge
transformations for special connections and/or field strengths). In these
spaces, the covariant divergence is an elliptic operator without zero modes
\cite{mon} and its Green function ${\vec \zeta}^{(A)}_{ab}(\vec \sigma ,\vec 
{\sigma}^{'};\tau )$ is globally defined

\begin{eqnarray}
&&{\hat {\vec D}}^{(A)}_{ab}(\tau ,\vec \sigma )\cdot {\vec \zeta}^{(A)}_{bc}
(\vec \sigma ,{\vec \sigma}^{'};\tau )
=- \delta_{ac} \delta^3(\vec \sigma -{\vec \sigma}^{'})\nonumber \\
&&{}\nonumber \\
&&{\vec \zeta}_{ab}^{(A)}(\vec \sigma ,{\vec \sigma}^{'};\tau )=\vec c (\vec 
\sigma -{\vec \sigma}^{'}) \zeta^{(A)}_{ab}(\vec \sigma ,{\vec \sigma}^{'};
\tau )=\vec c (\vec \sigma -{\vec \sigma}^{'}) (P\, e^{\int_{\sigma^{'}}
^{\sigma} d{\vec \sigma}^{"}\, \cdot {\vec A}_c(\tau ,{\vec \sigma}^{"}) {\hat 
T}^c}\, )_{ab},\nonumber \\
&&\vec c(\vec x)=\vec \partial c(\vec x)={{\vec \partial}\over {\triangle}}
\delta^3(\vec x)={{\vec x}\over {4\pi |\vec x|^3}},\quad \triangle =-{\vec 
\partial}^2,\quad c(\vec x)={1\over {\triangle}}\delta^3(\vec x)={{-1}\over
{4\pi |\vec x|}}.
\label{c2}
\end{eqnarray}

\noindent The path ordering is along the straigthline (flat geodesic) joining
$\vec y$ and $\vec x$ when $\Sigma_{\tau}$ is a hyperplane like $\Sigma_{W\,
\tau}$.

Therefore, we have

\begin{eqnarray}
{\vec \pi}_a(\tau ,\vec \sigma )&=& -{{\vec \partial}\over {\triangle}} \vec 
\partial \cdot {\vec \pi}_a(\tau ,\vec \sigma ) + {\vec \pi}_{a\perp}(\tau ,
\vec \sigma )=\nonumber \\
&=& {\vec \pi}_{a,D\perp}(\tau ,\vec \sigma )+ \int d^3\sigma^{'}\, {\vec 
\zeta}^{(A)}_{ab}(\vec \sigma ,{\vec \sigma}^{'};\tau ) [\Gamma_b(\tau ,{\vec 
\sigma}^{'})-\sum_{i=1}^NQ_{ib}(\tau )\delta^3({\vec \sigma}^{'}-{\vec \eta}
_i(\tau ))]\nonumber \\
&&{}\nonumber \\
&&\vec \partial \cdot {\vec \pi}_{a\perp}(\tau ,\vec \sigma ) = {\hat {\vec D}}
^{(A)}_{ab}(\tau ,\vec \sigma )\cdot {\vec \pi}_{b,D\perp}(\tau ,\vec \sigma ) 
= 0.
\label{c3}
\end{eqnarray}

It is shown in Eqs. (5-7), (5-8), (5-10), of the second paper in Ref.\cite{lusa}
that we have

\begin{eqnarray}
\vec \partial \cdot {\vec \pi}_a(\tau ,\vec \sigma ) &=& \int d^3\sigma^{'}\, 
{\vec \zeta}^{(A)}_{ab}(\vec \sigma ,{\vec \sigma}^{'};\tau )\nonumber \\
&&[c_{bef}{\vec A}_e(\tau ,{\vec \sigma}^{'})\cdot {\vec \pi}_{f\perp}(\tau ,
{\vec \sigma}^{'})+ \Gamma_b(\tau ,{\vec \sigma}^{'})-\sum_{i=1}^NQ_{ib}(\tau )
\delta^3({\vec \sigma}^{'}-{\vec \eta}_i(\tau ))], \nonumber \\
\pi^{i}_{a,D\perp}(\tau ,\vec \sigma )&=& \int d^3\sigma^{'}\, [\delta^{ij} 
\delta_{ab} \delta^3(\vec \sigma-{\vec \sigma}^{'})-\nonumber \\
&-&{{\partial^i_{\sigma}}\over {\triangle_{\sigma}}} {\vec \partial}_{\sigma}
\cdot {\vec \zeta}^{(A)}_{ac}(\vec \sigma,{\vec \sigma}^{'};\tau ) c_{ceb} A^j
_e(\tau ,{\vec \sigma}^{'})] \pi^{j}_{b\perp}(\tau ,{\vec \sigma}^{'})
\nonumber \\
&&{}\nonumber \\
&\Rightarrow& \pi^{i}_{a\perp}(\tau ,\vec \sigma )= P^{ij}_{\perp}(\vec \sigma)
\pi^{j}_{a,D\perp}(\tau ,\vec \sigma ),
\label{c4}
\end{eqnarray}

\noindent where $P^{ij}_{\perp}(\vec \sigma )=\delta^{ij}-\partial^i\partial^j
/\triangle$. Moreover, Eqs.(5-21) and (5-25) of that paper give

\begin{eqnarray}
{\vec \pi}_{a,D\perp}(\tau ,\vec \sigma )&=& (P\, e^{\Omega_s^{(\hat \gamma )}
(\eta^{(A)}(\tau ,\vec \sigma ))} )_{ab} {\check {\vec \pi}}_{b,D\perp}(\tau ,
\vec \sigma )=\nonumber \\
&=&(P\, e^{\Omega_s^{(\hat \gamma )}(\eta^{(A)}(\tau ,\vec \sigma ))} )_{ab} 
[{\check {\vec \pi}}_{b\perp}(\tau ,\vec \sigma ) - {{\vec \partial}\over 
{\triangle}} \vec \partial \cdot {\check {\vec \pi}}_{b,D\perp}(\tau ,
\vec \sigma )],\nonumber \\
&&{}\nonumber \\
{\check \pi}^{i}_{a\perp}(\tau ,\vec \sigma )&=& P^{ij}_{\perp}(\vec \sigma ) 
{\check \pi}^{j}_{a,D\perp}(\tau ,\vec \sigma )\nonumber \\
&&\lbrace {\check {\vec \pi}}_{a,D\perp}(\tau ,\vec \sigma ), \Gamma_b
(\tau ,{\vec \sigma}^{'})\rbrace =0.
\label{c5}
\end{eqnarray}

Therefore, the color SU(3) canonical pairs of Dirac's observables turn out to 
be ${\check {\vec A}}_{a\perp}(\tau ,\vec \sigma )$, ${\check {\vec \pi}}
_{a\perp}(\tau ,\vec \sigma )$. 
They satisfy the Poisson brackets

\begin{equation}
\lbrace {\check A}^i_{a\perp}(\tau ,\vec \sigma), {\check \pi}^{j}_{b\perp}
(\tau ,{\vec \sigma}^{'})\rbrace = -\delta_{ab} P^{ij}_{\perp}(\vec \sigma ) 
\delta^3(\vec \sigma -{\vec \sigma}^{'}).
\label{c6}
\end{equation}

\noindent Instead, the gauge sector is given by the pairs $A_{a\tau}(\tau ,\vec
\sigma )$, $\pi^{\tau}_a(\tau ,\vec \sigma )\approx 0$, $\eta^{(A)}_a(\tau ,\vec
\sigma )$, ${\tilde \Gamma}_a(\tau ,\vec \sigma )\approx 0$.

Let us now look for the Dirac observables of the particles. First of all,
the Grassmann variables are not gauge invariant because one has

\begin{eqnarray}
&&\lbrace \theta_{i\alpha}(\tau ),\Gamma_a(\tau \vec \sigma )\rbrace =
(T^a)_{\alpha\beta}\theta_{i\beta} \delta^3(\vec \sigma -{\vec \eta}_i(\tau )),
\nonumber \\
&&\lbrace \theta^{*}_{i\alpha}(\tau ),\Gamma_a(\tau \vec \sigma )\rbrace =-
\theta^{*}_{i\beta}(T^a)_{\beta\alpha} \delta^3(\vec \sigma -{\vec \eta}
_i(\tau )).
\label{c7}
\end{eqnarray}

The Grassmann Dirac observables are

\begin{eqnarray}
{\check \theta}_{i\alpha}(\tau )&=&
[P\, e^{\Omega^{(\hat \gamma )}_{sa}(\eta^{(A)}
(\tau ,-{\vec \eta}_i(\tau ))) T^a}]^{\dagger}_{\alpha\beta}\theta_{i\beta}
(\tau ),\nonumber \\
&&{}\nonumber \\
&&\Rightarrow \lbrace {\check \theta}_{i\alpha}(\tau ),\Gamma_a(\tau ,\vec
\sigma )\rbrace =0,\nonumber \\
{\check \theta}^{*}_{i\alpha}(\tau )&=&\theta_{i\beta}^{*}(\tau )
[P\, e^{\Omega^{(\hat \gamma )}_{sa}(\eta^{(A)}
(\tau ,-{\vec \eta}_i(\tau ))) T^a}]_{\beta\alpha},\nonumber \\
&&{}\nonumber \\
&&\Rightarrow \lbrace {\check \theta}^{*}_{i\alpha}(\tau ),\Gamma_a(\tau ,\vec
\sigma )\rbrace =0,\nonumber \\
&&\lbrace {\check \theta}_{i\alpha}(\tau ),{\check \theta}^{*}_{j\beta}(\tau )
\rbrace =-i\delta_{ij}\delta_{\alpha\beta},
\label{c8}
\end{eqnarray}

\noindent with the path ordering evaluated in the fundamental representation
of SU(3). The Dirac observables for the non-Abelian charges of the particles
are

\begin{eqnarray}
{\check Q}_{ia}&=&i{\check \theta}^{*}_{i\alpha}(T^a)_{\alpha\beta}{\check
\theta}_{i\beta}=\nonumber \\
&=&Q_{ib}[P\, e^{\Omega^{(\hat \gamma )}_{sa}(\eta^{(A)}
(\tau ,{\vec \eta}_i(\tau ))) {\hat T}^a}]_{ba},\nonumber \\
&&{}\nonumber \\
&&\lbrace {\check Q}_{ia},\Gamma_b(\tau ,\vec \sigma )\rbrace =0,\nonumber \\
&&\lbrace {\check Q}_{ia},{\check Q}_{jb}\rbrace =\delta_{ij} c_{abc}{\check
Q}_{ic},
\label{c9}
\end{eqnarray}

\noindent where in the second line the path ordering is evaluated in the adjoint
representation of SU(3), since one has used the identity

\begin{equation}
e^{u_bT^b} T^a e^{u_cT^c}=T^c\, (e^{-u_b{\hat T}^b})_{ca}.
\label{c10}
\end{equation}

By using Eqs.(\ref{c3}) and (\ref{c4}), for $\eta^{(A)}_a(\tau ,\vec \sigma )
=\Gamma_a(\tau ,\vec \sigma )=0$ 
[so that also $\vec \partial \eta^{(A)}_a(\tau ,\vec \sigma )=0$, 
$\Omega_s^{(\hat \gamma )}(\eta^{(A)}(\tau ,\vec \sigma ))=0$ as shown in the
second paper of Ref.\cite{lusa}], namely in the generalized Coulomb gauge [we 
still go on to use the notation $\lbrace .,.\rbrace$ for the new Dirac brackets 
with respect to the second class constraints $\eta^{(A)}_a(\tau ,\vec \sigma )
\approx 0$, ${\tilde \Gamma}_a(\tau ,\vec \sigma )\approx 0$; instead the
temporal variables $A_{a\tau}$ and $\pi^{\tau}_a$ simply decouple],
we get

\begin{eqnarray}
{\vec \pi}_a(\tau ,\vec \sigma ) &\rightarrow& {\hat {\vec \pi}}_a(\tau ,
\vec \sigma ) = {\check {\vec \pi}}_{a\perp}(\tau ,\vec \sigma ) -\nonumber \\
&-&{{\vec \partial}\over {\triangle}} \int d^3\sigma^{'}\, {\vec \partial}
_{\sigma}\cdot {\vec \zeta}^{({\check A}_{\perp})}_{ab}(\vec \sigma ,{\vec 
\sigma}^{'};\tau ) [c_{bce} {\check A}^h_{c\perp}(\tau ,{\vec \sigma}^{'})
{\check \pi}^{h}_{e\perp}(\tau ,{\vec \sigma}^{'})-\sum_{i=1}^N{\check Q}_{ib}
\delta^3({\vec \sigma}^{'} -{\vec \eta}_i(\tau ))],\nonumber \\
&&{}\nonumber \\
&&{\vec \zeta}^{({\check A}_{\perp})}_{ab}(\vec \sigma ,{\vec \sigma}^{'};
\tau ) = \vec c (\vec \sigma -{\vec \sigma}^{'}) (P\, e^{\int_{{\sigma}^{'}}
^{\sigma} d{\vec \sigma}^{"}\cdot {\check {\vec A}}_{c\perp}(\tau ,{\vec 
\sigma}^{"}){\hat T}^c_s} )_{ab}.
\label{c11}
\end{eqnarray}

While in the electromagnetic case it is possible to get the physical Hamiltonian
without imposing the Coulomb gauge-fixing ${\tilde \eta}_{em}(x)\approx 0$
as in Ref.\cite{lus1}
(namely it is obtained by a canonical decoupling of the gauge degrees of freedom
), this is too difficult in the non-Abelian case. 
Therefore, we shall evaluate the
physical quantities by imposing the generalized Coulomb gauge-fixings
$\eta^{(A)}_a(\tau ,\vec \sigma )\approx 0$. Conceptually, 
the canonical decoupling of the gauge
degrees of freedom gives the same results for the physical quantities.

As shown in Refs.\cite{lusa,lv1}, the Noether identities implied by the second
Noether theorem, applied to the color SU(3) gauge group, give the following
result for the weak improper conserved non-Abelian Noether charges $Q_a$ 
and for the strong improper conserved ones $Q_{(s)a}$

\begin{eqnarray}
Q_a&=& g_s^{-2} c_{abc} \int d^3\sigma \, F^{ok}_b(\tau ,\vec \sigma )A^k_c
(\tau ,\vec \sigma ) -\sum_{i=1}^NQ_{ia} {\buildrel \circ \over =}\nonumber \\
&{\buildrel \circ \over=}& Q_{(s)a}=\int d^2\vec \Sigma \cdot {\vec E}
_a(\tau ,\vec \sigma ),
\label{c12}
\end{eqnarray}

Then, we get [see Eqs.(6-33)-(6-35) in the second paper of 
Ref.\cite{lusa}] the following Dirac's observables

\begin{eqnarray}
Q_a&{\rightarrow}&_{\eta^{(A)}\rightarrow 0} {\check Q}_a=
{\check Q}^{(YM)}_a(\tau )-\sum_{i=1}^N{\check Q}_{ia}(\tau )=\nonumber \\
&=&\int d^3\sigma {\check \rho}_a(\tau ,\vec \sigma )=\int d^3\sigma
[{\check \rho}^{(YM)}_a(\tau ,\vec \sigma )+\sum_{i=1}^N{\check \rho}_{ia}
(\tau ,\vec \sigma )],\nonumber \\
&&{}\nonumber \\
{\check \rho}^{(YM)}_a(\tau ,\vec \sigma )&=&-
c_{abc}{\check {\vec A}}_{b\perp}(\tau ,\vec \sigma )\cdot {\check {\vec \pi}}
_{c\perp}(\tau ,\vec \sigma ),\nonumber \\
{\check \rho}_{ia}(\tau ,\vec \sigma )&=&-{\check Q}_{ia}(\tau ) 
\delta^3(\vec \sigma -{\vec \eta}_i(\tau )),\nonumber \\
&&{}\nonumber \\
&&\lbrace {\check Q}_a,{\check Q}_b\rbrace =c_{abc}{\check Q}_c,\quad
\lbrace {\check Q}_a^{(YM)}(\tau ),{\check Q}_b^{(YM)}(\tau )\rbrace =c_{abc}
{\check Q}_c^{(YM)}(\tau ),\nonumber \\
&&\lbrace {\check {\vec A}}_{a\perp}(\tau ,\vec \sigma ),{\check Q}_b\rbrace =
c_{abc}{\check {\vec A}}_{c\perp}(\tau ,\vec \sigma )\nonumber \\
&&\lbrace {\check {\vec \pi}}_{a\perp}(\tau ,\vec \sigma ),{\check Q}_b
\rbrace =c_{abc} {\check {\vec \pi}}_{c\perp}(\tau ,\vec \sigma ).
\label{c13}
\end{eqnarray}

While  particle's 3-positions ${\vec \eta}_i(\tau )$ are gauge invariant,
for particle's momenta we have

\begin{equation}
\lbrace \kappa^r_i(\tau ),\Gamma_a(\tau ,\vec \sigma )\rbrace =-Q_{ia}(\tau )
{{\partial}\over {\partial \eta_i^r}}\delta^3(\vec \sigma -{\vec \eta}_i(\tau 
)).
\label{c14}
\end{equation}

To find the Dirac observables ${\check {\vec \kappa}}_i(\tau )$
[namely to dress the particles with a color cloud], we note that,
anaguously to the electromagnetic case\cite{lus1}, we expect that the form of the
minimal coupling is mantained, namely

\begin{equation}
\vec{\kappa}_{i}(\tau )+Q_{ia}(\tau )\vec{A}_{a}(\tau,\vec{\eta}_{i}(\tau ))=
\check{\vec{\kappa}}_{i}(\tau )+\check{Q}_{ia}(\tau )\check{\vec{A}}_{a\perp}
(\tau,\vec{\eta}_{i}(\tau )).
\label{c15}
\end{equation}

\noindent Since we have

\begin{equation}
\check{Q}_{ib}(\tau )\check{\vec{A}}_{a\perp}(\tau ,{\vec \eta}_i(\tau ))=Q_{ia}
(\tau )[P\, e^{\Omega^{(\hat \gamma )}_{sa}(\eta^{(A)}
(\tau ,-{\vec \eta}_i(\tau ))) {\hat T}^a}]_{ab}
\check{\vec{A}}_{b\perp}(\tau ,{\vec \eta}_i(\tau )),
\label{c16}
\end{equation}

\noindent Eqs.(\ref{c1}) suggest the following expression for the gauge 
invariant momenta

\begin{equation}
\check{\vec{\kappa}}_{i}(\tau )=\vec{\kappa}_{i}(\tau )+\sum_aQ_{ia}(\tau )
\vec{\Theta}_{a}
(\eta^{(A)}(\tau,\vec{\eta}_{i}(\tau )),\vec \partial \eta^{(A)}
(\tau,\vec{\eta}_{i}(\tau ))).
\label{c17}
\end{equation}

\noindent Since we have

\begin{eqnarray}
\{\Theta_{a}^r(\eta^{(A)}(\tau,\vec \sigma ),\vec \partial \eta^{(A)}
(\tau,\vec \sigma )),\Gamma_{b}(\tau,\vec{\sigma}')\}&=&
=-B_{ub}(\eta^{(A)}(\tau ,{\vec \sigma}^{'}))
{ {\delta[A_{av}(\eta^{(A)})\partial^{r}\eta^{(A)}_{v}](\tau,\vec{\sigma})} 
\over {\delta\eta^{(A)}_{u}(\tau,\vec{\sigma}')}}=\nonumber \\
&=&c_{abc}\Theta_{c}^r(\eta^{(A)}(\tau,\vec \sigma ),\vec \partial \eta^{(A)}
(\tau,\vec \sigma ))\delta^3(\vec \sigma -{\vec \sigma}^{'})+\nonumber \\
&+&\delta_{ab}{{\partial \delta^3(\vec{\sigma}-\vec{\sigma}')}\over {\partial 
\sigma^r}},
\label{c18}
\end{eqnarray}

\noindent we get the gauge invariance of ${\check {\vec \kappa}}_i(\tau )$

\begin{eqnarray}
\{\check{\kappa}^{r}_{i}(\tau),\Gamma_{b}(\tau,\vec{\sigma})\}&=&
\{\kappa^{r}_{i}(\tau),\Gamma_{b}(\tau,\vec{\sigma})\}+\nonumber \\
&+&\{Q_{ia}(\tau )\Theta^{r}_{a}
(\eta^{(A)}(\tau,\vec{\eta}_{i}(\tau )),\vec \partial \eta^{(A)}
(\tau,\vec{\eta}_{i}(\tau ))),\Gamma_{b}(\tau ,\vec{\sigma})\}=\nonumber\\
&=&-Q_{ib}(\tau )\frac{\partial}{\partial\eta^{r}_{i}}
\delta^3(\vec{\sigma}-\vec{\eta}_{i}(\tau ))+\nonumber \\
&+&c_{abc}\Theta^r_{a}
(\eta^{(A)}(\tau,\vec{\eta}_{i}(\tau )),\vec \partial \eta^{(A)}
(\tau,\vec{\eta}_{i}(\tau )))
Q_{ic}(\tau )\delta^3(\vec{\sigma}-\vec{\eta}_{i}(\tau ))+\nonumber\\
&&+Q_{ia}(\tau )c_{abc}\Theta_{c}^r
(\eta^{(A)}(\tau,\vec{\eta}_{i}(\tau )),\vec \partial \eta^{(A)}
(\tau,\vec{\eta}_{i}(\tau )))
\delta^3(\vec{\eta}_{i}(\tau )-\vec{\sigma})+\nonumber \\
&+&Q_{ib}(\tau )\frac{\partial}{\partial\eta^{r}_{i}}\delta(\vec{\eta}_{i}
(\tau )-\vec{\sigma})=0.
\label{c19}
\end{eqnarray}

We can now rewrite the constraints ${\cal H}(\tau )\approx 0$, ${\vec {\cal H}}
_p(\tau )\approx 0$, $N_i\approx 0$ in terms of the Dirac observables with 
respect to the SU(3) gauge transformations. The Grassmann constraints become

\begin{equation}
{\check N}_i=\sum_{\alpha =1}^3 {\check \theta}^{*}_{i\alpha}(\tau ){\check
\theta}_{i\alpha}(\tau ) \approx 0.
\label{c20}
\end{equation}

As shown in Ref.\cite{lusa}, we have $\sum_a{\vec B}^2_a(\tau ,\vec \sigma )=
\sum_a{\check {\vec B}}^2_a(\tau ,\vec \sigma )$ with the chromomagnetic
field ${\check {\vec B}}_a(\tau ,\vec \sigma )$ built in terms of ${\check
{\vec A}}_{a\perp}(\tau ,\vec \sigma )$. For the chromoelectric field we have
from Eqs.(\ref{c11}) and using Eqs.(\ref{c13}) and (\ref{c2})

\begin{eqnarray}
\sum_a&& {\vec \pi}^2_a(\tau ,\vec \sigma ){|}_{\eta^{(A)}_b=\Gamma_b=0}=
\nonumber \\
&=&\sum_a [{\check {\vec \pi}}_{a\perp}(\tau ,\vec \sigma )+{{\vec \partial}
\over {\triangle}} \int d^3\sigma_1 {\vec \partial}_{\sigma}\cdot {\vec \zeta}
^{({\check A}_{\perp})}_{ab}(\vec \sigma ,{\vec \sigma}_1;\tau ){\check \rho}_b
(\tau ,{\vec \sigma}_1)\, ]^2=\nonumber \\
&=&\sum_a[{\check {\vec \pi}}_{a\perp}(\tau ,\vec \sigma )-{{\vec \partial}\over
{\triangle}} \int d^3\sigma_1 (\delta_{ab}\delta^3(\vec \sigma -{\vec \sigma}
_1)+c_{amn}{\check {\vec A}}_{m\perp}(\tau ,\vec \sigma )\cdot {\vec \zeta}
^{({\check A}_{\perp})}_{nb}(\vec \sigma ,{\vec \sigma}_1;\tau )){\check \rho}_b
(\tau ,{\vec \sigma}_1)\, ]^2,\nonumber \\
&&{}\nonumber \\
\Rightarrow && \int d^3\sigma \sum_a{\vec \pi}^2_a(\tau ,\vec \sigma )
{|}_{\eta^{(A)}_b=\Gamma_b=0}= \int d^3\sigma \sum_a{\check {\vec \pi}}^2
_{a\perp}(\tau ,\vec \sigma )+\nonumber \\
&+&\int d^3\sigma \sum_a [{{\vec \partial}\over {\triangle}} \int d^3\sigma_1
(\delta_{ab}\delta^3(\vec \sigma -{\vec \sigma}
_1)+c_{amn}{\check {\vec A}}_{m\perp}(\tau ,\vec \sigma )\cdot {\vec \zeta}
^{({\check A}_{\perp})}_{nb}(\vec \sigma ,{\vec \sigma}_1;\tau )){\check \rho}_b
(\tau ,{\vec \sigma}_1)\, ]^2.
\label{c21}
\end{eqnarray} 

To obtain the last line, we have done an integration by parts and used the 
transversality of ${\check {\vec \pi}}_{a\perp}(\tau ,\vec \sigma )$. The final 
result is

\begin{eqnarray}
g^2_s \int d^3\sigma &&\sum_a{\vec \pi}^2_a(\tau ,\vec \sigma )
{|}_{\eta^{(A)}_b=\Gamma_b=0}= \nonumber \\
&=&\int d^3\sigma \sum_a{\check {\vec \pi}}^2
_{a\perp}(\tau ,\vec \sigma )+ V[{\vec \eta}_i,{\check {\vec A}}_{a\perp},
{\check {\vec \pi}}_{a\perp}](\tau ),
\label{c22}
\end{eqnarray}

\noindent with the potential given by

\begin{eqnarray}
V[{\vec \eta}_i,{\check {\vec A}}_{a\perp},{\check {\vec \pi}}_{a\perp}](\tau )
&=&g^2_s \int d^3\sigma d^3\sigma_1d^3\sigma_2 \sum_{a,b,c}[{\vec \partial}
_{\sigma}{\vec \zeta}_{ca}^{({\check A}_{\perp})}(\vec \sigma ,{\vec \sigma}_1;
\tau ){\check \rho}_a(\tau ,{\vec \sigma}_1)]\nonumber \\
&\cdot& {1\over {\triangle_{\sigma}}} [{\vec \partial}
_{\sigma}{\vec \zeta}_{cb}^{({\check A}_{\perp})}(\vec \sigma ,{\vec \sigma}_2;
\tau ){\check \rho}_b(\tau ,{\vec \sigma}_2)]=\nonumber \\
&=&g^2_s \int d^3\sigma_1d^3\sigma_2 \sum_{a,b}
{\check \rho}_a(\tau ,{\vec \sigma}_1)
\, K_{ab}({\vec \sigma}_1,{\vec \sigma}_2;\tau )\, {\check \rho}_b(\tau ,
{\vec \sigma}_2)=\nonumber \\
&=&g^2_s \int d^3\sigma_1d^3\sigma_2 \sum_{a,b}
[{\check \rho}^{(YM)}_a(\tau ,{\vec 
\sigma}_1)-\sum_{i=1}^N{\check Q}_{ia}(\tau )\delta^3({\vec \sigma}_1-{\vec
\eta}_i(\tau ))]\nonumber \\ 
&&K_{ab}({\vec \sigma}_1,{\vec \sigma}_2;\tau )\, [{\check 
\rho}^{(YM)}_b(\tau ,{\vec \sigma}_2)-\sum_{j=1}^N{\check Q}_{jb}(\tau )
\delta^3({\vec \sigma}_2-{\vec \eta}_j(\tau ))],\nonumber \\
&&{}\nonumber \\
K_{ab}({\vec \sigma}_1,{\vec \sigma}_2;\tau )&=&K_{ba}({\vec \sigma}_2,{\vec 
\sigma}_1;\tau )=\nonumber \\
&=&\int d^3\sigma_3d^3\sigma_4
\{ \, { {\delta_{ab}\delta^3(\vec{\sigma}_{3}-\vec{\sigma}_{1})
\delta^3(\vec{\sigma}_{4}-\vec{\sigma}_{2})}\over {4\pi\mid\vec{\sigma}_{3}
-\vec{\sigma}_{4}\mid} }+\nonumber \\
&+& { {\delta^3(\vec{\sigma}_{4}-\vec{\sigma}_{2})[\check{\vec{A}}_{\perp}(\tau 
,\vec{\sigma}_{3})\cdot {\vec{\zeta}}^{({\check A}_{\perp})}(\vec{\sigma}_{3},
\vec{\sigma}_{1};
\tau )]_{ab}}\over {4\pi\mid\vec{\sigma}_{3}-\vec{\sigma}_{4}\mid} }+
(\vec{\sigma}_{1}\leftrightarrow \vec{\sigma}_{2})+\nonumber \\
&+&\frac{[\check{\vec{A}}_{\perp}(\tau ,\vec{\sigma}_{3})\cdot
{\vec{\zeta}}^{({\check A}_{\perp})}(\vec{\sigma}_{3},
\vec{\sigma}_{1};\tau )]_{au}
      [\check{\vec{A}}_{\perp}(\tau ,\vec{\sigma}_{4})\cdot
{\vec{\zeta}}^{({\check A}_{\perp})}(\vec{\sigma}_{4},
\vec{\sigma}_{2};\tau )]_{bu}}
     {4\pi\mid\vec{\sigma}_{3}-\vec{\sigma}_{4}\mid} .
\label{c23}
\end{eqnarray}

\noindent The first line of this equation agrees with Eq.(6-25) of the second
paper in Refs.\cite{lusa}, while, from Eq.(6-27) of that paper, we get that 
$K_{ab}({\vec \sigma}_1,{\vec \sigma}_2;\tau )=-G^{({\check A}_{\perp})}
_{\triangle ,ab}({\vec \sigma}_1,{\vec \sigma}_2;\tau )$, where $G^{({\check A}
_{\perp})}_{\triangle ,ab}$ is the Green function of ${\hat {\vec D}}
^{({\check A}_{\perp})}_{ac}(\tau ,\vec \sigma )\cdot {\hat {\vec D}}
^{({\check A}_{\perp})}_{cb}(\tau ,\vec \sigma )$ [see Eqs.(3-16), (3-20),
(3-21), (3-25) of that paper].

The constraint giving the invariant mass of the system takes the form

\begin{eqnarray}
{\check {\cal H}}(\tau )&=&\epsilon_s-\sum_{i}\eta_{i}\sqrt{m_{i}^{2}+
(\check{\vec{\kappa}}_{i}(\tau)+\sum_{a}{\check Q}_{ai}(\tau )\check{\vec{A}}
_{a\perp}(\tau,\vec{\eta}_{i}(\tau )))^{2}}-\nonumber\\
&-&\sum_{a}\int d^{3}\sigma \sum_a \:[
\frac{g^{2}_{s}\check{\vec{\pi}}^{2}_{a\perp}(\tau,\vec{\sigma})}{2}+
\frac{\check{\vec{B}}_{a\perp}^{2}(\tau,\vec{\sigma})}{2g^{2}_{s}}]-
\frac{1}{2}V[\vec{\eta}_{i},\check{\vec{A}}_{a\perp},\check{\vec{\pi}}
_{a\perp}](\tau )=\nonumber \\
&=&\epsilon_s-H_{rel}\approx 0.
\label{c24}
\end{eqnarray}

We see that, since ${\check \rho}_a={\check \rho}^{(YM)}_a+\sum_{i=1}^N{\check
\rho}_{ia}$, there is a universal interaction kernel $K_{ab}({\vec \sigma}_1,
{\vec \sigma}_2;\tau )$ which creates the particle-particle, the particle-
field and the field-field interaction between the corresponding color charge
densities. This interaction kernel contains 3 kinds of instantaneous (i.e. at
equal $\tau$ on the Wigner hyperplane) interactions:

i) a Coulomb interaction;

ii) an interaction mediated by an arbitrary center (over whose spatial location
is integrated) : one color density has a
Coulomb interaction with the transverse potential at the center, which 
simultaneously interacts with the other color density through an
instantaneous ``Wilson line" along the geodesic (the straightline) on the
Wigner hyperplane, i.e. [${\check {\vec A}}_{\perp}={\check {\vec A}}_{a\perp}
{\hat T}^a$ with $({\hat T}^c)_{ab}=c_{cab}$]:

$[{\check {\vec A}}_{\perp}(\tau ,{\vec {\bar \sigma}}) \cdot {\vec \zeta}
^{({\check A}_{\perp})}({\vec {\bar \sigma}},\vec \sigma ;\tau )]_{ab}=
{\check {\vec A}}_{c\perp}(\tau ,\vec \sigma ) c_{cad} \cdot \vec c ({\vec {\bar
\sigma}}-\vec \sigma )\, (P\, e^{\int^{{\vec {\bar \sigma}}}_{\vec \sigma}
d{\vec \sigma}^{'} \cdot {\check {\vec A}}_{e\perp}(\tau ,{\vec \sigma}^{'})
{\hat T}^c}\, )_{db}$,

\noindent where ${\vec {\bar \sigma }}$ is the position of the center and 
$\vec \sigma$ that of the color density;

iii) an interaction mediated by two arbitrary centers (over whose spatial
location is integrated): each color density interacts, through a Wilson line
along the geodesic, with the transverse potential at one center and the two
centers have a mutual Coulomb interaction.

If we rescale the transverse potentials and electric fields [${\check {\vec
A}}_{a\perp}=g_s{\check {\vec {\tilde A}}}_{a\perp}$, ${\check {\vec \pi}}
_{a\perp}=g^{-1}_s{\check {\vec {\tilde \pi}}}_{a\perp}$, $\Rightarrow \,
{\check \rho}^{(YM)}_a=-c_{abc}{\check {\vec A}}_{b\perp}\cdot {\check {\vec 
\pi}}_{c\perp}=c_{abc}{\check {\vec {\tilde A}}}_{b\perp}\cdot {\check {\vec
{\tilde \pi}}}_{c\perp}$], we get

\begin{eqnarray}
{\check {\cal H}}(\tau )&=&\epsilon_s-\sum_{i}\eta_{i}\sqrt{m_{i}^{2}+
(\check{\vec{\kappa}}_{i}(\tau)+g_s\sum_{a}{\check Q}_{ai}(\tau )
{\check {\vec {\tilde A}}}
_{a\perp}(\tau,\vec{\eta}_{i}(\tau )))^{2}}-\nonumber\\
&-&\sum_{a}\int d^{3}\sigma {1\over 2}\sum_a \:[
{\check {\vec {\tilde \pi}}}^{2}_{a\perp}(\tau,\vec{\sigma})+
{\check {\vec {\tilde B}}}_{a}^{2}(\tau,\vec{\sigma})]-
\frac{1}{2}V[\vec{\eta}_{i},g_s{\check {\vec {\tilde A}}}_{a\perp},g^{-1}_s
{\check {\vec {\tilde \pi}}}_{a\perp}](\tau )\approx 0.
\label{c25}
\end{eqnarray}

To get the constraint defining the intrinsic rest frame, i.e. ${\check {\vec
{\cal H}}}_p(\tau )\approx 0$ of Eq.(\ref{b41}), we must consider the term
$\sum_a{\vec \pi}_a(\tau ,\vec \sigma ) \times {\vec B}_a(\tau ,\vec \sigma )=
\sum_a {\vec \pi}_a(\tau ,\vec \sigma ) \times {\check {\vec B}}_a(\tau ,\vec 
\sigma )$. From Eq.(\ref{c11}) we get

\begin{eqnarray}
\int d^3\sigma \sum_a && {\vec \pi}_a(\tau ,\vec \sigma )\times {\check {\vec 
B}}_a(\tau ,\vec \sigma ){|}_{\eta^{(A)}_b=\Gamma_b=0}=\nonumber \\
&=& \int d^3\sigma \sum_a [{\check {\vec \pi}}_{a\perp}(\tau ,\vec \sigma )+
{{\vec \partial}\over {\triangle}} \int d^3\sigma_1 {\vec \partial}_{\sigma}
\cdot {\vec \zeta}_{ab}^{({\check A}_{\perp})}(\vec \sigma ,{\vec \sigma}_1;
\tau ) {\check \rho}_b(\tau ,{\vec \sigma}_1)] \times {\check {\vec B}}_a(\tau 
,\vec \sigma ).
\label{c26}
\end{eqnarray}

\noindent Remembering that $({\vec \pi}_a\times {\vec B}_a)_r=\epsilon_{rmn}
\pi^m_aB^n_a=F_{ars}\pi^s_a$, we get

\begin{eqnarray}
\int d^3\sigma \sum_a && [{\vec \pi}_a(\tau ,\vec \sigma )\times {\check {\vec 
B}}_a(\tau ,\vec \sigma )]_r\, {|}_{\eta^{(A)}_b=\Gamma_b=0}=\nonumber \\
&=& \int d^3\sigma \sum_a {\check F}_{a\, rs}(\tau ,\vec \sigma ){\check 
\pi}^s_{a\perp}(\tau ,\vec \sigma )+\nonumber \\
&+&\int d^3\sigma \sum_a[\partial_r{\check A}_{a\perp s}-\partial_s{\check A}
_{a\perp r}+c_{abc}{\check A}_{b\perp r}{\check A}_{c\perp s}](\tau ,\vec
\sigma )\cdot \nonumber \\
&\cdot& {{\partial^s}\over {\triangle}} \int d^3\sigma_1 {\vec \partial}
_{\sigma}\cdot {\vec \zeta}_{ad}^{({\check A}_{\perp})}(\vec \sigma ,{\vec 
\sigma}_1;\tau ){\check \rho}_d(\tau ,{\vec \sigma}_1)=\nonumber \\
&=&\int d^3\sigma \sum_a {\check F}_{a\perp rs}(\tau ,\vec \sigma ){\check 
\pi}^s_{a\perp}(\tau ,\vec \sigma )+\nonumber \\
&+&\int d^3\sigma \sum_a{\check A}_{b\perp r}(\tau ,\vec \sigma ) [\delta_{ba}
\triangle -c_{bca}{\check A}_{c\perp s}(\tau ,\vec \sigma )
\partial_s]\cdot \nonumber \\
&\cdot& {1\over {\triangle}} \int d^3\sigma_1 {\vec \partial}
_{\sigma}\cdot {\vec \zeta}_{ad}^{({\check A}_{\perp})}(\vec \sigma ,{\vec 
\sigma}_1;\tau ){\check \rho}_d(\tau ,{\vec \sigma}_1).
\label{c27}
\end{eqnarray}

\noindent We have done a first integration by parts to use the transversality
of ${\check {\vec A}}_{a\perp}$ and then a second one on $\partial_s{\check A}
_{a\perp r}$. In the last expression we recognize the Faddeev-Popov operator 
${\tilde K}_{ba}^{({\check A}_{\perp})}(\tau ,\vec \sigma )=
-\vec \partial \cdot 
{\hat {\vec D}}^{({\check A}_{\perp})}_{ba}(\tau ,\vec \sigma )=\delta_{ba}
\triangle +c_{bac} {\check {\vec A}}_{c\perp}(\tau ,\vec \sigma )\cdot \vec
\partial=\delta_{ba}\triangle -c_{bca} {\check {\vec A}}_{c\perp}(\tau ,\vec 
\sigma )\cdot \vec \partial$. 
As shown in Eqs.(3-11) and (3-17) of the second paper in Refs.
\cite{lusa}, we have ${\tilde K}_{ab}^{({\check A}_{\perp})}(\tau ,\vec \sigma 
) {1\over {\triangle_{\sigma}}} {\vec \partial}_{\sigma}\cdot {\vec \zeta}_{bc}
^{({\check A}_{\perp})}(\vec \sigma ,{\vec \sigma}_1;\tau )=-\delta_{ac}
\delta^3(\vec \sigma -{\vec \sigma}_1)$. Therefore, we get

\begin{eqnarray}
\int d^3\sigma \sum_a && [{\vec \pi}_a(\tau ,\vec \sigma )\times {\check {\vec 
B}}_a(\tau ,\vec \sigma )]_r\, {|}_{\eta^{(A)}_b=\Gamma_b=0}=\nonumber \\
&=& \int d^3\sigma \sum_a [{\check F}_{a\, rs}{\check \pi}^s_{a\perp}-
{\check A}_{a\perp r}{\check \rho}_a](\tau ,\vec \sigma )=\nonumber \\
&=&\int \sum_a [(\partial_r{\check A}_{a\perp s}-\partial_s{\check A}_{a\perp r}
+c_{abc}{\check A}_{b\perp r}{\check A}_{c\perp s}){\check \pi}^s_{a\perp}-
\nonumber \\
&-&{\check A}_{a\perp r} (-c_{abc}{\check {\vec A}}_{b\perp}\cdot {\check {\vec
\pi}}_{c\perp}+\sum_{i=1}^N{\check \rho}_{ia})](\tau ,\vec \sigma )=\nonumber \\
&=&\sum_{i=1}^N\sum_a {\check Q}_{ia}(\tau ){\check A}_{a\perp r}(\tau ,{\vec 
\eta}_i(\tau ))+\int d^3\sigma \sum_a[(\partial_r{\check A}_{a\perp s}-
\partial_s{\check A}_{a\perp r}){\check \pi}^s_{a\perp}](\tau ,\vec \sigma ).
\label{c28}
\end{eqnarray}

Then, the constraints ${\check {\vec {\cal H}}}_p(\tau )\approx 0$ have the form

\begin{eqnarray}
{\check {\cal H}}_{p\, r}(\tau )&=&\sum_{i=1}^N {\check \kappa}_{ir}(\tau )+\int
d^3\sigma \sum_a[(\partial_r{\check A}_{a\perp s}-\partial_s{\check A}
_{a\perp r}) {\check \pi}^s_{a\perp}](\tau ,\vec \sigma )=\nonumber \\
&=&{\check \kappa}_{+ r}(\tau )+\int
d^3\sigma \sum_a[(\partial_r{\check A}_{a\perp s}-\partial_s{\check A}
_{a\perp r}) {\check \pi}^s_{a\perp}](\tau ,\vec \sigma )\approx 0.
\label{c29}
\end{eqnarray}

As in the electromagnetic case of Ref.\cite{lus1}, there is no interaction term
in this constraint (neither minimal coupling to particles nor self-coupling of 
the color field). As shown there, this is the requirement for being in an
instant form of the dynamics. The other requirement is that the rest-frame
spin tensor ${\bar S}^{rs}_s$ must also not depend on the interaction. From
Eq.(\ref{b46}), rewritten in terms of ${\check {\vec \eta}}_i$, ${\check {\vec 
\kappa}}_i$, one has

\begin{eqnarray}
\bar{S}^{rs}(\tau)&=&\sum_{i}\eta^{r}_{i}\kappa^{s}_{i}+\sum_{i}\eta^{r}_{i}
\sum_{a}Q_{ia}A^{s}_{a}(\vec{\eta}_{i})
-(r\leftrightarrow s)+\nonumber\\
&+&\int d^{3}\sigma\:\sigma^{r}F^{r}_{au}(\vec{\sigma})\pi^{u}_{a}(\vec{\sigma})
-(r\leftrightarrow s)=
\nonumber\\
&=&\sum_{i}\eta^{r}_{i}\check{\kappa}^{s}_{i}
+\sum_{i}\eta^{r}_{i}\sum_{a}\check{Q}_{ia}
\check{A}^{s}_{a\perp}(\vec{\eta}_{i})-(r\leftrightarrow s)+\nonumber\\
&+&\int d^{3}\sigma\:\sigma^{r}\check{F}^{s}_{au}(\vec{\sigma})
\check{\pi}^{u}_{a}(\vec{\sigma})-(r\leftrightarrow s).
\label{c30}
\end{eqnarray}

By doing various integrations by part, discarding
terms symmetric in $(r \leftrightarrow s)$ and using the Faddeev-Popov operator,
we obtain

\begin{eqnarray}
&& \int d^{3}\sigma\:\sigma_{r}\check{F}_{asu}(\vec{\sigma})
                        \check{\pi}^{u}_{a}(\vec{\sigma})
-(r\leftrightarrow s)=\nonumber\\
&=&\int d^{3}\sigma\:\sigma_{r}\check{F}_{asu}(\vec{\sigma})
                        \check{\pi}^{u}_{a\perp}(\vec{\sigma})+\nonumber\\
&+&\int d^{3}\sigma\:\sigma_{r}\check{F}_{asu}(\vec{\sigma})
   \frac{\partial^{u}}{\Delta}
   \int d^{3}\sigma'\:
\vec{\partial}\cdot\check{\vec{\zeta}}_{ab}(\vec{\sigma},\vec{\sigma}')
{\check \rho}_{b}(\vec{\sigma}')-
(r\leftrightarrow s)=\nonumber \\
&=&\int d^{3}\sigma\:\sigma_{r}\check{F}_{asu}(\vec{\sigma})
                        \check{\pi}^{u}_{a\perp}(\vec{\sigma})+\nonumber\\
&-&\int d^{3}\sigma\: \partial^{u}(\check{F}_{asu}\sigma^{r})
   \int d^{3}\sigma'\:\frac{
\vec{\partial}\cdot\check{\vec{\zeta}}_{ab}(\vec{\sigma},\vec{\sigma}')}{\Delta}
\rho^{(T)}_{b}(\vec{\sigma}')
-(r\leftrightarrow s)=\nonumber\\
&=&\int d^{3}\sigma\:\sigma_{r}\check{F}_{asu}(\vec{\sigma})
                        \check{\pi}^{u}_{a\perp}(\vec{\sigma})+\nonumber\\
&-&\int d^{3}\sigma\:[-\partial_{s}\check{A}_{ar\perp}(\vec{\sigma})-
                      \partial_{r}\check{A}_{as\perp}(\vec{\sigma})-
\partial^{u}\partial_{u}(A_{as\perp}(\vec{\sigma})\sigma^{r})+
c_{ahk}\check{A}_{ku\perp}(\vec{\sigma})\partial^{u}
      (\check{A}_{hs\perp}(\vec{\sigma})\sigma^{r})]\cdot\nonumber\\
&\cdot&\int d^{3}\sigma'\:\frac{
\vec{\partial}\cdot\check{\vec{\zeta}}_{ab}(\vec{\sigma},\vec{\sigma}')}{\Delta}
{\check \rho }_{b}(\vec{\sigma}')
-(r\leftrightarrow s)=\nonumber \\
&=&\int d^{3}\sigma\:\sigma_{r}\check{F}_{asu}(\vec{\sigma})
                        \check{\pi}^{u}_{a\perp}(\vec{\sigma})+\nonumber\\
&+&\int d^{3}\sigma\:\int d^{3}\sigma'\:[
\partial_{u}\partial^{u}(A_{as\perp}(\vec{\sigma})\sigma_{r})-
c_{hka}\check{A}_{ku\perp}(\vec{\sigma})\partial^{u}
      (\check{A}_{hs\perp}(\vec{\sigma})\sigma^{r})]\frac{
\vec{\partial}\cdot\check{\vec{\zeta}}_{ab}(\vec{\sigma},\vec{\sigma}')}{\Delta}
{\check \rho}_{b}(\vec{\sigma}')+\nonumber\\
&-&(r\leftrightarrow s)=\nonumber\\
&=&\int d^{3}\sigma\:\sigma_{r}\check{F}_{asu}(\vec{\sigma})
                        \check{\pi}^{u}_{a\perp}(\vec{\sigma})+\nonumber\\
&+&\int d^{3}\sigma\:\int d^{3}\sigma'\:\check{A}_{hs\perp}\sigma_{r}
(\vec{\sigma})
[\partial_{u}\partial^{u}\delta_{ha}+
c_{hka}A_{ku\perp}(\vec{\sigma})\partial^{u}]\frac{
\vec{\partial}\cdot\check{\vec{\zeta}}_{ab}(\vec{\sigma},\vec{\sigma}')}{\Delta}
{\check \rho}_{b}(\vec{\sigma}')+\nonumber\\
&-&(r\leftrightarrow s)=\nonumber \\
&=&\int d^{3}\sigma\:\sigma_{r}\check{F}_{asu}(\vec{\sigma})
                        \check{\pi}^{u}_{a\perp}(\vec{\sigma})
 -\int d^{3}\sigma\:\check{A}_{hs\perp}(\vec{\sigma})\sigma_{r}
                          {\check \rho}_{h}(\vec{\sigma})
-(r\leftrightarrow s).
\label{c31}
\end{eqnarray}

In conclusion, we get [in the last line we use ${\check {\vec {\cal H}}}_p(\tau 
)\approx 0$ and go to relative variables]

\begin{eqnarray}
{\bar S}^{rs}_s&=&\sum_{i=1}^N\eta^r_i(\tau ){\check \kappa}^s_i(\tau )+\int
d^3\sigma \sigma^r \sum_a[\partial^s{\check A}_{a\perp u}(\tau ,\vec \sigma )-
\partial_u{\check A}^s_{a\perp}(\tau ,\vec \sigma )] {\check \pi}^u_{a\perp}
(\tau ,\vec \sigma )-\nonumber \\
&-& (r \leftrightarrow s) \approx \nonumber \\
&\approx& \sum_{\bar a=1}^{N-1}\rho^r_{\bar a}(\tau ){\check \pi}^s_{\bar a}
(\tau )+ \int d^3\sigma (\sigma^r-\eta^r_{+}(\tau )) 
\sum_a[\partial^s{\check A}_{a\perp u}(\tau ,\vec \sigma )-
\partial_u{\check A}^s_{a\perp}(\tau ,\vec \sigma )] {\check \pi}^u_{a\perp}
(\tau ,\vec \sigma )-\nonumber \\
&-&(r \leftrightarrow s) ,
\label{c32}
\end{eqnarray}

\noindent as expected in an instant form also in the non-Abelian case.

\vfill\eject

\section{The reduced  Hamilton-Dirac equations}

To write the reduced Hamilton equations, it is convenient to introduce the
gauge-fixing

\begin{equation}
\chi=T_{s}-\tau\approx 0,
\label{d1}
\end{equation}

\noindent whose conservation in time requires $\lambda(\tau)=-1$ in 
Eq.(\ref{b47}). Then, we can eliminate the pair of variables  $(T_{s},
\epsilon_{s})$ and describe the evolution in terms of the rest-frame time.
The Hamiltonian for this evolution will be

\begin{equation}
\hat{H}_{D}=H_{rel}-\vec{\lambda}(\tau)\cdot {\check {\vec {\cal H}}}_p(\tau)+
\sum_{i=1}^N\mu_i(\tau ){\check N}_i,
\label{d2}
\end{equation}

\noindent with

\begin{eqnarray}
H_{rel}&=&\sum_{i}\eta_{i}\sqrt{m_{i}^{2}+
(\check{\vec{\kappa}}_{i}(\tau)+\sum_{a}\check{Q}_{ia}(\tau )
\check{\vec{A}}_{a\perp}(\tau,\vec{\eta}_{i}(\tau )))^{2}}
+\nonumber\\
&+&\sum_{a}\int d^{3}\sigma\:
[\frac{g^{2}_{s}\check{\vec{\pi}}^{2}_{a\perp}(\tau,\vec{\sigma})}{2}+
\frac{\check{\vec{B}}^{2}_{a\perp}(\tau,\vec{\sigma})}{2g^{2}_{s}}]+
\frac{1}{2}V[{\vec{\eta}}_{i},\check{\vec{A}}_{a\perp},
\check{\vec{\pi}}_{a\perp}](\tau ).
\label{d3}
\end{eqnarray}

\noindent If we would know the correct form of the gauge-fixing to eliminate
the intrinsic center-of-mass degrees of freedom associated with the constraints
${\check {\vec {\cal H}}}_p(\tau )\approx 0$, we could put $\vec \lambda (\tau )
=0$ and rewrite the invariant mass $H_{rel}$ only in terms of relative 
variables. See the analogous discussion for the electromagnetic case in
Ref.\cite{alba}

At this stage of reduction, we get the following Hamilton equations of motion
[$\tau \equiv T_s$; $P^{rs}_{\perp}(\vec \sigma )=\delta^{rs}+\partial^r
\partial^s/\triangle$; ${\vec \eta}_i=(\eta^r_i)$; $\partial /\partial {\vec
\eta}_i=(\partial /\partial \eta^r_i)$; the symbol ${\buildrel \circ \over =}$ 
means evaluated on the solutions of the equations of motion]

\begin{eqnarray}
{d\over {d\tau}}{\vec{\eta}}_{i}(\tau)&\, {\buildrel \circ \over =}\,& \eta_{i}
\frac{
\check{\vec{\kappa}}_{i}(\tau)+\sum_{a}\check{Q}_{ia}(\tau )
\check{\vec{A}}_{a\perp}(\tau,\vec{\eta}_{i}(\tau ))}
{\sqrt{m_{i}^{2}+(
\check{\vec{\kappa}}_{i}(\tau)+\sum_{a}\check{Q}_{ia}(\tau )
\check{\vec{A}}_{a\perp}(\tau,\vec{\eta}_{i}(\tau )))^{2}}}
-\vec{\lambda}(\tau)\nonumber\\
{d\over {d\tau}}{\check{\vec{\kappa}}}_{i}(\tau)&\, {\buildrel \circ \over =}\,&
-\sum_{u}({{d{\eta}^{u}_{i}(\tau)}\over {d\tau}}+\lambda^{u}(\tau))\sum_{a}
{\check Q}_{ia}(\tau ){{\partial \check{A}^{u}_{a\perp}(\tau,\vec{\eta}
_{i}(\tau ))}\over {\partial {\vec \eta}_i}}-\nonumber \\
&-&g^{2}_{s}\sum_{a,b}\check{Q}_{ia}(\tau )\int d^{3}\sigma'
{{\partial K_{ab}(\vec{\eta}_{i}(\tau ),\vec{\sigma'};\tau )}\over {\partial
{\vec \eta}_i}} {\check \rho}_b(\tau ,\vec{\sigma'}),\nonumber \\
&&{}\nonumber \\
{d\over {d\tau}} {\check \theta}_{i\alpha}(\tau )\, &{\buildrel \circ \over 
=}\,& \sum_a(T^a)_{\alpha\beta}{\check \theta}_{i\beta}(\tau ) [{\dot {\vec 
\eta}}_i(\tau )\cdot {\check {\vec A}}_{b\perp}(\tau ,{\vec \eta}_i(\tau ))-
\nonumber \\
&-&g^2_s \int d^3\sigma K_{ab}({\vec \eta}_i(\tau ),\vec \sigma ;\tau ){\check 
\rho}_b(\tau ,\vec \sigma )]-i\mu_i(\tau ){\check \theta}_{i\alpha}(\tau ),
\nonumber \\
{d\over {d\tau}} {\check \theta}^{*}_{i\alpha}(\tau )\, &{\buildrel \circ \over 
=}\,& -\sum_a {\check \theta}^{*}_{i\beta}(\tau )(T^a)_{\beta\alpha} [{\dot 
{\vec \eta}}_i(\tau )\cdot {\check {\vec A}}_{b\perp}(\tau ,{\vec \eta}_i(\tau 
))-\nonumber \\
&-&g^2_s \int d^3\sigma K_{ab}({\vec \eta}_i(\tau ),\vec \sigma ;\tau ){\check 
\rho}_b(\tau ,\vec \sigma )]+i\mu_i(\tau ){\check \theta}^{*}_{i\alpha}(\tau ),
\nonumber \\
{d\over {d\tau}}{\check Q}_{ia}(\tau )\, &{\buildrel \circ \over =}\,& c_{acd}
{\check Q}_{id}(\tau )[{\dot {\vec \eta}}_i(\tau )\cdot {\check {\vec A}}
_{c\perp}(\tau ,{\vec \eta}_i(\tau ))-\nonumber \\
&-&g^2_s \int d^3\sigma K_{cb}({\vec \eta}_i(\tau ),\vec \sigma ;\tau ){\check 
\rho}_b(\tau ,\vec \sigma )],
\nonumber \\
&&{},
\label{d4}
\end{eqnarray}

\begin{eqnarray}
{{\partial}\over {\partial \tau}}
{\check{A}}_{a\perp r}(\tau,\vec{\sigma})&\, {\buildrel \circ \over =}\,&
-g^{2}_{s}\check{\pi}_{a\perp r}(\tau,\vec{\sigma})
-[\vec{\lambda}(\tau)\cdot\vec{\partial}]
\check{A}_{a\perp r}(\tau,\vec{\sigma})+\nonumber\\
&+&g^{2}_{s} P^{rs}_{\perp}(\vec{\sigma}) \int d^{3}\sigma'\:c_{abu}
[\check{A}_{u\perp}^{s}(\tau ,\vec{\sigma})K_{bc}(\vec{\sigma},\vec{\sigma}';
\tau ))]{\check \rho}_{c}(\tau ,\vec{\sigma'})\nonumber \\
{{\partial}\over {\partial \tau}}
{\check{\pi}}^{r}_{a\perp}(\tau,\vec{\sigma})&\, {\buildrel \circ \over =}\,&
g^{-2}_sP^{rs}_{\perp}(\vec \sigma ){\tilde K}_{ab}(\check{A}_{\perp})(\tau ,
\vec \sigma )\check{A}^{s}_{b\perp}(\tau,\vec{\sigma})-
[\vec{\lambda}(\tau)\cdot\vec{\partial}]\check{\pi}^{r}_{a\perp}
(\tau,\vec{\sigma})-\nonumber\\
&-&\sum_{i=1}^N \check{Q}_{ia}(\tau )P^{rs}_{\perp}(\vec{\sigma})
({{d\eta^s_i(\tau )}\over {d\tau}}+\lambda^s(\tau ))
\delta^3(\vec{\sigma}-\vec{\eta}_{i}(\tau ))+
\nonumber\\
&+&\frac{1}{2} P^{rs}_{\perp}(\vec \sigma )
{{\delta V[\vec{\eta}_{i},\check{\vec{A}}_{a\perp},\check{\vec{\pi}}_{a\perp}]
(\tau )}\over {\delta {\check A}_{a\perp s}(\tau ,\vec \sigma )}},\nonumber \\
&&{}\nonumber \\
{d\over {d \tau}}{\check Q}^{(YM)}_a(\tau )&\, {\buildrel \circ \over =}\,& -
{d\over {d \tau}}\sum_{i=1}^N{\check Q}_{ia}(\tau )=-c_{acd}\sum_{i=1}^N
{\check Q}_{id}(\tau )[{\dot {\vec \eta}}_i(\tau )\cdot {\check {\vec A}}
_{c\perp}(\tau ,{\vec \eta}_i(\tau ))-\nonumber \\
&-&g^2_s \int d^3\sigma K_{cb}({\vec \eta}_i(\tau ),\vec \sigma ;\tau ){\check 
\rho}_b(\tau ,\vec \sigma )],
\label{d5}
\end{eqnarray}

\noindent since ${d\over {d\tau}} {\check Q}_a\, {\buildrel \circ \over =}\, 0$
[it is not known (see also Ref.\cite{lusa}) how to check this formula by a 
direct calculation].

In Eqs.(\ref{d5}) we have

\begin{eqnarray}
{1\over 2}
{{\delta V[\vec{\eta}_{i},\check{\vec{A}}_{a\perp},\check{\vec{\pi}}_{a\perp}]
(\tau )}\over {\delta {\check A}_{a\perp s}(\tau ,\vec \sigma )}}&=&
-g^2_sP^{rs}_{\perp}(\vec \sigma ) c_{abv}{\check \pi}^s_{v\perp}(\tau ,\vec 
\sigma ) \int d^3\sigma^{'} K_{bc}(\vec \sigma ,{\vec \sigma}^{'};\tau ){\check
\rho}_c(\tau ,{\vec \sigma}^{'})+\nonumber \\
&&+g^2_sP^{rs}_{\perp}(\vec \sigma ) \int d^3\sigma_1d^3\sigma_2 {\check \rho}
_b(\tau ,{\vec \sigma}_1) \nonumber \\
&&\{ \int {{d^3\sigma^{'}}\over {4\pi |{\vec \sigma}^{'}
-{\vec \sigma}_2|}} [-\delta^3(\vec \sigma -{\vec \sigma}^{'}) c_{abe}\zeta
^{({\check A}_{\perp})s}_{ec}({\vec \sigma}^{'},{\vec \sigma}_1;\tau )+
\nonumber \\
&&+c_{bef}{\check {\vec A}}_{f\perp}(\tau ,{\vec \sigma}^{'})\cdot {\vec \zeta}
^{({\check A}_{\perp})}_{em}({\vec \sigma}^{'},\vec \sigma ;\tau ) c_{man}
\zeta^{({\check A}_{\perp})s}_{nc}(\vec \sigma ,{\vec \sigma}_1;\tau )]+
({\vec \sigma}_1\leftrightarrow {\vec \sigma}_2)+\nonumber \\
&&+\int {{d^3\sigma_3d^3\sigma_4}\over {4\pi |{\vec \sigma}_3-{\vec \sigma}_4|}}
\{ [-\delta^3(\vec \sigma -{\vec \sigma}_3) c_{abe} \zeta^{({\check A}_{\perp})
s}_{eu}({\vec \sigma}_3,{\vec \sigma}_1;\tau )+\nonumber \\
&&+c_{bef} {\check {\vec A}}
_{f\perp}(\tau ,{\vec \sigma}_3)\cdot {\vec \zeta}^{({\check A}_{\perp})}_{em}
({\vec \sigma}_3,\vec \sigma ;\tau ) c_{man} \zeta^{({\check A}_{\perp})s}
_{nu}(\vec \sigma ,{\vec \sigma}_1;\tau )] \cdot \nonumber \\
&&c_{ers}{\check {\vec A}}_{s\perp}(\tau ,{\vec \sigma}_4)\cdot {\vec \zeta}
^{({\check A}_{\perp})}_{ru}({\vec \sigma}_4,{\vec \sigma}_2;\tau )+\nonumber \\
&&c_{bef}
{\check {\vec A}}_{f\perp}(\tau ,{\vec \sigma}_3)\cdot {\vec \zeta}^{({\check A}
_{\perp})}_{eu}({\vec \sigma}_3,{\vec \sigma}_1;\tau )\nonumber \\
&&[-\delta^3(\vec \sigma -{\vec \sigma}_4) c_{acm} \zeta^{({\check A}
_{\perp})s}_{mu}({\vec \sigma}_4,{\vec \sigma}_2;\tau )+\nonumber \\
&&+c_{cmn}{\check {\vec A}}
_{n\perp}(\tau ,{\vec \sigma}_4)\cdot {\vec \zeta}^{({\check A}_{\perp})}
_{mr}({\vec \sigma}_4,\vec \sigma ;\tau ) c_{ras} \zeta^{({\check A}_{\perp})s}
_{su}(\vec \sigma ,{\vec \sigma}_2;\tau )]\}\, \}{\check \rho}_c(\tau ,{\vec 
\sigma}_2),\nonumber \\
&&{}
\label{d6}
\end{eqnarray}

\noindent because from Eq.(\ref{c2}) we get

\begin{eqnarray}
{\hat {\vec D}}^{({\check A}_{\perp})}_{ab}(\tau ,{\vec \sigma}_1) &\cdot&
{{\delta {\vec \zeta}^{({\check A}_{\perp})}_{bc}({\vec \sigma}_1,{\vec \sigma}
_2;\tau )}\over {\delta {\check A}_{u\perp s}(\tau ,\vec \sigma )}}=-
P^{st}_{\perp}(\vec \sigma )
\delta^3(\vec \sigma -{\vec \sigma}_1) c_{aub} \zeta^{({\check A}_{\perp}) t}
_{bc}({\vec \sigma}_1,{\vec \sigma}_2;\tau ),\nonumber \\
&&{}\nonumber \\
\Rightarrow && {{\delta {\vec \zeta}^{({\check A}_{\perp})}_{bc}({\vec \sigma}_1
,{\vec \sigma}_2;\tau )}\over {\delta {\check A}_{u\perp s}(\tau ,\vec \sigma 
)}}=P^{st}_{\perp}(\vec \sigma )
{\vec \zeta}^{({\check A}_{\perp})}_{be}({\vec \sigma}_1,\vec \sigma ;\tau )
c_{eud} \zeta^{({\check A}_{\perp}) t}_{dc}(\vec \sigma ,{\vec \sigma}_2;\tau ).
\label{d7}
\end{eqnarray}

These equations of motion have to be supplemented with the constraints

\begin{eqnarray}
{\check {\cal H}}_{p\, r}(\tau )&=&\sum_{i=1}^N {\check \kappa}_{ir}(\tau )+\int
d^3\sigma \sum_a[(\partial_r{\check A}_{a\perp s}-\partial_s{\check A}
_{a\perp r}) {\check \pi}^s_{a\perp}](\tau ,\vec \sigma )\approx 0,
\nonumber \\
{\check N}_i&=&\sum_{\alpha =1}^3\theta^{*}_{i\alpha}(\tau )\theta_{i\alpha}
(\tau )\approx 0.
\label{d8}
\end{eqnarray}

The first line of Eq.(\ref{d4}) can  been inverted to get

\begin{eqnarray}
\check{\vec{\kappa}}_{i}(\tau)\, &{\buildrel \circ \over =}&\, \eta_{i}
m_{i}\frac{\dot{\vec{\eta}}_{i}(\tau)+\vec{\lambda}(\tau)}
{\sqrt{1-
(\dot{\vec{\eta}}_{i}(\tau)+\vec{\lambda}(\tau))^{2}}}
-\sum_{a}{\check Q}_{ia}(\tau )\check{\vec{A}}_{a\perp}(\tau,\vec{\eta}_{i}
(\tau )).
\label{d9}
\end{eqnarray}

The first line of Eqs.(\ref{d5}) coincides with Eq.(6-24) of the second paper
in Refs.\cite{lusa} if in this equation ${\check {\vec \pi}}_{a\perp} 
\rightarrow -{\check {\vec \pi}}_{a\perp}$, $i{\check \psi}^{\dagger} T^c 
\check \psi \rightarrow \sum_{i=1}^N{\check Q}_{ic}$ and if we note (see
Ref.\cite{lusa}) that ${\hat D}^{({\check A}_{\perp})k}_{vb}(\tau ,\vec \sigma )
P^{kn}_{\perp}(\vec \sigma ) \zeta^{({\check A}_{\perp})n}_{bc}(\vec \sigma ,
{\vec \sigma}^{'};\tau )=0$, $K_{ab}({\vec \sigma}_1,{\vec \sigma}_2;\tau )=-
G^{({\check A}_{\perp})}_{\triangle ,ab}({\vec \sigma}_1,{\vec \sigma}_2;
\tau )$ and (see Eq.(3-17) of that paper)
${\vec \zeta}^{({\check A}_{\perp})}_{ac}(\vec \sigma ,{\vec \sigma}
^{'};\tau )=-{\hat {\vec D}}^{({\check A}_{\perp})}_{ab}(\tau ,\vec \sigma )
G^{({\check A}_{\perp})}_{\triangle ,bc}(\vec \sigma ,{\vec \sigma}^{'};\tau )=
{\hat {\vec D}}^{({\check A}_{\perp})}_{ab}(\tau ,\vec \sigma )
K_{bc}(\vec \sigma ,{\vec \sigma}^{'};\tau )$.
Therefore, Eq.(6-21) of that paper gives its inversion in the
form

\begin{eqnarray}
{\check \pi}^r_{a\perp}(\tau ,\vec \sigma )&=&g^{-2}_s{\check E}^r_{a\perp}(\tau
,\vec \sigma )\, {\buildrel \circ \over =}\,
-g^{-2}_s P^{rs}_{\perp}(\vec \sigma ) \int d^3\sigma_1 {\cal P}^{({\check A}
_{\perp})st}_{ab}(\vec \sigma ,{\vec \sigma}_1;\tau )\cdot \nonumber \\
&&[{{\partial }\over {\partial \tau}}+
\vec \lambda (\tau )\cdot {{\partial}\over {\partial {\vec \sigma}_1}}]
{\check A}^t_{b\perp}(\tau ,{\vec \sigma}_1)-\nonumber \\
&&-P^{rs}_{\perp}(\vec \sigma ) \int d^3\sigma_1 \zeta^{({\check A}_{\perp})s}
_{ab}(\vec \sigma ,{\vec \sigma}_1;\tau ) \sum_{i=1}^N{\check \rho}_{ib}(\tau ,
{\vec \sigma}_1)=\nonumber \\
&=&-g^{-2}_sP^{rs}_{\perp}(\vec \sigma ) \int d^3\sigma_1 [\delta^{st}\delta
_{ab}\delta^3(\vec \sigma -{\vec \sigma}_1)+\nonumber \\
&+&{\hat D}^{({\check A}_{\perp}) s}_{ad}(\tau ,\vec \sigma )K_{de}(\vec \sigma 
,{\vec \sigma}_1;\tau ){\hat D}^{({\check A}_{\perp}) t}_{eb}(\tau ,{\vec
\sigma}_1)]\, [{{\partial }\over {\partial \tau}}+
\vec \lambda (\tau )\cdot {{\partial}\over {\partial {\vec \sigma}_1}}]
{\check A}^t_{b\perp}(\tau ,{\vec \sigma}_1)+\nonumber \\
&+&P^{rs}_{\perp}(\vec \sigma ) c_{adc}{\check A}^s_{d\perp}(\tau ,\vec \sigma )
\sum_{i=1}^N K_{cb}(\vec \sigma ,{\vec \eta}_i(\tau );\tau ) {\check Q}_{ib}
(\tau ),
\label{d10}
\end{eqnarray}

\noindent where [see Eq.(3-30) of Ref.\cite{lusa}] ${\cal P}^{({\check A}
_{\perp})rs}_{ab}(\vec \sigma ,{\vec \sigma}_1;\tau )=\delta^{rs}\delta_{ab}
\delta^3(\vec \sigma -{\vec \sigma}_1)-{\hat D}_{ad}^{({\check A}_{\perp})r}
(\tau ,\vec \sigma ) G^{({\check A}_{\perp})}_{\triangle ,dc}(\vec \sigma ,{\vec
\sigma}_1;\tau ) {\hat D}^{({\check A}_{\perp})s}_{cb}(\tau ,{\vec \sigma}_1)=
\delta^{rs}\delta_{ab}\delta^3(\vec \sigma -{\vec \sigma}_1)+\zeta^{({\check A}
_{\perp})r}_{ac}(\vec \sigma ,{\vec \sigma}_1;\tau ){\hat D}^{({\check A}
_{\perp})s}_{cb}(\tau ,{\vec \sigma}_1)$ is a projector giving an explicit
realization of the Mitter-Viallet abstract metric [it satisfies
$\int d^3\sigma_1 {\cal P}^{({\check A}_{\perp})ij}_{ab}(\vec \sigma ,{\vec 
\sigma}_1;\tau ) {\cal P}^{({\check A}_{\perp})jk}_{bc}({\vec \sigma}_1,{\vec 
\sigma}^{'};\tau )={\cal P}^{({\check A}_{\perp})ik}_{ac}(\vec \sigma ,{\vec
\sigma}^{'};\tau )$; $\int d^3\sigma_1 {\cal P}^{({\check A}_{\perp})ij}_{ab}
(\vec \sigma ,{\vec \sigma}_1;\tau ){\hat D}^{({\check A}_{\perp})j}_{bc}(\tau 
,{\vec \sigma}_1)=\int d^3\sigma_1{\hat D}^{({\check A}_{\perp})i}_{ab}(\tau 
,{\vec \sigma}_1) {\cal P}^{({\check A}_{\perp})ij}_{bc}({\vec \sigma}_1,\vec 
\sigma ;\tau )=0$].

By using also Eq.(6-23) of Ref.\cite{lusa}, namely ${\hat D}
^{({\check A}_{\perp})r}_{ac}(\tau ,\vec \sigma )P^{rs}_{\perp}(\vec \sigma )
{\hat D}^{({\check A}_{\perp}) s}_{cb}(\tau ,\vec \sigma )=0$ (when acting on 
functions of $\vec \sigma$), we get

\begin{eqnarray}
{\check \rho}_a(\tau ,\vec \sigma )&=&g^{-2}_sc_{abc}{\check A}^r_{b\perp}(\tau 
,\vec \sigma ) P^{rs}_{\perp}(\vec \sigma ) \int d^3\sigma_1 {\cal P}
^{({\check A}_{\perp}) st}_{cd}(\vec \sigma ,{\vec \sigma}_1;\tau )
[{{\partial }\over {\partial \tau}}+
\vec \lambda (\tau )\cdot {{\partial}\over {\partial {\vec \sigma}_1}}]
{\check A}^t_{b\perp}(\tau ,{\vec \sigma}_1)+\nonumber \\
&+&\sum_{i=1}^N\int d^3\sigma_1 [\delta_{ad}\delta^3(\vec \sigma -{\vec \sigma}
_1)+c_{abc}{\check A}^r_{b\perp}(\tau ,\vec \sigma )P^{rs}_{\perp}(\vec \sigma 
)\zeta^{({\check A}_{\perp}) s}_{cd}(\vec \sigma ,{\vec \sigma}_1;\tau )]
{\check \rho}_{id}(\tau ,{\vec \sigma}_1)=\nonumber \\
&=&g^{-2}_sc_{abc}{\check {\vec A}}_{b\perp}(\tau ,\vec \sigma )
\cdot [{{\partial }\over {\partial \tau}}+
\vec \lambda (\tau )\cdot {{\partial}\over {\partial \vec \sigma}}]
{\check {\vec A}}_{c\perp}(\tau ,\vec \sigma)+\sum_{i=1}^N{\check \rho}_{ia}
(\tau ,\vec \sigma )=\nonumber \\
&=&{\check \rho}^{(YM)}_{a}(\tau ,\vec \sigma )+\sum_{i=1}^N{\check \rho}_{ia}
(\tau ,\vec \sigma ),\nonumber \\
&&{}\nonumber \\
{\check \rho}^{(YM)}_a(\tau ,\vec \sigma )&=&-g^{-2}_s c_{abc}{\check {\vec 
A}}_{b\perp}(\tau ,\vec \sigma )\cdot {\check {\vec E}}_{c\perp}(\tau ,\vec
\sigma ).
\label{d11}
\end{eqnarray}

Therefore by putting $\vec \lambda (\tau )={\dot {\vec g}}(\tau )$ like in
Ref.\cite{alba}
, the equations of motion for ${\check {\vec A}}_{a\perp}(\tau ,\vec
\sigma )$ become

\begin{eqnarray}
[{{\partial }\over {\partial \tau}}&+&
\vec \lambda (\tau )\cdot {{\partial}\over {\partial \vec \sigma}}]
\{ -g^{-2}_s P^{rs}_{\perp}(\vec \sigma )
\int d^3\sigma_1 {\cal P}^{({\check A}
_{\perp})st}_{ab}(\vec \sigma ,{\vec \sigma}_1;\tau )\cdot \nonumber \\
&&[{{\partial}\over {\partial \tau}}+
{\dot {\vec g}}(\tau )\cdot {{\partial}
\over {\partial {\vec \sigma}_1}}]
{\check A}^t_{b\perp}(\tau ,{\vec \sigma}_1)-\nonumber \\
&&-P^{rs}_{\perp}(\vec \sigma ) \int d^3\sigma_1 \zeta^{({\check A}_{\perp})s}
_{ab}(\vec \sigma ,{\vec \sigma}_1;\tau ) \sum_{i=1}^N{\check \rho}_{ib}(\tau ,
{\vec \sigma}_1) \}-\nonumber \\
&&-g^{-2}_s
P^{rs}_{\perp}(\vec \sigma ) [\delta_{ab}\triangle +c_{abc}{\check {\vec
A}}_{c\perp}(\tau ,\vec \sigma )\cdot \vec \partial ]{\check A}^s_{b\perp}(\tau
,\vec \sigma )\, {\buildrel \circ \over =}\, \nonumber \\
&&{\buildrel \circ \over =}\, \sum_{i=1}^N {\check Q}_{ia}(\tau ) P^{rs}
_{\perp}(\vec \sigma ) [{\dot \eta}^s_i(\tau )+\lambda^s(\tau )]\delta^3(\vec
\sigma -{\vec \eta}_i(\tau ))+\nonumber \\
&&-g^2_sP^{rs}_{\perp}(\vec \sigma ) c_{abv}{\check \pi}^s_{v\perp}(\tau ,\vec 
\sigma ) \int d^3\sigma^{'} K_{bc}(\vec \sigma ,{\vec \sigma}^{'};\tau ){\check
\rho}_c(\tau ,{\vec \sigma}^{'})+\nonumber \\
&&+g^2_sP^{rs}_{\perp}(\vec \sigma ) \int d^3\sigma_1d^3\sigma_2 {\check \rho}
_b(\tau ,{\vec \sigma}_1) \nonumber \\
&&\{ \int {{d^3\sigma^{'}}\over {4\pi |{\vec \sigma}^{'}
-{\vec \sigma}_2|}} [-\delta^3(\vec \sigma -{\vec \sigma}^{'}) c_{abe}\zeta
^{({\check A}_{\perp})s}_{ec}({\vec \sigma}^{'},{\vec \sigma}_1;\tau )+
\nonumber \\
&&+c_{bef}{\check {\vec A}}_{f\perp}(\tau ,{\vec \sigma}^{'})\cdot {\vec \zeta}
^{({\check A}_{\perp})}_{em}({\vec \sigma}^{'},\vec \sigma ;\tau ) c_{man}
\zeta^{({\check A}_{\perp})s}_{nc}(\vec \sigma ,{\vec \sigma}_1;\tau )]+
({\vec \sigma}_1\leftrightarrow {\vec \sigma}_2)+\nonumber \\
&&+\int {{d^3\sigma_3d^3\sigma_4}\over {4\pi |{\vec \sigma}_3-{\vec \sigma}_4|}}
\{ [-\delta^3(\vec \sigma -{\vec \sigma}_3) c_{abe} \zeta^{({\check A}_{\perp})
s}_{eu}({\vec \sigma}_3,{\vec \sigma}_1;\tau )+\nonumber \\
&&+c_{bef} {\check {\vec A}}
_{f\perp}(\tau ,{\vec \sigma}_3)\cdot {\vec \zeta}^{({\check A}_{\perp})}_{em}
({\vec \sigma}_3,\vec \sigma ;\tau ) c_{man} \zeta^{({\check A}_{\perp})s}
_{nu}(\vec \sigma ,{\vec \sigma}_1;\tau )] \cdot \nonumber \\
&&c_{ers}{\check {\vec A}}_{s\perp}(\tau ,{\vec \sigma}_4)\cdot {\vec \zeta}
^{({\check A}_{\perp})}_{ru}({\vec \sigma}_4,{\vec \sigma}_2;\tau )+\nonumber \\
&&c_{bef}
{\check {\vec A}}_{f\perp}(\tau ,{\vec \sigma}_3)\cdot {\vec \zeta}^{({\check A}
_{\perp})}_{eu}({\vec \sigma}_3,{\vec \sigma}_1;\tau )\nonumber \\
&&[-\delta^3(\vec \sigma -{\vec \sigma}_4) c_{acm} \zeta^{({\check A}
_{\perp})s}_{mu}({\vec \sigma}_4,{\vec \sigma}_2;\tau )+\nonumber \\
&&+c_{cmn}{\check {\vec A}}
_{n\perp}(\tau ,{\vec \sigma}_4)\cdot {\vec \zeta}^{({\check A}_{\perp})}
_{mr}({\vec \sigma}_4,\vec \sigma ;\tau ) c_{ras} \zeta^{({\check A}_{\perp})s}
_{su}(\vec \sigma ,{\vec \sigma}_2;\tau )]\}\, \}{\check \rho}_c(\tau ,{\vec 
\sigma}_2),
\label{d12}
\end{eqnarray}

After some manipulations its final form is

\begin{eqnarray}
P^{rs}_{\perp}(\vec \sigma )&& \{ 
[{{\partial}\over {\partial \tau}}+{\dot {\vec g}}(\tau )\cdot {{\partial}
\over {\partial \vec \sigma}}] \int d^3\bar \sigma {\cal P}^{({\check A}
_{\perp}) st}_{ad}(\vec \sigma ,{\vec {\bar \sigma}};\tau )
[{{\partial}\over {\partial \tau}}+{\dot {\vec g}}(\tau )\cdot {{\partial}
\over {\partial {\vec {\bar \sigma}} }}]+\nonumber \\
&+&\delta^{st} \int d^3\bar \sigma \delta^3(\vec \sigma -{\vec {\bar \sigma}})
{\tilde K}_{ad}^{({\check A}_{\perp})}(\tau ,{\vec {\bar 
\sigma}})\, \} {\check A}^t
_{d\perp}(\tau ,{\vec {\bar \sigma}})\, {\buildrel \circ \over =}\, \nonumber \\
&&{\buildrel \circ \over =}\, -g^2_s P^{rs}_{\perp}(\vec \sigma )
\{ \sum_{i=1}^N {\check Q}_{ia}(\tau )  [{\dot \eta}^s_i(\tau )+{\dot g}^s(\tau 
)]\delta^3(\vec \sigma -{\vec \eta}_i(\tau ))+\nonumber \\
&+& c_{abd}{\check A}^s_{d\perp}(\tau ,\vec \sigma )\int d^3\sigma^{'}
K_{bc}(\vec \sigma ,{\vec \sigma}^{'};\tau ) {\check \rho}_c(\tau ,{\vec 
\sigma}^{'}) \}-\nonumber \\
&-&g^4_s P^{rs}_{\perp}(\vec \sigma ) {\check A}^s_{d\perp}(\tau ,\vec \sigma )
\int d^3\sigma_1d^3\sigma_2 {\check \rho}_b(\tau ,{\vec \sigma}_1 [F_{abcd}
(\vec \sigma ,{\vec \sigma}_1,{\vec \sigma}_2;\tau )+\nonumber \\
&+&c_{auv}c_{vde} K_{eb}(\vec \sigma ,{\vec \sigma}_1;\tau ) K_{uc}(\vec 
\sigma ,{\vec \sigma}_2;\tau )] {\check \rho}_c(\tau ,{\vec \sigma}_2),
\nonumber \\
&&{}\nonumber \\
&&or \nonumber \\
&&{}\nonumber \\
P^{rs}_{\perp}(\vec \sigma )&& \{ 
[{{\partial}\over {\partial \tau}}+{\dot {\vec g}}(\tau )\cdot {{\partial}
\over {\partial \vec \sigma}}] \int d^3\bar \sigma [\delta^{st}\delta_{ad}
\delta^3(\vec \sigma -{\vec {\bar \sigma}})+\nonumber \\
&+&{\hat D}^{({\check A}_{\perp}) s}_{au}(\tau ,\vec \sigma ) K_{uv}(\vec
\sigma ,{\vec {\bar \sigma}};\tau ) {\hat D}^{({\check A}_{\perp}) t}_{vd}(\tau
,{\vec {\bar \sigma}})] 
[{{\partial}\over {\partial \tau}}+{\dot {\vec g}}(\tau )\cdot {{\partial}
\over {\partial {\vec {\bar \sigma}} }}]+\nonumber \\
&+&\delta^{st} \int d^3\bar \sigma \delta^3(\vec \sigma -{\vec {\bar \sigma}})
[\delta_{ad}\bar \triangle +c_{adc}{\check {\vec A}}_{c\perp}(\tau ,{\vec {\bar
\sigma}})\cdot {{\partial}\over {\partial {\vec {\bar \sigma}} }}]\, \} 
{\check A}^t_{d\perp}(\tau ,{\vec {\bar \sigma}})
\, {\buildrel \circ \over =}\, \nonumber \\
&&{\buildrel \circ \over =}\, -g^2_s P^{rs}_{\perp}(\vec \sigma )
\{ \sum_{i=1}^N {\check Q}_{ia}(\tau )  [{\dot \eta}^s_i(\tau )+{\dot g}^s(\tau 
)]\delta^3(\vec \sigma -{\vec \eta}_i(\tau ))+\nonumber \\
&+& c_{abd}{\check A}^s_{d\perp}(\tau ,\vec \sigma )\int d^3\sigma^{'}
K_{bc}(\vec \sigma ,{\vec \sigma}^{'};\tau ) {\check \rho}_c(\tau ,{\vec 
\sigma}^{'}) \}-\nonumber \\
&-&g^4_s P^{rs}_{\perp}(\vec \sigma ) {\check A}^s_{d\perp}(\tau ,\vec \sigma )
\int d^3\sigma_1d^3\sigma_2 {\check \rho}_b(\tau ,{\vec \sigma}_1)[F_{abcd}
(\vec \sigma ,{\vec \sigma}_1,{\vec \sigma}_2;\tau )+\nonumber \\
&+&c_{auv}c_{vde} K_{eb}(\vec \sigma ,{\vec \sigma}_1;\tau ) K_{uc}(\vec 
\sigma ,{\vec \sigma}_2;\tau )] {\check \rho}_c(\tau ,{\vec \sigma}_2),
\nonumber \\
&&{}\nonumber \\
&&or\nonumber \\
&&{}\nonumber \\
P^{rs}_{\perp}(\vec \sigma )&& \{ \delta^{st}[\delta_{ad} ( 
({{\partial}\over {\partial \tau}}+{\dot {\vec g}}(\tau )\cdot {{\partial}
\over {\partial \vec \sigma}})^2+\triangle )+c_{adc}{\check {\vec A}}_{c\perp}
(\tau ,\vec \sigma )\cdot {{\partial}\over {\partial \vec \sigma}}]\cdot
\nonumber \\
&&\int d^3\bar \sigma \delta^3(\vec \sigma -{\vec {\bar \sigma}})
+[{{\partial}\over {\partial \tau}}+{\dot {\vec g}}(\tau )\cdot {{\partial}
\over {\partial \vec \sigma}}]{\hat D}^{({\check A}_{\perp}) s}_{au}(\tau 
,\vec \sigma ) \int d^3\bar \sigma \nonumber \\
&&K_{uv}(\vec \sigma ,{\vec {\bar \sigma}};\tau ) {\hat D}^{({\check A}
_{\perp}) t}_{vd}(\tau ,{\vec {\bar \sigma}})] 
[{{\partial}\over {\partial \tau}}+{\dot {\vec g}}(\tau )\cdot {{\partial}
\over {\partial {\vec {\bar \sigma}} }}]\, \} {\check A}^t
_{d\perp}(\tau ,{\vec {\bar \sigma}})\, {\buildrel \circ \over =}\, \nonumber \\
&&{\buildrel \circ \over =}\, -g^2_s P^{rs}_{\perp}(\vec \sigma )
\{ \sum_{i=1}^N {\check Q}_{ia}(\tau )  [{\dot \eta}^s_i(\tau )+{\dot g}^s(\tau 
)]\delta^3(\vec \sigma -{\vec \eta}_i(\tau ))+\nonumber \\
&&+ c_{abd}{\check A}^s_{d\perp}(\tau ,\vec \sigma )\int d^3\sigma^{'}
K_{bc}(\vec \sigma ,{\vec \sigma}^{'};\tau ) \nonumber \\
&&[{\check \rho}_c^{(YM)}(\tau ,{\vec \sigma}^{'})+\sum_{i=1}^N{\check \rho}
_{ic}(\tau ,{\vec \sigma}^{'})] \}-\nonumber \\
&&-g^4_s P^{rs}_{\perp}(\vec \sigma ) {\check A}^s_{d\perp}(\tau ,\vec \sigma )
\int d^3\sigma_1d^3\sigma_2 [{\check \rho}^{(YM)}_b(\tau ,{\vec \sigma}_1)+
\sum_{i=1}^N{\check \rho}_{ib}(\tau ,{\vec \sigma}_1)]\nonumber \\
&&[F_{abcd}(\vec \sigma ,{\vec \sigma}_1,{\vec \sigma}_2;\tau )+
c_{auv}c_{vde} K_{eb}(\vec \sigma ,{\vec \sigma}_1;\tau ) K_{uc}(\vec 
\sigma ,{\vec \sigma}_2;\tau )] \nonumber \\
&&[{\check \rho}^{(YM)}_c(\tau ,{\vec \sigma}_2)+\sum_{j=1}^N{\check \rho}_{jc}
(\tau ,{\vec \sigma}_2)],\nonumber \\
&&{}\nonumber \\
F_{abcd}(\vec \sigma ,{\vec \sigma}_1,{\vec \sigma}_2;\tau )&=&-c_{abe}c_{edf}
({{ K_{fc}(\vec \sigma ,{\vec \sigma}_1;\tau )}\over {4\pi |\vec \sigma -{\vec
\sigma}_2|}}+{{ K_{fc}(\vec \sigma ,{\vec \sigma}_2;\tau )}\over {4\pi |\vec
\sigma -{\vec \sigma}_1|}})+\nonumber \\
&+&\int {{d^3\sigma^{'}}\over {4\pi}} c_{bef}{\check {\vec A}}_{f\perp}(\tau ,
{\vec \sigma}^{'})\cdot {\vec \zeta}^{({\check A}_{\perp})}_{em}({\vec \sigma}
^{'},\vec \sigma ;\tau ) \nonumber \\
&&c_{man}c_{ndk}({{ K_{kc}(\vec \sigma ,{\vec \sigma}_1;
\tau )}\over {4\pi |{\vec \sigma}^{'}-{\vec \sigma}_2|}}+{{ K_{kc}(\vec \sigma 
,{\vec \sigma}_2;\tau )}\over {4\pi |{\vec \sigma}^{'}-{\vec \sigma}_1|}})-
\nonumber \\
&-&\int {{d^3\sigma^{'}}\over {4\pi |\vec \sigma -{\vec \sigma}^{'}|}} [c_{abe}
c_{edf} K_{fu}(\vec \sigma ,{\vec \sigma}_1;\tau ) c_{ers}{\check {\vec A}}
_{s\perp}(\tau ,{\vec \sigma}^{'})\cdot {\vec \zeta}^{({\check A}_{\perp})}
_{ru}({\vec \sigma}^{'},{\vec \sigma}_2;\tau )+\nonumber \\
&+&c_{bef} {\check {\vec A}}_{f\perp}(\tau ,{\vec \sigma}^{'})\cdot {\vec
\zeta}^{({\check A}_{\perp})}_{eu}({\vec \sigma}^{'},{\vec \sigma}_1;\tau )
c_{acm}c_{mdn} K_{nu}(\vec \sigma ,{\vec \sigma}_2;\tau )]+\nonumber \\
&+&\int {{d^3\sigma_1d^3\sigma_2}\over {4\pi |{\vec \sigma}_3-{\vec \sigma}_4|}}
[ c_{bef}{\check {\vec A}}_{f\perp}(\tau ,{\vec \sigma}_3)\cdot {\vec \zeta}
^{({\check A}_{\perp})}_{em}({\vec \sigma}_3,\vec \sigma ;\tau ) c_{man}c_{ndk}
K_{ku}(\vec \sigma ,{\vec \sigma}_1;\tau )\nonumber \\
&&c_{ers}{\check {\vec A}}_{s\perp}(\tau ,{\vec \sigma}_4)\cdot {\vec \zeta}
^{({\check A}_{\perp})}_{ru}({\vec \sigma}_4,{\vec \sigma}_2;\tau )+\nonumber \\
&+&c_{bef}{\check {\vec A}}_{f\perp}(\tau ,{\vec \sigma}_3)\cdot {\vec \zeta}
^{({\check A}_{\perp})}_{eu}({\vec \sigma}_3,{\vec \sigma}_1;\tau )\nonumber \\
&&c_{cmn}{\check {\vec A}}_{u\perp}(\tau ,{\vec \sigma}_4)\cdot {\vec \zeta}
^{({\check A}_{\perp})}_{mr}({\vec \sigma}_4,\vec \sigma ;\tau ) c_{ras}c_{sdk}
 K_{ku}(\vec \sigma ,{\vec \sigma}_2;\tau )],
\label{d13}
\end{eqnarray}

\noindent where the identity $P^{rs}_{\perp}(\vec \sigma )\zeta^{({\check A}
_{\perp}) s}_{ab}(\vec \sigma ,{\vec \sigma}^{'};\tau )=$
$P^{rs}_{\perp}(\vec \sigma ){\hat D}^{({\check A}_{\perp}) s}_{ac}(\tau ,\vec 
\sigma )K_{cb}(\vec \sigma ,{\vec \sigma}^{'};\tau )$
$=P^{rs}_{\perp}(\vec \sigma )c_{acb}{\check A}^s
_{c\perp}(\tau \sigma )K_{cb}(\vec \sigma ,{\vec \sigma}^{'};\tau )$ was used
in Eq.(\ref{d6}) to obtain $F_{abcd}$.

The equations of motion for the particles are

\begin{eqnarray}
{d\over {d\tau}}&& [\eta_im_i
\frac{\dot{\vec{\eta}}_{i}(\tau)+\dot{\vec{g}}(\tau)}{\sqrt{1-
(\dot{\vec{\eta}}_{i}(\tau)+\dot{\vec{g}}(\tau))^{2}}}
-\sum_{a}{\check Q}_{ia}(\tau )\check{\vec{A}}_{a\perp}(\tau,\vec{\eta}_{i}
(\tau ))]\, {\buildrel \circ \over =}\, \nonumber \\
&&{\buildrel \circ \over =}\, -[{\dot \eta}^u_i(\tau )+{\dot g}^u(\tau )]\sum_a
{\check Q}_{ia}(\tau ){{\partial {\check A}^u_{a\perp}(\tau ,{\vec \eta}
_i(\tau ))}\over {\partial {\vec \eta}_i}}-\nonumber \\
&-&g^2_s \sum_{a,b} {\check Q}_{ia}(\tau ) \int d^3\sigma
{{\partial K_{ab}({\vec \eta}_i(\tau ),{\vec \sigma};\tau )}\over
{\partial {\vec \eta}_i}} {\check \rho}_b(\tau ,\vec \sigma),\nonumber \\
&&{}\nonumber \\
&&or\nonumber \\
&&{}\nonumber \\
{d\over {d\tau}}&& [\eta_im_i
\frac{\dot{\vec{\eta}}_{i}(\tau)+\dot{\vec{g}}(\tau)}{\sqrt{1-
(\dot{\vec{\eta}}_{i}(\tau)+\dot{\vec{g}}(\tau))^{2}}}]\, {\buildrel \circ \over
=}\, \sum_a {\check {\vec A}}_{a\perp}(\tau ,{\vec \eta}_i(\tau )) c_{acd}
{\check Q}_{id}(\tau )\nonumber \\
&&[{\dot {\vec \eta}}_i(\tau )\cdot {\check {\vec A}}_{c\perp}(\tau ,{\vec 
\eta}_i(\tau ))-g^2_s \int d^3\sigma K_{cb}({\vec \eta}_i(\tau ),\vec \sigma ;
\tau ){\check \rho}_b(\tau ,\vec \sigma )]+\nonumber \\
&&+\sum_a{\check Q}_{ia}(\tau ) [{{\partial {\check {\vec A}}_{a\perp}(\tau ,
{\vec \eta}_i(\tau ))}\over {\partial \tau}}+{\dot {\vec \eta}}_i(\tau )\cdot
{{\partial {\check {\vec A}}_{a\perp}(\tau ,{\vec \eta}_i(\tau ))}\over 
{\partial {\vec \eta}_i}}]-\nonumber \\
&&-[{\dot \eta}^u_i(\tau )+{\dot g}^u(\tau )]\sum_a
{\check Q}_{ia}(\tau ){{\partial {\check A}^u_{a\perp}(\tau ,{\vec \eta}_i(\tau 
))}\over {\partial {\vec \eta}_i}}-\nonumber \\
&&-g^2_s \sum_{a,b}{\check Q}_{ia}(\tau ) \int d^3\sigma 
{{\partial K_{ab}({\vec \eta}_i(\tau ),{\vec \sigma};\tau )}\over
{\partial {\vec \eta}_i}} {\check \rho}_b(\tau ,\vec \sigma),\nonumber \\
&&{}\nonumber \\
&&or\nonumber \\
&&{}\nonumber \\
{d\over {d\tau}}&&[\eta_im_i {{{\dot \eta}^r_i(\tau )+{\dot g}^r(\tau )}\over
{\sqrt{1-({\dot {\vec \eta}}_i(\tau )+{\dot {\vec g}}(\tau ))^2}}}]\,
{\buildrel \circ \over =}\nonumber \\
&&{\buildrel \circ \over =}\, \sum_a{\check Q}_{ia}(\tau ) \{ {\check E}^r
_{a\perp}(\tau ,{\vec \eta}_i(\tau ))+[({\dot {\vec \eta}}_i(\tau )+{\dot 
{\vec g}}(\tau )) \times {\check {\vec B}}_a(\tau ,{\vec \eta}_i(\tau ))]^r \}-
\nonumber \\
&&-\sum_a {\check Q}_{ia}(\tau )P^{rs}_{\perp}({\vec \eta}_i) c_{amd}{\check A}
^s_{m\perp}(\tau ,{\vec \eta}_i(\tau )) \nonumber \\
&&\int d^3\sigma K_{de}({\vec \eta}
_i(\tau ),\vec \sigma ;\tau ) c_{enb}{\check {\vec A}}_{n\perp}(\tau ,\vec 
\sigma )\cdot {\check {\vec E}}_{b\perp}(\tau ,\vec \sigma )+\nonumber \\
&&+g^2_s\sum_{a,b} \{ {\check Q}_{ia}(\tau )\sum_{j=1}^N [c_{adc}{\check A}^r
_{d\perp}(\tau ,{\vec \eta}_i(\tau ))K_{cb}({\vec \eta}_i(\tau ),{\vec \eta}
_j(\tau );\tau )+\nonumber \\
&&+{{\partial K_{ab}({\vec \eta}_i(\tau ),{\vec \eta}_j(\tau );
\tau )}\over {\partial \eta^r_i}}]{\check Q}_{jb}(\tau )-\nonumber \\
&&-{\check Q}_{ia}(\tau ) \int d^3\sigma [c_{adc}{\check A}^r
_{d\perp}(\tau ,{\vec \eta}_i(\tau ))K_{cb}({\vec \eta}_i(\tau ),\vec \sigma;
\tau )+\nonumber \\
&&+{{\partial K_{ab}({\vec \eta}_i(\tau ),\vec \sigma;
\tau )}\over {\partial \eta^r_i}}]{\check \rho}^{(YM)}_b(\tau ,\vec \sigma ) 
\} -\nonumber \\
&&-\sum_a{\check Q}_{ia}(\tau ){\dot {\vec g}}(\tau )\cdot [{{\partial {\check
A}^r_{a\perp}(\tau ,{\vec \eta}_i(\tau ))}\over {\partial {\vec \eta}_i}}-
c_{acb}{\check {\vec A}}_{c\perp}(\tau ,{\vec \eta}_i(\tau )){\check A}^r
_{c\perp}(\tau ,{\vec \eta}_i(\tau ))]\, \} ,
\label{d14}
\end{eqnarray}

\noindent where Eqs. (\ref{d4}), (\ref{d5}), (\ref{d10}), (\ref{d11}), the
definition ${\check F}_a^{rs}=\partial^r{\check A}^s_{a\perp}-\partial^s{\check 
A}^r_{a\perp}+c_{abc}{\check A}^r_{b\perp}{\check A}^s_{c\perp}=\epsilon^{rst}
{\check B}^t_a$ and
${\hat D}^{({\check A}_{\perp}) s}_{ac}(\tau ,\vec \sigma )P^{sr}_{\perp}
(\vec \sigma ){\hat D}^{({\check A}_{\perp}) r}_{cb}(\tau ,\vec 
\sigma )=0$ have been used. Let us remark that Eqs.(\ref{d13}) and (\ref{d14})
are a system of integrodifferential equations due to the presence of the
transverse projectors and a completely open problem is how to define an initial
data problem for them.

Eq.(\ref{d14}) is the non-Abelian version of the particle equations of motion
given in Ref.\cite{alba}, which should produce the Abraham-Lorentz-Dirac
equation if we were able to get a non-Abelian Lienard-Wiechert potential from
Eq.(\ref{d13}). The last term would be absent in the gauge $\vec \lambda (\tau )
={\dot {\vec g}}(\tau )=0$. This equation
shows explicitely the chromo-electric [${\check {\vec E}}
_{a\perp}$] and chromo-magnetic [${\check {\vec B}}_a$, ${\check {\vec A}}
_{a\perp}$] forces acting on the particles. Moreover,  the explicit
interparticle instantaneous color forces are given by

\begin{eqnarray}
g^2_s\sum_{a,b}&& {\check Q}_{ia}(\tau ) \sum_{j=1}^N {\check Q}_{jb}(\tau )
[c_{adc}{\check A}^r_{d\perp}(\tau ,{\vec \eta}_i(\tau ))K_{cb}({\vec \eta}
_i(\tau ),{\vec \eta}_j(\tau );\tau )+{{\partial K_{ab}({\vec \eta}_i(\tau ),
{\vec \eta}_j(\tau );\tau )}\over {\partial \eta^r_i}}]=\nonumber \\
&=&g^2_s \sum_{a,b} {\check Q}_{ia}(\tau ) \sum_{j=1}^N {\check Q}_{jb}(\tau )
\nonumber \\
&&[c_{adc}{\check A}^r_{d\perp}(\tau ,{\vec \eta}_i(\tau ))
\int d^3\sigma_3d^3\sigma_4
\{ \, { {\delta_{cb}\delta^3(\vec{\sigma}_{3}-\vec{\eta}_i(\tau ))
\delta^3(\vec{\sigma}_{4}-\vec{\eta}_j(\tau ))}\over {4\pi\mid\vec{\sigma}_{3}
-\vec{\sigma}_{4}\mid} }+\nonumber \\
&+& { {\delta^3(\vec{\sigma}_{4}-\vec{\eta}_j(\tau ))[\check{\vec{A}}_{\perp}
(\tau ,\vec{\sigma}_{3})\cdot {\vec{\zeta}}^{({\check A}_{\perp})}
(\vec{\sigma}_{3},\vec{\eta}_i(\tau );
\tau )]_{cb}}\over {4\pi\mid\vec{\sigma}_{3}-\vec{\sigma}_{4}\mid} }+
(\vec{\eta}_i\leftrightarrow \vec{\eta}_j)+\nonumber \\
&+&\frac{[\check{\vec{A}}_{\perp}(\tau ,\vec{\sigma}_{3})\cdot
{\vec{\zeta}}^{({\check A}_{\perp})}(\vec{\sigma}_{3},
\vec{\eta}_i(\tau );\tau )]_{cu}
      [\check{\vec{A}}_{\perp}(\tau ,\vec{\sigma}_{4})\cdot
{\vec{\zeta}}^{({\check A}_{\perp})}(\vec{\sigma}_{4},
\vec{\eta}_j(\tau );\tau )]_{bu}}
     {4\pi\mid\vec{\sigma}_{3}-\vec{\sigma}_{4}\mid} +\nonumber \\
&&+{{\partial}\over {\partial \eta^r_i}}
\int d^3\sigma_3d^3\sigma_4
\{ \, { {\delta_{cb}\delta^3(\vec{\sigma}_{3}-\vec{\eta}_i(\tau ))
\delta^3(\vec{\sigma}_{4}-\vec{\eta}_j(\tau ))}\over {4\pi\mid\vec{\sigma}_{3}
-\vec{\sigma}_{4}\mid} }+\nonumber \\
&+& { {\delta^3(\vec{\sigma}_{4}-\vec{\eta}_j(\tau ))[\check{\vec{A}}_{\perp}
(\tau ,\vec{\sigma}_{3})\cdot {\vec{\zeta}}^{({\check A}_{\perp})}
(\vec{\sigma}_{3},\vec{\eta}_i(\tau );
\tau )]_{cb}}\over {4\pi\mid\vec{\sigma}_{3}-\vec{\sigma}_{4}\mid} }+
(\vec{\eta}_i\leftrightarrow \vec{\eta}_j)+\nonumber \\
&+&\frac{[\check{\vec{A}}_{\perp}(\tau ,\vec{\sigma}_{3})\cdot
{\vec{\zeta}}^{({\check A}_{\perp})}(\vec{\sigma}_{3},
\vec{\eta}_i(\tau );\tau )]_{cu}
      [\check{\vec{A}}_{\perp}(\tau ,\vec{\sigma}_{4})\cdot
{\vec{\zeta}}^{({\check A}_{\perp})}(\vec{\sigma}_{4},
\vec{\eta}_j(\tau );\tau )]_{bu}}
     {4\pi\mid\vec{\sigma}_{3}-\vec{\sigma}_{4}\mid}\, ].
\label{d15}
\end{eqnarray}

It is evident that there are divergences due to the self-energies for i=j.
They will be discussed in the case N=2 in the next Section, because their 
regularization is connected with the existence of asymptotic freedom already 
at this pseudoclassical level.

After an integration by parts, the constraints ${\check {\cal H}}_{pr}(\tau )
\approx 0$ become

\begin{eqnarray}
\sum_{i=1}^N[\eta_im_i
\frac{\dot{\vec{\eta}}_{i}(\tau)+\dot{\vec{g}}(\tau)}{\sqrt{1-
(\dot{\vec{\eta}}_{i}(\tau)+\dot{\vec{g}}(\tau))^{2}}}
&-&\sum_{a}{\check Q}_{ia}(\tau )\check{\vec{A}}_{a\perp}(\tau,\vec{\eta}_{i}
(\tau ))]+\nonumber \\
&+&g^{-2}_s \int d^3\sigma [(\vec \partial{\check A}^s_{a\perp}){\check E}^s
_{a\perp}](\tau ,\vec \sigma )\, {\buildrel \circ \over =}\, 0,
\label{d16}
\end{eqnarray}

\noindent and the expression of the constant invariant mass $H_{rel}$ becomes

\begin{eqnarray}
E_{rel}&=&\sum_{i=1}^N \frac{\eta_im_i}{\sqrt{1-
(\dot{\vec{\eta}}_{i}(\tau)+\dot{\vec{g}}(\tau))^{2}}}+\nonumber \\
&+&{1\over 2} V[{\vec \eta}_i, {\check {\vec A}}_{a\perp}, {\check {\vec E}}
_{a\perp}]+\sum_a {1\over {2g^2_s}} \int d^3\sigma [{\check {\vec E}}^2_{a\perp}
+{\check {\vec B}}^2_a](\tau ,\vec \sigma ).
\label{d17}
\end{eqnarray}

To find the classical theory implied by the pseudoclassical one we follow
Ref.\cite{lus2}, where the Berezin-Marinov distribution function [see in
Refs.\cite{casalb}] for the Grassmann variables of phase space is defined
for the case of SU(3) [in this discussion we will denote $\theta$ the Dirac
observables $\check \theta$]. 
For each particle i, on the space of analytic functions
of the $\theta_{i\alpha}$'s, namely of the functions $f(\theta_i)=\alpha_i+\beta
_{i\alpha}\theta_{i\alpha}+\gamma_{i\alpha}\epsilon_{\alpha\beta\gamma}\theta
_{i\beta}\theta_{i\gamma}+\delta_i\epsilon_{\alpha\beta\gamma}\theta_{i\alpha}
\theta_{i\beta}\theta_{i\gamma}$, the more general density function $\rho_i
(\theta_i,\theta^{*}_i)$ satisfying the normalization condition $\int \rho_i
d\mu_i=1$ [$d\mu_i=\prod_{\alpha =1}^3d\theta_{i\alpha}d\theta^{*}_{i\alpha}$ is
the Berezin integration measure on the Grassmann sector of phase space] and
the positivity condition $\int \rho_i f^{*}f d\mu_i \geq 0$ for any analytic
function f, is [$c_i=c_i(\tau )$, $W_{i\alpha\beta}=W_{i\alpha\beta}(\tau )$,
$V_{i\alpha\beta}=V_{i\alpha\beta}(\tau)$]

\begin{eqnarray}
\rho_i(\theta_i,\theta^{*}_i)&=&c_i+\sum_{\alpha\beta}\theta^{*}_{i\alpha}
W_{i\alpha\beta}\theta_{i\beta}+2(Tr\, V_i)(\sum_{\alpha}\theta^{*}_{i\alpha}
\theta_{i\alpha})^2-4(\sum_{\alpha\beta}\theta^{*}_{i\alpha}V_{i\alpha\beta}
\theta_{i\beta})(\sum_{\gamma}\theta^{*}_{i\gamma}\theta_{i\gamma})+
\nonumber \\
&+&{1\over 6}
(\sum_{\alpha}\theta^{*}_{i\alpha}\theta_{i\alpha})^3,\nonumber \\
&&{}\nonumber \\
c_i&=&c_i^{*} > 0,\nonumber \\
W_i&=&W^{\dagger}_i=w_{io}\openone +w_{ia}{1\over 2}\lambda_a,\quad definite\, 
positive,\nonumber \\
V_i&=&V^{\dagger}_i=v_{io}\openone +v_{ia}{1\over 2}\lambda_a,\quad definite\,
positive,\nonumber \\
&&{}\nonumber \\
&&w_{io}={1\over 3} Tr\, W_i,\quad\quad w_{ia}=Tr\, (\lambda_aW_i),\nonumber \\
&&v_{io}={1\over 3} Tr\, V_i,\quad\quad v_{ia}=Tr\, (\lambda_aV_i),
\label{d18}
\end{eqnarray}

\noindent since one gets $\int \rho_i f^{*}f d\mu_i=|\alpha |^2+36c_i|\delta |^2
+4\beta^{*}V_i\beta +4\gamma^{8}W_i\gamma$. Other useful formulas are
[${1\over 2}d_{abc}=Tr\, [\{ {1\over 2}\lambda_a,
{1\over 2}\lambda_b\} {1\over 2}\lambda_c]\,$]

\begin{eqnarray}
&&\int \rho_i \theta^{*}_{i\alpha}\theta_{i\beta} d\mu_i = 4 V_{i\beta\alpha},
\nonumber \\
&&\int \rho_i \theta^{*}_{i\alpha}\theta_{i\beta}\theta^{*}_{i\gamma}\theta
_{i\delta} d\mu_i = W_{i\lambda\tau} \epsilon_{\alpha\gamma\lambda}\epsilon
_{\beta\delta\tau},\nonumber \\
&&\int \rho_i \theta^{*}_{i\alpha}\theta_{i\beta}\theta^{*}_{i\gamma}\theta
_{i\delta}\theta^{*}_{i\lambda}\theta_{i\tau} d\mu_i =c_i \epsilon_{\alpha\gamma
\lambda} \epsilon_{\beta\delta\tau},\nonumber \\
&&{}\nonumber \\
&&< N_i >= < \sum_{\alpha}\theta^{*}_{i\alpha}\theta_{i\alpha} > = 12 v_{io},
\nonumber \\
&&< Q_{ia} > = < \sum_{\alpha\beta}\theta^{*}_{i\alpha} ({1\over 2}\lambda_a)
_{\alpha\beta}\theta_{i\beta} > = 2 v_{ia},\nonumber \\
&&< N^2_i > = < (\sum_{\alpha}\theta^{*}_{i\alpha}\theta_{i\alpha})^2 > =
6 w_{io},\nonumber \\
&&< (\sum_{\alpha\beta}\theta^{*}_{i\alpha} ({1\over 2}\lambda_a)_{\alpha\beta}
\theta_{i\beta} )(\sum_{\gamma}\theta^{*}_{i\gamma}\theta_{i\gamma}) > = -
{1\over 2} w_{ia},\nonumber \\
&&< N^3_i > = < (\sum_{\alpha}\theta^{*}_{i\alpha}\theta_{i\alpha} )^3 > = 6
c_i,\nonumber \\
&&{}\nonumber \\
&&< Q_{ia}Q_{ib} =
< ((\sum_{\alpha\beta}\theta^{*}_{i\alpha} ({1\over 2}\lambda_a)_{\alpha\beta}
\theta_{i\beta} )(\sum_{\gamma\delta}\theta^{*}_{i\gamma} ({1\over 2}\lambda_b)
_{\gamma\delta}\theta_{i\delta} ) > =\nonumber \\
&&={1\over 2} (d_{abc}w_{ic}-w_{io}\delta
_{ab}),\nonumber \\
&&< Q_{ia}Q_{ib}Q_{ic} > =
< (\sum_{\alpha\beta}\theta^{*}_{i\alpha} ({1\over 2}\lambda_a)_{\alpha\beta}
\theta_{i\beta} )(\sum_{\gamma\delta}\theta^{*}_{i\gamma} ({1\over 2}\lambda
_b)_{\gamma\delta}\theta_{i\delta} )(\sum_{\mu\nu}\theta^{*}_{i\mu} ({1\over 2}
\lambda_c)_{\mu\nu}\theta_{i\nu} ) > =\nonumber \\
&&= {1\over 2} c_i d_{abc},\nonumber \\
&&< Q_{ia}Q_{ib}N_i > =
< (\sum_{\alpha\beta}\theta^{*}_{i\alpha} ({1\over 2}\lambda_a)_{\alpha\beta}
\theta_{i\beta} )(\sum_{\gamma\delta}\theta^{*}_{i\gamma} ({1\over 2}\lambda_b)
_{\gamma\delta}\theta_{i\delta} )(\sum_{\mu}\theta^{*}_{i\mu} 
\theta_{i\mu} ) > =\nonumber \\
&&= - {1\over 2} c_i \delta_{ab}.
\label{d19}
\end{eqnarray}

Therefore, the constraints $N_i\approx 0$ imply at the classical level
[at the quantum level in general $c_i\not= 0$ due to ordering problems]

\begin{eqnarray}
&&c_i=v_{io}=w_{io}=w_{ia}=0,\nonumber \\
&&{}\nonumber \\
&&\Rightarrow \rho_i(\theta_i,\theta^{*}_i)=-4 v_{ia}(\tau ) Q_{ia}(\tau ) N_i+
{1\over 6} N_i^3 \approx 0,\nonumber \\
&&{}\nonumber \\
&&< Q_{ia} > = 2v_{ia},\quad\quad < Q_{ia}Q_{ib} > = < Q_{ia}Q_{ib}Q_{ic} > =0,
\nonumber \\
&&< N_i > = < N^2_i > = < N^3_i > = < Q_{ia}Q_{ib}N_i > = 0.
\label{d20}
\end{eqnarray}

The classical observable associated to an even Grassmann-valued function $g$
is

\begin{equation}
< g > = \int g \prod_{i=1}^N \rho_i  d\mu_i=\int g \rho d\mu,\quad\quad \rho=
\prod_{i=1}^N\rho_i,\quad d\mu =\prod_{i=1}^Nd\mu_i.
\label{d21}
\end{equation}

The distribution function $\rho =\prod_{i=1}^N \rho_i$ must satisfy the 
Liouville equation

\begin{eqnarray}
{{\partial \rho}\over {\partial \tau}}&+&\lbrace \rho ,{\hat H}_D\rbrace \,
{\buildrel \circ \over =}\, 0,\nonumber \\
&&{}\nonumber \\
&&\sum_{k=1}^N \rho_1...\rho_{k-1} [{{\partial \rho_k}\over {\partial \tau}}+
\lbrace \rho_k,{\hat H}_D\rbrace ]\rho_{k+1}...\rho_N\, {\buildrel \circ \over
=}\, 0,\nonumber \\
&&\Downarrow \nonumber \\
{{\partial \rho}\over {\partial \tau}}&+&\lbrace \rho ,H_{rel}\rbrace \,
{\buildrel \circ \over =}\, 0,\nonumber \\
\{ \rho &,& {\check {\vec H}}_p(\tau )\} \approx 0,\nonumber \\
\{ \rho &,& N_i=\sum_{\alpha}\theta^{*}_{i\alpha}\theta_{i\alpha} \} \approx 0,
\quad \quad i=1,..,N,
\label{d22}
\end{eqnarray}

\noindent with ${\hat H}_D$ the Hamiltonian of Eq.(\ref{d2}). The last two 
equations are identically satisfied. Therefore, by using Eqs.(\ref{d18}) and
(\ref{d4}) for ${\dot {\check Q}}_{ia}$, we get the equations [valid in a
neighbourhood of the region $N_i\approx 0$, where however we will put $N^3_i
\equiv 0$ in the sense of Dirac's strong equality]

\begin{eqnarray}
\sum_{k=1}^N&& \rho_1...\rho_{k-1}[-4 \{ {\dot V}_{k\alpha\beta}+[V_k,T^a]
_{\alpha\beta} [{\dot {\vec \eta}}_k(\tau )\cdot {\check {\vec A}}_{a\perp}
(\tau ,{\vec \eta}_k(\tau ))-\nonumber \\
&&-g^2_s\int d^3\sigma K_{ab}({\vec \eta}_k(\tau ),\vec \sigma ;\tau ){\check
\rho}^{(YM)}_b(\tau ,\vec \sigma )+\nonumber \\
&&+g^2_s\sum_{h=1}^NK_{ab}({\vec \eta}_k(\tau 
),{\vec \eta}_h(\tau );\tau ) i\theta^{*}_{h\gamma}(T^b)_{\gamma\delta}\theta
_{h\delta}]\} \theta^{*}_{k\alpha}\theta_{k\beta} N_k] \rho_{k+1}...\rho_N=
\nonumber \\
&&=\sum_{k=1}^N \rho_1...\rho_{k-1}[-4 \{ {\dot V}_{k\alpha\beta}+[V_k,T^a]
_{\alpha\beta} [{\dot {\vec \eta}}_k(\tau )\cdot {\check {\vec A}}_{a\perp}
(\tau ,{\vec \eta}_k(\tau ))-\nonumber \\
&&-g^2_s\int d^3\sigma K_{ab}({\vec \eta}_k(\tau ),\vec \sigma ;\tau ){\check
\rho}^{(YM)}_b(\tau ,\vec \sigma )] \rho_{k+1}...\rho_N +\nonumber \\
&&+(terms\, containing\, factors\, N^3_h\, )
\, {\buildrel \circ \over =}\, 0,
\label{d23}
\end{eqnarray}

\noindent so that, with $N^3_h\equiv 0$ for every h, the equations for 
$v_{ka}(\tau )$ are

\begin{eqnarray}
{{dv_{ka}(\tau )}\over {d\tau}}&-&c_{abc} v_{kc}(\tau ) [{\dot {\vec \eta}}
_k(\tau )\cdot {\check {\vec A}}_{b\perp}(\tau ,{\vec \eta}_k(\tau ))-
\nonumber \\
&-&g^2_s \int d^3\sigma K_{ab}({\vec \eta}_k(\tau ),\vec \sigma ;\tau ){\check
\rho}_b^{(YM)}(\tau ,\vec \sigma )]
\, {\buildrel \circ \over =}\, 0,
\label{d24}
\end{eqnarray}

\noindent in accord with the mean value of the last of Eqs.(\ref{d4}) and with
Ref.\cite{lus2} except for the last term which is absent when the field is 
considered external.

In analogy with the electromagnetic case (see Ref.\cite{lus2}), we could try
to find solutions to the field equations (\ref{d13}) necessarily of the form
${\check {\vec A}}_{a\perp}(\tau ,\vec \sigma )={\check {\vec A}}^{(o)}_{a\perp}
(\tau ,\vec \sigma )+ (terms\, at\, least\, linear\, in\, the\, {\check Q}_{ia}
(\tau )'s\, )$ [${\check {\vec A}}^{(o)}_{a\perp}(\tau ,\vec \sigma )$ satisfies
Eq.(\ref{d13}) with ${\check Q}_{ia}(\tau )=0$, namely in absence of particle 
sources], to put them in the particle equations (\ref{d14}) and to take the
mean value ($< D > =\int \rho D d\mu$) of the resulting equations to get the 
real ``classical" equations. In the electromagnetic case, this procedure 
eliminates infinities from self-energies and causal pathologies (runaway 
solutions or preaccelerations) from the classical equations (see also 
Ref.\cite{alba} for the rest-frame analysis of these problems), which, instead,
would be present by taking immediately the mean value of Eqs.(\ref{d13}) and
(\ref{d14}). However, it is not possible to attack the former procedure in the
non-Abelian case in absence of an analogue of the Lienard-Wiechert potential
[but see in any case next Section for the N=2 case].

\vfill\eject

\section
{The Quark Model and the Pseudoclassical Asymptotic Freedom for N=2}

In the nonrelativistic quark model, in which no SU(3) color Yang-Mills field 
appears, one assumes that the physical states are color singlets. Since
Eq.(\ref{d3}) gives the invariant mass of N colored relativistic scalar 
particles together with the SU(3) color Yang-Mills field, this is the
starting point to try to extract a pseudoclassical basis for the missing
relativistic quark model.

A first step is to study what happens if we add 8 extra constraints implying
the vanishing of the total color charge, so that only global color singlets
are allowed for the particles+Yang-Mills field system:

\begin{equation}
{\check Q}_a={\check Q}^{(YM)}_a+\sum_{i=1}^N{\check Q}_{ia}=0.
\label{e1}
\end{equation}

Then, we ask that these 8 conditions be fulfilled by the separate vanishing
of the particle and field contributions to the color charge

\begin{eqnarray}
\sum_{i}\check{Q}_{ia}&=&0\nonumber \\
\check{Q}^{(YM)}_{a}&=&0.
\label{e2}
\end{eqnarray}

The first condition defines the relativistic scalar quark model: the particles
by themselves are a color singlet independently from the SU(3) color field. 
For N=1, Eqs.(\ref{e2}) plus the constraint $\check N=\sum_{\alpha}{\check
\theta}^{*}_{\alpha}{\check \theta}_{\alpha}\approx 0$ gives 9 conditions on
6 Grassmann variables: therefore a single pseudoclassical scalar quark cannot 
be a color singlet. For N=2, besides the constraints ${\check N}_1=
\sum_{\alpha}{\check \theta}^{*}_{1\alpha}{\check \theta}_{1\alpha}\approx 0$
and ${\check N}_2=\sum_{\alpha}{\check \theta}^{*}_{2\alpha}{\check \theta}
_{2\alpha}\approx 0$, one has

\begin{equation}
\check{Q}_{1a}+\check{Q}_{2a}=0,
\label{e3}
\end{equation}

\noindent namely 10 conditions on 12 Grasmmann variables.

The condition ${\check Q}^{(YM)}_a=0$ replaces the Abelian condition 
${\vec A}_{\perp}(\tau ,\vec \sigma )={\vec \pi}_{\perp}(\tau ,\vec \sigma )=0$
of absence of radiation [see Ref.\cite{lus1}]. Since in the non-Abelian case
we do not know how to solve the equations of motion and since the superposition
principle does not hold, we can only ask that there is no color flux on the
surface at space infinity. This requirement also implies that the 
pseudoclassical solutions of the SU(3) Yang-Mills equations are restricted to
those configurations which are color singlets like the glueball states at the
quantum level.

Due to the Gauss laws,
the condition ${\check Q}_a(\tau )=0$ can be imposed by choosing suitable
boundary conditions on the transverse SU(3) fields (see Eqs.(2-40) of
Ref.\cite{lusa}): ${\check {\vec A}}_{a\perp}(\tau ,\vec \sigma ){\rightarrow}
_{|\vec \sigma |\rightarrow \infty}\, O({|\vec \sigma |}^{-(2+\epsilon )})$,
${\check {\vec \pi}}_{a\perp}(\tau ,\vec \sigma ){\rightarrow}
_{|\vec \sigma |\rightarrow \infty}\, O({|\vec \sigma |}^{-(2+\epsilon )})$
with $\epsilon > 0$ (for $\epsilon \rightarrow 0$ one gets ${\check Q}_a\not=
0$). With these boundary conditions the requirement of color singlets for the
whole theory becomes automatically the same requirement in field-independent
quark models, if the equations of motion imply ${\check Q}^{(YM)}_a(\tau )\,
{\buildrel \circ \over =}\, 0$.

Eqs.(\ref{e2}), (\ref{d4}) and (\ref{d5}) imply

\begin{eqnarray}
&&{d\over {d\tau}} {\check Q}^{(YM)}_a(\tau )\, {\buildrel \circ \over =}\, 0,
\nonumber \\
&&{d\over {d\tau}} ({\check Q}_{1a}(\tau )+{\check Q}_{2a}(\tau ))\, {\buildrel
\circ \over =}\, c_{acd}{\check Q}_{1d}(\tau ) \{ {\dot {\vec \eta}}_1(\tau )
\cdot {\check {\vec A}}_{c\perp}(\tau ,{\vec \eta}_1(\tau ))-{\dot {\vec \eta}}
_2(\tau )\cdot {\check {\vec A}}_{c\perp}(\tau ,{\vec \eta}_2(\tau ))-
\nonumber \\
&&-g^2_s\int d^3\sigma [K_{cb}({\vec \eta}_1(\tau ),\vec \sigma ;\tau )-
K_{cb}({\vec \eta}_2(\tau ),\vec \sigma ;\tau )]{\check \rho}^{(YM)}_b(\tau ,
\vec \sigma )+\nonumber \\
&&+g^2_s[K_{cb}({\vec \eta}_1(\tau ),{\vec \eta}_1(\tau );\tau )+
K_{cb}({\vec \eta}_2(\tau ),{\vec \eta}_2(\tau );\tau )-\nonumber \\
&&-K_{cb}({\vec \eta}_1(\tau ),{\vec \eta}_2(\tau );\tau )-
K_{cb}({\vec \eta}_2(\tau ),{\vec \eta}_1(\tau );\tau )]{\check Q}_{1b}(\tau )\,
{\buildrel \circ \over =}\, 0.
\label{ee3}
\end{eqnarray}

In the gauge $\vec \lambda (\tau )={\dot {\vec g}}(\tau )=0$, the equations of
motion for the field and the particles and the equations defining the 
rest-frame become [in Eq.(\ref{ea3}) there is the same F-function of
Eq.(\ref{d13})]

\begin{eqnarray}
P^{rs}_{\perp}(\vec \sigma )&& \{ \delta^{st}[\delta_{ad} ( 
{{\partial}\over {\partial \tau}}^2
+\triangle )+c_{adc}{\check {\vec A}}_{c\perp}
(\tau ,\vec \sigma )\cdot {{\partial}\over {\partial \vec \sigma}}]\cdot
\nonumber \\
&&\int d^3\bar \sigma \delta^3(\vec \sigma -{\vec {\bar \sigma}})
+{{\partial}\over {\partial \tau}}
{\hat D}^{({\check A}_{\perp}) s}_{au}(\tau 
,\vec \sigma ) \int d^3\bar \sigma \nonumber \\
&&K_{uv}(\vec \sigma ,{\vec {\bar \sigma}};\tau ) {\hat D}^{({\check A}
_{\perp}) t}_{vd}(\tau ,{\vec {\bar \sigma}})] 
{{\partial}\over {\partial \tau}}\, \} {\check A}^t
_{d\perp}(\tau ,{\vec {\bar \sigma}})\, {\buildrel \circ \over =}\, \nonumber \\
&&{\buildrel \circ \over =}\, -g^2_s P^{rs}_{\perp}(\vec \sigma )
\{ {\check Q}_{1a}(\tau )  [{\dot \eta}^s_1(\tau )\delta^3(\vec \sigma -
-{\vec \eta}_1(\tau ))-\nonumber \\
&&-{\dot \eta}^s_2(\tau )\delta^3(\vec \sigma -{\vec 
\eta}_2(\tau ))]
+ c_{abd}{\check A}^s_{d\perp}(\tau ,\vec \sigma )\int d^3\sigma^{'}
K_{bc}(\vec \sigma ,{\vec \sigma}^{'};\tau ) \nonumber \\
&&[{\check \rho}_c^{(YM)}(\tau ,{\vec \sigma}^{'})-{\check Q}_{1c}(\tau )
[\delta^3({\vec \sigma}^{'}-{\vec \eta}_1(\tau ))-\delta^3({\vec \sigma}^{'}-
{\vec \eta}_2(\tau ))] \}-\nonumber \\
&&-g^4_s P^{rs}_{\perp}(\vec \sigma ) {\check A}^s_{d\perp}(\tau ,\vec \sigma )
\int d^3\sigma_1d^3\sigma_2 \nonumber \\
&&[{\check \rho}^{(YM)}_b(\tau ,{\vec \sigma}_1)-{\check Q}_{1b}(\tau )
[\delta^3({\vec \sigma}_1-{\vec \eta}_1(\tau ))-\delta^3({\vec \sigma}_1-
{\vec \eta}_2(\tau ))]\nonumber \\
&&[F_{abcd}(\vec \sigma ,{\vec \sigma}_1,{\vec \sigma}_2;\tau )+
c_{auv}c_{vde} K_{eb}(\vec \sigma ,{\vec \sigma}_1;\tau ) K_{uc}(\vec 
\sigma ,{\vec \sigma}_2;\tau )] \nonumber \\
&&[{\check \rho}^{(YM)}_c(\tau ,{\vec \sigma}_2)-{\check Q}_{1c}(\tau )
[\delta^3({\vec \sigma}_2-{\vec \eta}_1(\tau ))-\delta^3({\vec \sigma}_2-
{\vec \eta}_2(\tau ))],
\label{ea3}
\end{eqnarray}

\begin{eqnarray}
{d\over {d\tau}}&&[\eta_1m_1 {{{\dot \eta}^r_1(\tau )}\over
{\sqrt{1-{\dot {\vec \eta}}_1^2(\tau )}}}]\,
{\buildrel \circ \over =}\nonumber \\
&&{\buildrel \circ \over =}\, \sum_a{\check Q}_{1a}(\tau ) \{ {\check E}^r
_{a\perp}(\tau ,{\vec \eta}_1(\tau ))+[{\dot {\vec \eta}}_1(\tau )
\times {\check {\vec B}}_a(\tau ,{\vec \eta}_1(\tau ))]^r \}-
\nonumber \\
&&-\sum_a {\check Q}_{1a}(\tau )P^{rs}_{\perp}({\vec \eta}_1) c_{amd}{\check A}
^s_{m\perp}(\tau ,{\vec \eta}_1(\tau )) \nonumber \\
&&\int d^3\sigma K_{de}({\vec \eta}
_1(\tau ),\vec \sigma ;\tau ) c_{enb}{\check {\vec A}}_{n\perp}(\tau ,\vec 
\sigma )\cdot {\check {\vec E}}_{b\perp}(\tau ,\vec \sigma )+\nonumber \\
&&+g^2_s\sum_{a,b} \{ {\check Q}_{1a}(\tau ){\check Q}_{1b}(\tau )
[c_{adc}{\check A}^r_{d\perp}(\tau ,{\vec \eta}_1(\tau ))(K_{cb}({\vec \eta}
_1(\tau ),{\vec \eta}_1(\tau );\tau )-\nonumber \\
&&-K_{cb}({\vec \eta}_1(\tau ),{\vec \eta}_2(\tau );\tau )\, )+{{\partial}\over
{\partial \sigma^r}}{|}_{\vec \sigma ={\vec \eta}_1}\, (K_{ab}(\vec \sigma ,
{\vec \eta}_1(\tau );\tau )-K_{ab}(\vec \sigma ,{\vec \eta}_2(\tau );\tau )\,)]-
\nonumber \\
&&-{\check Q}_{1a}(\tau ) \int d^3\sigma [c_{adc}{\check A}^r
_{d\perp}(\tau ,{\vec \eta}_1(\tau ))K_{cb}({\vec \eta}_1(\tau ),\vec \sigma;
\tau )+\nonumber \\
&&+{{\partial K_{ab}({\vec \eta}_1(\tau ),\vec \sigma;
\tau )}\over {\partial \eta^r_1}}]{\check \rho}^{(YM)}_b(\tau ,\vec \sigma ) 
\} ,\nonumber \\
&&{}\nonumber \\
{d\over {d\tau}}&&[\eta_2m_2 {{{\dot \eta}^r_2(\tau )}\over
{\sqrt{1-{\dot {\vec \eta}}_2^2(\tau )}}}]\,
{\buildrel \circ \over =}\nonumber \\
&&{\buildrel \circ \over =}\, -\sum_a{\check Q}_{1a}(\tau ) \{ {\check E}^r
_{a\perp}(\tau ,{\vec \eta}_2(\tau ))+[{\dot {\vec \eta}}_2(\tau )
\times {\check {\vec B}}_a(\tau ,{\vec \eta}_2(\tau ))]^r \}+
\nonumber \\
&&+\sum_a {\check Q}_{1a}(\tau )P^{rs}_{\perp}({\vec \eta}_1) c_{amd}{\check A}
^s_{m\perp}(\tau ,{\vec \eta}_2(\tau )) \nonumber \\
&&\int d^3\sigma K_{de}({\vec \eta}
_2(\tau ),\vec \sigma ;\tau ) c_{enb}{\check {\vec A}}_{n\perp}(\tau ,\vec 
\sigma )\cdot {\check {\vec E}}_{b\perp}(\tau ,\vec \sigma )-\nonumber \\
&&-g^2_s\sum_{a,b} \{ {\check Q}_{1a}(\tau ){\check Q}_{1b}(\tau )
[c_{adc}{\check A}^r_{d\perp}(\tau ,{\vec \eta}_2(\tau ))(K_{cb}({\vec \eta}
_1(\tau ),{\vec \eta}_1(\tau );\tau )-\nonumber \\
&&-K_{cb}({\vec \eta}_2(\tau ),{\vec \eta}_2(\tau );\tau )\, )+{{\partial}\over
{\partial \sigma^r}}{|}_{\vec \sigma ={\vec \eta}_2}\, (K_{ab}(\vec \sigma ,
{\vec \eta}_1(\tau );\tau )-K_{ab}(\vec \sigma ,{\vec \eta}_2(\tau );\tau )\,)]-
\nonumber \\
&&-{\check Q}_{1a}(\tau ) \int d^3\sigma [c_{adc}{\check A}^r
_{d\perp}(\tau ,{\vec \eta}_2(\tau ))K_{cb}({\vec \eta}_2(\tau ),\vec \sigma;
\tau )+\nonumber \\
&&+{{\partial K_{ab}({\vec \eta}_2(\tau ),\vec \sigma;
\tau )}\over {\partial \eta^r_2}}]{\check \rho}^{(YM)}_b(\tau ,\vec \sigma ) 
\} ,
\label{eb3}
\end{eqnarray}

\begin{eqnarray}
&&\eta_1m_1{{ {\dot {\vec \eta}}_1(\tau )}\over
{\sqrt{1-{\dot {\vec \eta}}_1^2(\tau )} }}+
\eta_2m_2{{ {\dot {\vec \eta}}_2(\tau )}\over
{\sqrt{1-{\dot {\vec \eta}}_2^2(\tau )} }}-
\nonumber \\
&&-\sum_a {\check Q}_{1a}(\tau ) [{\check {\vec A}}_{a\perp}(\tau ,{\vec
\eta}_1(\tau ))-{\check {\vec A}}_{a\perp}(\tau ,{\vec \eta}_2(\tau ))+
g^{-2}_s\int d^3\sigma \{ \vec \partial {\check A}^{s}_{a\perp}{\check
E}^{s}_{a\perp} \} (\tau ,\vec \sigma )\, {\buildrel \circ \over =}\, 0.
\label{ec3}
\end{eqnarray}

\noindent Note that in Eqs.(\ref{eb3}) the Coulomb interaction inside the
kernel K does not contribute to the terms ${\check Q}_{1a}(\tau ){\check Q}
_{1b}(\tau )$ due to $c_{adb}{\check Q}_{1a}{\check Q}_{1b}=\delta_{ab}{\check 
Q}_{1a}{\check Q}_{1b}=0$.

The invariant mass of the system is [we choose $\eta_1=\eta_2=+1$, i.e. the 
quark and antiquark have positive energies and the antiquark is distinguished by
the opposite color charge as a classical antiparticle moving forward in $\tau$
(see Ref.\cite{st,fey,cos} and its bibliography)]

\begin{eqnarray}
H_{rel}&=&\sqrt{m_1^2+({\check {\vec \kappa}}_1(\tau )+\sum_a{\check Q}_{1a}
(\tau ){\check {\vec A}}_{a\perp}(\tau ,{\vec \eta}_1(\tau ))\, )^2}+
\nonumber \\
&+&\sqrt{m_2^2+({\check {\vec \kappa}}_2(\tau )-\sum_a{\check Q}_{1a}
(\tau ){\check {\vec A}}_{a\perp}(\tau ,{\vec \eta}_2(\tau ))\, )^2}+
\nonumber \\
&+&{1\over 2} V[{\vec \eta}_i,{\check {\vec A}}_{a\perp},{\check {\vec \pi}}
_{a\perp}](\tau )+{1\over 2}\sum_a \int d^3\sigma [g^2_s{\check {\vec \pi}}^2
_{a\perp}+g^{-2}_s{\check {\vec B}}^2_{a\perp}](\tau ,\vec \sigma ).
\label{ed3}
\end{eqnarray}

In the case N=2, which should correspond at the quantum level
to a meson configuration formed from a scalar quark and a scalar antiquark, 
plus glue [there are not sea-quarks, because there is no pair production at
this pseudoclassical level], 
if ${\check N}_1 \rightarrow \sum_{\alpha =0}^3b^{\dagger}_{1\alpha}
b_{1\alpha} -1$ and ${\check N}_2 \rightarrow \sum_{\alpha =0}^3b^{\dagger}
_{2\alpha}b_{2\alpha} -2$, the first condition in Eq.(\ref{e2}) and the
observation that the SU(3)
Yang-Mills fields, solutions of the field equations (\ref{ea3}), 
will depend on the Grassmann variables of the particles
only through the color charges ${\check Q}_{ia}(\tau )$, allow us to write the
following developments [from now on we shall not write $\sum_a$ for repeated
color indices]

\begin{eqnarray}
\check{\vec{A}}_{a\perp}(\tau,\vec{\sigma})&=&
\check{\vec{A}}_{a\perp}^{(0)}(\tau,\vec{\sigma})+
\check{Q}_{1u}(\tau )\check{\vec{A}}_{au\perp}^{(1)}(\tau,\vec{\sigma})+
\check{Q}_{1u}(\tau )\check{Q}_{1v}(\tau )\check{\vec{A}}_{auv\perp}^{(2)}
(\tau,\vec{\sigma}),\nonumber \\
\check{\vec{\pi}}_{a\perp}(\tau,\vec{\sigma})&=&
\check{\vec{\pi}}^{(0)}_{a\perp}(\tau,\vec{\sigma})+
\check{Q}_{1u}(\tau )\check{\vec{\pi}}_{au\perp}^{(1)}(\tau,\vec{\sigma})+
\check{Q}_{1u}(\tau )\check{Q}_{1v}(\tau )\check{\vec{\pi}}_{auv\perp}^{(2)}
(\tau,\vec{\sigma}).
\label{e4}
\end{eqnarray}

As a consequence we have [we suppres the $\tau$-dependence  and also the
$\vec \sigma$-dependence when possible with the replacement ${\vec \sigma}_i
\rightarrow i$]

\begin{eqnarray}
{\check \rho}_a(\vec \sigma )&=&{\check \rho}^{(YM)}_a(\vec \sigma )+\sum_{i=1}
^2{\check \rho}_{ia}(\vec \sigma )=\nonumber \\
&=&{\check \rho}^{(YM)(o)}_a(\vec \sigma )+{\check Q}_{1u}{\check \rho}
^{(YM)(1)}_{au}(\vec \sigma )+{\check Q}_{1u}{\check Q}_{1v}{\check \rho}
^{(YM)(2)}_{auv}(\vec \sigma )+\sum_{i=1}^2{\check \rho}_{ia}(\vec \sigma )=
\nonumber \\
&=&c_{abc}({\check {\vec A}}^{(o)}_{b\perp}\cdot {\check {\vec \pi}}^{(o)}
_{c\perp})(\vec \sigma )+{\check Q}_{1u} [c_{abc}({\check {\vec A}}^{(o)}
_{b\perp}\cdot {\check {\vec \pi}}^{(1)}_{cu\perp}+{\check {\vec A}}^{(1)}
_{bu\perp}\cdot {\check {\vec \pi}}^{(o)}_{c\perp})(\vec \sigma )-\nonumber \\
&-&\delta_{au}(\delta^3(\vec \sigma -
{\vec \eta}_1)-\delta^3(\vec \sigma -{\vec \eta}_2))]+\nonumber \\
&+&{\check Q}_{1u}{\check Q}_{1v}c_{abc}[{\check {\vec A}}^{(o)}_{b\perp}\cdot
{\check {\vec \pi}}^{(2)}_{buv\perp}+{\check {\vec A}}^{(1)}_{bu\perp}\cdot
{\check {\vec \pi}}^{(1)}_{cv\perp}+{\check {\vec A}}^{(2)}_{buv\perp}\cdot
{\check {\vec \pi}}^{(o)}_{c\perp}](\vec \sigma ),\nonumber \\
&&{}\nonumber \\
{\check \rho}_a({\vec \sigma}_1){\check \rho}_b({\vec \sigma}_2)&=&c_{amn}
{\check {\vec A}}^{(o)}_{m\perp}(1)\cdot {\check {\vec \pi}}^{(o)}_{c\perp}(1)
c_{brs}{\check {\vec A}}^{(o)}_{r\perp}(2)\cdot {\check {\vec \pi}}^{(o)}
_{s\perp}(2)+\nonumber \\
&+&{\check Q}_{1u} [(\, c_{amn}({\check {\vec A}}^{(o)}_{m\perp}(1)\cdot
{\check {\vec \pi}}^{(1)}_{nu\perp}(1)+{\check {\vec A}}^{(1)}_{mu\perp}(1)
\cdot {\check {\vec \pi}}^{(o)}_{n\perp}(1))-\nonumber \\
&-&\delta_{au}(\delta^3({\vec \sigma}
_1-{\vec \eta}_1)-\delta^3({\vec \sigma}_2-{\vec \eta}_2))\, )c_{brs}
{\check {\vec A}}^{(o)}_{r\perp}(2)\cdot {\check {\vec \pi}}^{(o)}_{s\perp}(2)+
\nonumber \\
&+&c_{amn}{\check {\vec A}}^{(o)}_{m\perp}(1)\cdot {\check {\vec \pi}}^{(o)}
_{n\perp}(1) (c_{brs}({\check {\vec A}}^{(o)}_{r\perp}(2)\cdot {\check {\vec 
\pi}}^{(1)}_{su\perp}(2)+{\check {\vec A}}^{(1)}_{ru\perp}(2)\cdot {\check 
{\vec \pi}}^{(o)}_{s\perp}(2))-\nonumber \\
&-&\delta_{bu}(\delta^3({\vec \sigma}_2-{\vec
\eta}_1)-\delta^3({\vec \sigma}_2-{\vec \eta}_2))\, ]+\nonumber \\
&+&{\check Q}_{1u}{\check Q}_{1v}[\, c_{amn}{\check {\vec A}}^{(o)}_{m\perp}(1)
\cdot {\check {\vec \pi}}^{(o)}_{n\perp}(1) c_{brs}({\check {\vec A}}^{(o)}
_{r\perp}(2)\cdot {\check {\vec \pi}}^{(2)}_{suv\perp}(2)+\nonumber \\
&+&{\check {\vec A}}
^{(1)}_{ru\perp}(2)\cdot {\check {\vec \pi}}^{(1)}_{sv\perp}(2)+{\check {\vec
A}}^{(2)}_{ruv\perp}(2)\cdot {\check {\vec \pi}}^{(o)}_{s\perp}(2))+\nonumber \\
&+&c_{amn}(({\check {\vec A}}^{(o)}
_{m\perp}(1)\cdot {\check {\vec \pi}}^{(2)}_{nuv\perp}(1)+{\check {\vec A}}
^{(1)}_{mu\perp}(1)\cdot {\check {\vec \pi}}^{(1)}_{nv\perp}(1)+\nonumber \\
&+&{\check {\vec
A}}^{(2)}_{muv\perp}(1)\cdot {\check {\vec \pi}}^{(o)}_{n\perp}(1)) c_{brs}
{\check {\vec A}}^{(o)}_{r\perp}(2)\cdot {\check {\vec \pi}}^{(o)}_{s\perp}(2)+
\nonumber \\
&+& (\, c_{amn}({\check {\vec A}}^{(o)}_{m\perp}(1)\cdot {\check {\vec \pi}}
^{(1)}_{nu\perp}(1)+{\check {\vec A}}^{(1)}_{mu\perp}(1)\cdot {\check {\vec 
\pi}}^{(o)}_{n\perp}(1))\nonumber \\
&-&-\delta_{au}(\delta^3({\vec \sigma}_1-{\vec \eta}_1)-
\delta^3({\vec \sigma}_1-{\vec \eta}_2))\, )\nonumber \\
&\cdot& (\, c_{brs}({\check {\vec A}}^{(o)}_{r\perp}(2)\cdot {\check {\vec \pi}}
^{(1)}_{sv\perp}(2)+{\check {\vec A}}^{(1)}_{rv\perp}(2)\cdot {\check {\vec 
\pi}}^{(o)}_{s\perp}(2))-\nonumber \\
&-&\delta_{av}(\delta^3({\vec \sigma}_2-{\vec \eta}_1)-
\delta^3({\vec \sigma}_2-{\vec \eta}_2))\, )\, ]=\nonumber \\
&=&R_{(1,2)ab}({\vec \sigma}_1,{\vec \sigma}_2;\tau )+
R_{(1)ab}({\vec \sigma}_1,{\vec \sigma}_2;\tau )+
R_{(2)ab}({\vec \sigma}_1,{\vec \sigma}_2;\tau )+
R_{ab}({\vec \sigma}_1,{\vec \sigma}_2;\tau ),\nonumber \\
&&{}\nonumber \\
R_{(1,2)ab}({\vec \sigma}_1,{\vec \sigma}_2;\tau )&=&{\check Q}_{1a}{\check Q}
_{1b}[\delta^3(\vec{\sigma}_{1}-\vec{\eta}_{1})-
\delta^3(\vec{\sigma}_{1}-\vec{\eta}_{2})]
[\delta^3(\vec{\sigma}_{2}-\vec{\eta}_{1})-
\delta^3(\vec{\sigma}_{1}-\vec{\eta}_{2})],\nonumber \\
R_{(i)ab}({\vec \sigma}_1,{\vec \sigma}_2;\tau )&=&(-)^i {\check Q}_{1u}[\delta
_{au}\delta^3({\vec \sigma}_1-{\vec \sigma}_i)c_{brs} {\check {\vec A}}^{(o)}
_{r\perp}(2)\cdot {\check {\vec \pi}}^{(o)}_{s\perp}(2)+\nonumber \\
&+&\delta_{bu} \delta^3
({\vec \sigma}_2-{\vec \eta}_i) c_{amn} {\check {\vec A}}^{(o)}_{m\perp}(1)
\cdot {\check {\vec \pi}}^{(o)}_{n\perp}(1)={\check Q}_{1u}(\tau )
R_{(i)abu}({\vec \sigma}_1,{\vec \sigma}_2;\tau ),\nonumber \\
R_{ab}({\vec \sigma}_1,{\vec \sigma}_2;\tau )&=&
R_{ab}^{(o)}({\vec \sigma}_1,{\vec \sigma}_2;\tau )+{\check Q}_{1u}
R_{abu}^{(1)}({\vec \sigma}_1,{\vec \sigma}_2;\tau )+{\check Q}_{1u}{\check Q}
_{1v}R_{abuv}^{(2)}({\vec \sigma}_1,{\vec \sigma}_2;\tau )=\nonumber \\
&=&c_{amn}
{\check {\vec A}}^{(o)}_{m\perp}(1)\cdot {\check {\vec \pi}}^{(o)}_{c\perp}(1)
c_{brs}{\check {\vec A}}^{(o)}_{r\perp}(2)\cdot {\check {\vec \pi}}^{(o)}
_{s\perp}(2)+\nonumber \\
&+&{\check Q}_{1u} [(\, c_{amn}({\check {\vec A}}^{(o)}_{m\perp}(1)\cdot
{\check {\vec \pi}}^{(1)}_{nu\perp}(1)+{\check {\vec A}}^{(1)}_{mu\perp}(1)
\cdot {\check {\vec \pi}}^{(o)}_{n\perp}(1))\, )c_{brs}
{\check {\vec A}}^{(o)}_{r\perp}(2)\cdot {\check {\vec \pi}}^{(o)}_{s\perp}(2)+
\nonumber \\
&+&c_{amn}{\check {\vec A}}^{(o)}_{m\perp}(1)\cdot {\check {\vec \pi}}^{(o)}
_{n\perp}(1) (c_{brs}({\check {\vec A}}^{(o)}_{r\perp}(2)\cdot {\check {\vec 
\pi}}^{(1)}_{su\perp}(2)+{\check {\vec A}}^{(1)}_{ru\perp}(2)\cdot {\check 
{\vec \pi}}^{(o)}_{s\perp}(2))\, ]+\nonumber \\
&+&{\check Q}_{1u}{\check Q}_{1v}[\, c_{amn}{\check {\vec A}}^{(o)}_{m\perp}(1)
\cdot {\check {\vec \pi}}^{(o)}_{n\perp}(1) c_{brs}({\check {\vec A}}^{(o)}
_{r\perp}(2)\cdot {\check {\vec \pi}}^{(2)}_{suv\perp}(2)+\nonumber \\
&+&{\check {\vec A}}
^{(1)}_{ru\perp}(2)\cdot {\check {\vec \pi}}^{(1)}_{sv\perp}(2)+{\check {\vec
A}}^{(2)}_{ruv\perp}(2)\cdot {\check {\vec \pi}}^{(o)}_{s\perp}(2))+\nonumber \\
&+&c_{amn}(({\check {\vec A}}^{(o)}
_{m\perp}(1)\cdot {\check {\vec \pi}}^{(2)}_{nuv\perp}(1)+{\check {\vec A}}
^{(1)}_{mu\perp}(1)\cdot {\check {\vec \pi}}^{(1)}_{nv\perp}(1)+\nonumber \\
&+&{\check {\vec
A}}^{(2)}_{muv\perp}(1)\cdot {\check {\vec \pi}}^{(o)}_{n\perp}(1)) c_{brs}
{\check {\vec A}}^{(o)}_{r\perp}(2)\cdot {\check {\vec \pi}}^{(o)}_{s\perp}(2)+
\nonumber \\
&+& (\, c_{amn}({\check {\vec A}}^{(o)}_{m\perp}(1)\cdot {\check {\vec \pi}}
^{(1)}_{nu\perp}(1)+{\check {\vec A}}^{(1)}_{mu\perp}(1)\cdot {\check {\vec 
\pi}}^{(o)}_{n\perp}(1))\, )\nonumber \\
&\cdot& (\, c_{brs}({\check {\vec A}}^{(o)}_{r\perp}(2)\cdot {\check {\vec \pi}}
^{(1)}_{sv\perp}(2)+{\check {\vec A}}^{(1)}_{rv\perp}(2)\cdot {\check {\vec 
\pi}}^{(o)}_{s\perp}(2))\, )\, ].
\label{e5}
\end{eqnarray}

\noindent In the last lines we separated the particle-particle {$R_{(1,2)ab}$],
the particle-field [$R_{(i)ab}$] and field-field [$R_{ab}$] terms.

If we write

\begin{eqnarray}
{\vec \zeta}^{({\check A}_{\perp})}_{ab}({\vec \sigma}_1,{\vec \sigma}_2;\tau )
&=&{\vec \zeta}^{({\check A}^{(o)}_{\perp})}_{ab}({\vec \sigma}_1,{\vec \sigma}
_2;\tau )+{\check Q}_{1u}{\vec \zeta}^{({\check A}^{(o)}_{\perp},{\check A}
^{(1)}_{\perp})}_{abu}({\vec \sigma}_1,{\vec \sigma}_2;\tau )+\nonumber \\
&+&{\check Q}_{1u}{\check 
Q}_{1v}{\vec \zeta}^{({\check A}^{(o)}_{\perp},{\check A}^{(1)}_{\perp},{\check
A}^{(2)}_{\perp})}_{abuv}({\vec \sigma}_1,{\vec \sigma}_2;\tau ),
\label{e6}
\end{eqnarray}

\noindent where ${\vec \zeta}^{({\check A}^{(o)}_{\perp})}_{ab}({\vec \sigma}_1
,{\vec \sigma}_2;\tau )=\vec c ({\vec \sigma}_1 -{\vec \sigma}_2) (P\, e^{\int
_{\sigma_2}^{\sigma_1} d{\vec \sigma}^{"}\, \cdot {\check {\vec A}}^{(o)}_c
(\tau ,{\vec \sigma}^{"}) {\hat T}^c}\, )_{ab}$ and ${\vec \zeta}^{({\check A}
^{(o)}_{\perp},{\check A}^{(1)}_{\perp})}_{abu}$ and ${\vec \zeta}
^{({\check A}^{(o)}_{\perp},{\check A}^{(1)}_{\perp},{\check A}^{(2)}
_{\perp})}_{abuv}$ are functions which could be evaluated by using the 
definition of Wilson path-ordering, then we can write the following 
decompositions of the interaction kernel $K_{ab}({\vec \sigma}_1,{\vec \sigma}_2
;\tau )$

\begin{eqnarray}
K_{ab}({\vec \sigma}_1,{\vec \sigma}_2;\tau )&=&
K_{ab}^{(o)}({\vec \sigma}_1,{\vec \sigma}_2;\tau )+{\check Q}_{1u}K_{abu}
^{(1)}({\vec \sigma}_1,{\vec \sigma}_2;\tau )+{\check Q}_{1u}{\check Q}_{1v}
K_{abuv}^{(2)}({\vec \sigma}_1,{\vec \sigma}_2;\tau ),\nonumber \\
&&{}\nonumber \\
K_{ab}^{(o)}({\vec \sigma}_1,{\vec \sigma}_2;\tau )&=&\int d^3\sigma_1d^3\sigma
_2 [ {{\delta_{ab}\delta^3({\vec \sigma}_3-{\vec \sigma}_1)\delta^3({\vec 
\sigma}_4-{\vec \sigma}_2)}\over {4\pi |{\vec \sigma}_3-{\vec \sigma}_4|}}+
\nonumber \\
&+&{{\delta^3({\vec \sigma}_4-{\vec \sigma}_2)[{\check {\vec A}}^{(o)}(\tau ,
{\vec \sigma}_3)\cdot {\vec \zeta}^{({\check A}^{(o)}_{\perp})}]_{ab}({\vec
\sigma}_3,{\vec \sigma}_1;\tau )+({\vec \sigma}_1 \leftrightarrow {\vec 
\sigma}_2)}\over {4\pi |{\vec \sigma}_3-{\vec \sigma}_4|}}+
\nonumber \\
&+&{{[{\check {\vec A}}^{(o)}_{\perp}(\tau ,{\vec \sigma}_3)\cdot {\vec \zeta}
^{({\check A}^{(o)}_{\perp})}({\vec \sigma}_3,{\vec \sigma}_1;\tau )]_{au}
[{\check {\vec A}}^{(o)}_{\perp}(\tau ,{\vec \sigma}_4)\cdot {\vec \zeta}
^{({\check A}^{(o)}_{\perp})}({\vec \sigma}_4,{\vec \sigma}_2;\tau )]_{bu} }
\over {4\pi |{\vec \sigma}_3-{\vec \sigma}_4|}}\, ],\nonumber \\
K_{abu}^{(1)}({\vec \sigma}_1,{\vec \sigma}_2;\tau )&=&\int {{d^3\sigma_3
d^3\sigma_4}\over {4\pi |{\vec \sigma}_3-{\vec \sigma}_4|}} \cdot\nonumber \\
&\cdot& [\, \delta^3({\vec \sigma}_4-{\vec \sigma}_2) [{\check {\vec A}}^{(o)}
_{\perp}(\tau ,{\vec \sigma}_3)\cdot {\vec \zeta}^{({\check A}^{(o)}_{\perp},
{\check A}^{(1)}_{\perp})}_u({\vec \sigma}_3,{\vec \sigma}_1;\tau )+\nonumber \\
&+&{\check
{\vec A}}^{(1)}_{u\perp}(\tau ,{\vec \sigma}_3)\cdot {\vec \zeta}^{({\check
A}_{\perp}^{(o)})}({\vec \sigma}_3,{\vec \sigma}_1;\tau )\, ]_{ab}+ ({\vec 
\sigma}_1 \leftrightarrow {\vec \sigma}_2)+\nonumber \\
&+&[{\check {\vec A}}^{(o)}_{\perp}(\tau ,{\vec \sigma}_3)\cdot {\vec \zeta}
^{({\check A}^{(o)}_{\perp})}({\vec \sigma}_3,{\vec \sigma}_1;\tau )\, ]_{ae}
[{\check {\vec A}}^{(o)}_{\perp}(\tau ,{\vec \sigma}_4)\cdot {\vec \zeta}_u
^{({\check A}^{(o)}_{\perp},{\check A}^{(1)}_{\perp})}({\vec \sigma}_4,{\vec 
\sigma}_2;\tau )+\nonumber \\
&+&{\check {\vec A}}^{(1)}_{u\perp}(\tau ,{\vec \sigma}_4)\cdot
{\vec \zeta}^{({\check A}^{(o)}_{\perp})}({\vec \sigma}_4,{\vec \sigma}_2;\tau 
) ]_{be}+\nonumber \\
&+&[{\check {\vec A}}^{(o)}_{\perp}(\tau ,{\vec \sigma}_3)\cdot {\vec \zeta}_u
^{({\check A}^{(o)}_{\perp},{\check A}^{(1)}_{\perp})}({\vec \sigma}_3,{\vec 
\sigma}_1;\tau )+{\check {\vec A}}^{(1)}_{u\perp}(\tau ,{\vec \sigma}_3)\cdot
{\vec \zeta}^{({\check A}^{(o)}_{\perp})}({\vec \sigma}_3,{\vec \sigma}_1;\tau 
)]_{ae}\nonumber \\
&&[{\check {\vec A}}^{(o)}_{\perp}(\tau ,{\vec \sigma}_4)\cdot {\vec \zeta}
^{({\check A}^{(o)}_{\perp})}({\vec \sigma}_4,{\vec \sigma}_2;\tau )\, ]_{be}
\, ],\nonumber \\
K_{abuv}^{(2)}({\vec \sigma}_1,{\vec \sigma}_2;\tau )&=&
\int {{d^3\sigma_3d^3\sigma_4}\over {4\pi |{\vec \sigma}_3-{\vec \sigma}_4|}}
\nonumber \\
&\cdot& [\, \delta^3({\vec \sigma}_4-{\vec \sigma}_2)[{\check {\vec A}}^{(o)}
_{\perp}(\tau ,{\vec \sigma}_3)\cdot {\vec \zeta}_{uv}^{({\check A}^{(o)}
_{\perp},{\check A}^{(1)}_{\perp},{\check A}^{(2)}_{\perp})}({\vec \sigma}_3,
{\vec \sigma}_1;\tau )+{\check {\vec A}}^{(1)}_{u\perp}(\tau ,{\vec \sigma}_3)
\cdot \nonumber \\
&\cdot& {\vec \zeta}_v^{({\check A}^{(o)}_{\perp},{\check A}^{(1)}_{\perp})}
({\vec \sigma}_3,{\vec \sigma}_1;\tau )+{\check {\vec A}}^{(2)}_{uv\perp}(\tau 
,{\vec \sigma}_3)\cdot {\vec \zeta}^{({\check A}^{(o)}_{\perp})}({\vec \sigma}
_3,{\vec \sigma}_1;\tau )]_{ab}+({\vec \sigma}_1\leftrightarrow {\vec \sigma}_2)
+\nonumber \\
&+&[{\check {\vec A}}^{(o)}_{\perp}(\tau ,{\vec \sigma}_3)\cdot {\vec \zeta}
^{({\check A}^{(o)}_{\perp})}({\vec \sigma}_3,{\vec \sigma}_1;\tau )\, ]_{ae}
[{\check {\vec A}}^{(o)}
_{\perp}(\tau ,{\vec \sigma}_4)\cdot {\vec \zeta}_{uv}^{({\check A}^{(o)}
_{\perp},{\check A}^{(1)}_{\perp},{\check A}^{(2)}_{\perp})}({\vec \sigma}_4,
{\vec \sigma}_2;\tau )+\nonumber \\
&+&{\check {\vec A}}^{(1)}_{u\perp}(\tau ,{\vec \sigma}_4)
\cdot {\vec \zeta}_v^{({\check A}^{(o)}_{\perp},{\check A}^{(1)}_{\perp})}
({\vec \sigma}_4,{\vec \sigma}_2;\tau )+{\check {\vec A}}^{(2)}_{uv\perp}(\tau 
,{\vec \sigma}_4)\cdot {\vec \zeta}^{({\check A}^{(o)}_{\perp})}({\vec \sigma}
_4,{\vec \sigma}_2;\tau )]_{be}+\nonumber \\
&+&[{\check {\vec A}}^{(o)}_{\perp}(\tau ,{\vec \sigma}_3)\cdot {\vec \zeta}_u
^{({\check A}^{(o)}_{\perp},{\check A}^{(1)}_{\perp})}({\vec \sigma}_3,{\vec 
\sigma}_1;\tau )+{\check {\vec A}}^{(1)}_{u\perp}(\tau ,{\vec \sigma}_3)\cdot
{\vec \zeta}^{({\check A}^{(o)}_{\perp})}({\vec \sigma}_3,{\vec \sigma}_1;\tau 
)]_{ae}\nonumber \\
&&[{\check {\vec A}}^{(o)}_{\perp}(\tau ,{\vec \sigma}_4)\cdot {\vec \zeta}_v
^{({\check A}^{(o)}_{\perp},{\check A}^{(1)}_{\perp})}({\vec \sigma}_4,{\vec 
\sigma}_2;\tau )+{\check {\vec A}}^{(1)}_{v\perp}(\tau ,{\vec \sigma}_4)\cdot
{\vec \zeta}^{({\check A}^{(o)}_{\perp})}({\vec \sigma}_4,{\vec \sigma}_2;\tau 
)]_{be}+\nonumber \\
&+&[{\check {\vec A}}^{(o)}
_{\perp}(\tau ,{\vec \sigma}_3)\cdot {\vec \zeta}_{uv}^{({\check A}^{(o)}
_{\perp},{\check A}^{(1)}_{\perp},{\check A}^{(2)}_{\perp})}({\vec \sigma}_3,
{\vec \sigma}_1;\tau )+{\check {\vec A}}^{(1)}_{u\perp}(\tau ,{\vec \sigma}_3)
\cdot {\vec \zeta}_v^{({\check A}^{(o)}_{\perp},{\check A}^{(1)}_{\perp})}
({\vec \sigma}_3,{\vec \sigma}_1;\tau )+\nonumber \\
&+&{\check {\vec A}}^{(2)}_{uv\perp}(\tau 
,{\vec \sigma}_3)\cdot {\vec \zeta}^{({\check A}^{(o)}_{\perp})}({\vec \sigma}
_3,{\vec \sigma}_1;\tau )]_{ae}
[{\check {\vec A}}^{(o)}_{\perp}(\tau ,{\vec \sigma}_4)\cdot {\vec \zeta}
^{({\check A}^{(o)}_{\perp})}({\vec \sigma}_4,{\vec \sigma}_2;\tau )\, ]_{be}
\, ],
\label{e7}
\end{eqnarray}

\noindent and of the potential V in particle-particle [$V_{PP}$], particle-
field [$V_{(i)PF}$] and field-field [$V_{FF}$] contributions

\begin{eqnarray}
\frac{1}{2}V[\vec{\eta}_{i},\check{\vec{A}}_{a\perp},\check{\vec{\pi}}
_{a\perp}](\tau )&=&V_{PP}[\vec{\eta}_{i},\vec{\eta}_{j},{\check {\vec A}}
^{(o)}_{a\perp}](\tau )+\sum_{i=1}^2V_{(i)PF}[\vec{\eta}_{i},{\check {\vec A}}
^{(o)}_{a\perp},{\check {\vec \pi}}^{(o)}_{a\perp},{\check {\vec A}}^{(1)}
_{au\perp}](\tau )+\nonumber \\
&+&V_{FF}[{\check {\vec A}}^{(k)}_{\perp},{\check {\vec \pi}}
^{(k)}_{\perp}](\tau ),\nonumber \\
&&{}\nonumber \\
V_{PP}&=&g^2_s \int d^3\sigma_1d^3\sigma_2 \sum_{a,b}[R_{(1,2)ab}K_{ab}]({\vec
\sigma}_1,{\vec \sigma}_2;\tau )=\nonumber \\
&=&g^2_s\int d^3\sigma_1d^3\sigma_2 \sum_{a,b}{\check Q}_{1a}{\check Q}_{1b}
[\delta^3({\vec \sigma}_1-{\vec \eta}_1)-\delta^3({\vec \sigma}_1-{\vec \eta}_2)
]\nonumber \\
&&[\delta^3({\vec \sigma}_2-{\vec \eta}_1)-\delta^3({\vec \sigma}_2-{\vec \eta}
_2)] K^{(o)}_{ab}({\vec \sigma}_1,{\vec \sigma}_2;\tau ), \nonumber \\
V_{(i)PF}&=&g^2_s\int d^3\sigma_1d^3\sigma_2 \sum_{a,b} [R_{(i)ab}K_{ab}]
({\vec \sigma}_1,{\vec \sigma}_2;\tau )=\nonumber \\
&=&g^2_s \int d^3\sigma_1d^3\sigma_2 \sum_{a,b}{\check Q}_{1u} R_{(i)abu}({\vec
\sigma}_1,{\vec \sigma}_2;\tau )[K^{(o)}_{ab}+{\check Q}_{1v}K^{(1)}_{abv}]
({\vec \sigma}_1,{\vec \sigma}_2;\tau ),\nonumber \\
V_{FF}&=&g^2_s \int d^3\sigma_1d^3\sigma_2 {\check \rho}_a^{(YM)}(\tau,{\vec 
\sigma}_1) K_{ab}({\vec \sigma}_1,{\vec \sigma}_2;\tau ) {\check \rho}_b^{(YM)}
(\tau ,{\vec \sigma}_2)=\nonumber \\
&=&g^2_s \int d^3\sigma_1d^3\sigma_2 \sum_{a,b} [R^{(o)}_{ab}K^{(o)}_{ab}+
{\check Q}_{1u}(R^{(o)}_{ab}K^{(1)}_{abu}+R^{(1)}_{abu}K^{(o)}_{ab})+
\nonumber \\
&+&{\check Q}_{1u}{\check Q}_{1v}(R^{(o)}_{ab}K^{(2)}_{abuv}+R^{(1)}_{abu}K
^{(1)}_{abv}+R^{(2)}_{abuv}K^{(o)}_{ab})]({\vec \sigma}_1,{\vec \sigma}_2;\tau )
.
\label{e8}
\end{eqnarray}

The explicit form of the particle-particle potential is

\begin{eqnarray}
V_{PP}&=&\frac{1}{2}g^2_s \sum_{a,b}\check{Q}_{1a}\check{Q}_{1b}
\int d^{3}\sigma_{1}\int d^{3}\sigma_{2}
[\delta^3(\vec{\sigma}_{1}-\vec{\eta}_{1}(\tau ))
   \delta^3(\vec{\sigma}_{2}-\vec{\eta}_{1}(\tau ))+\nonumber\\
&+& \delta^3(\vec{\sigma}_{1}-\vec{\eta}_{2}(\tau ))
   \delta^3(\vec{\sigma}_{2}-\vec{\eta}_{2}(\tau ))-2
   \delta^3(\vec{\sigma}_{1}-\vec{\eta}_{1}(\tau ))
   \delta^3(\vec{\sigma}_{2}-\vec{\eta}_{2}(\tau ))]\cdot\nonumber\\
&\cdot&\frac{1}{4\pi}\int d^{3}\sigma_3\int d^{3}\sigma_4\{
\frac{\delta_{ab}\delta^3({\vec{\sigma}}_3-\vec{\sigma}_{1})
                 \delta^3(\vec{\sigma}_4-\vec{\sigma}_{2})}
{\mid\vec{\sigma}_3-\vec{\sigma}_4\mid}+\nonumber\\
&+&\frac{c_{auv}}{4\pi}[\frac{\delta^3(\vec{\sigma}_3-\vec{\sigma}_{1})
(\vec{\sigma}_4-\vec{\sigma}_{2})\cdot\check{\vec{A}}^{(0)}_{v\perp}(\tau ,
\vec{\sigma}_4){\zeta}^{({\check A}^{(o)}_{\perp})}
_{ub}(\vec{\sigma}_{2},\vec{\sigma}_4;\tau )}
{\mid\vec{\sigma}_3-\vec{\sigma}_4\mid\mid\vec{\sigma}_4-\vec{\sigma}_{2}\mid
^{3}}+\nonumber\\
&+&                     \frac{\delta^3(\vec{\sigma}_4-\vec{\sigma}_{2})
(\vec{\sigma}_3-\vec{\sigma}_{1})\cdot\check{\vec{A}}^{(0)}_{v\perp}(\tau ,
\vec{\sigma}_3){\zeta}^{({\check A}^{(o)}_{\perp})}
_{ub}(\vec{\sigma}_{1},\vec{\sigma}_3;\tau )}
{\mid\vec{\sigma}_3-\vec{\sigma}_4\mid\mid\vec{\sigma}_3-\vec{\sigma}_{1}\mid
^{3}}]+\frac{c_{arv}c_{bst}}{(4\pi)^{2}}\cdot \nonumber\\
&\cdot& \frac{
(\vec{\sigma}_4-\vec{\sigma}_{2})\cdot\check{\vec{A}}^{(0)}_{v\perp}(\tau ,
\vec{\sigma}_4){\zeta}^{({\check A}^{(o)}_{\perp})}
_{ru}(\vec{\sigma}_{2},\vec{\sigma}_4;\tau )
(\vec{\sigma}_3-\vec{\sigma}_{1})\cdot\check{\vec{A}}^{(0)}_{t\perp}(\tau ,
\vec{\sigma}_3){\zeta}^{({\check A}^{(o)}_{\perp})}
_{su}(\vec{\sigma}_{1},\vec{\sigma}_3;\tau )}{
\mid\vec{\sigma}_3-\vec{\sigma}_{1}\mid^{3}
\mid\vec{\sigma}_3-\vec{\sigma}_4\mid
\mid\vec{\sigma}_4-\vec{\sigma}_{2}\mid^{3}}\}=\nonumber \\
&=&\frac{1}{2} g^2_s \sum_{a,b}\check{Q}_{1a}\check{Q}_{1b}
\int d^{3}\sigma_{1}\int d^{3}\sigma_{2}
[\delta^3(\vec{\sigma}_{1}-\vec{\eta}_{1}(\tau ))
   \delta^3(\vec{\sigma}_{2}-\vec{\eta}_{1}(\tau ))+\nonumber\\
&+&   \delta^3(\vec{\sigma}_{1}-\vec{\eta}_{2}(\tau ))
   \delta^3(\vec{\sigma}_{2}-\vec{\eta}_{2}(\tau ))-2
   \delta^3(\vec{\sigma}_{1}-\vec{\eta}_{1}(\tau ))
   \delta^3(\vec{\sigma}_{2}-\vec{\eta}_{2}(\tau ))]\cdot\nonumber\\
&\cdot&\frac{1}{4\pi}\int d^{3}\sigma_3\int d^{3}\sigma_4\{
\frac{c_{auv}}{4\pi}[\frac{\delta^3(\vec{\sigma}_3-\vec{\sigma}_{1})
(\vec{\sigma}_4-\vec{\sigma}_{2})\cdot\check{\vec{A}}^{(0)}_{v\perp}(\tau ,
\vec{\sigma}_4){\zeta}^{({\check A}^{(o)}_{\perp})}
_{ub}(\vec{\sigma}_{2},\vec{\sigma}_4;\tau )}
{\mid\vec{\sigma}_3-\vec{\sigma}_4\mid\mid\vec{\sigma}_4-\vec{\sigma}_{2}\mid
^{3}}+\nonumber\\
&+&                     \frac{\delta^3(\vec{\sigma}_4-\vec{\sigma}_{2})
(\vec{\sigma}_3-\vec{\sigma}_{1})\cdot\check{\vec{A}}^{(0)}_{v\perp}(\tau ,
\vec{\sigma}_3){\zeta}^{({\check A}^{(o)}_{\perp})}
_{ub}(\vec{\sigma}_{1},\vec{\sigma}_3;\tau )}
{\mid\vec{\sigma}_3-\vec{\sigma}_4\mid\mid\vec{\sigma}_3-\vec{\sigma}_{1}\mid
^{3}}]+\frac{c_{arv}c_{bst}}{(4\pi)^{2}}\cdot \nonumber\\
&\cdot& \frac{
(\vec{\sigma}_4-\vec{\sigma}_{2})\cdot\check{\vec{A}}^{(0)}_{v\perp}(\tau ,
\vec{\sigma}_4){\zeta}^{({\check A}^{(o)}_{\perp})}
_{ru}(\vec{\sigma}_{2},\vec{\sigma}_4;\tau )
(\vec{\sigma}_3-\vec{\sigma}_{1})\cdot\check{\vec{A}}^{(0)}_{t\perp}(\tau ,
\vec{\sigma}_3){\zeta}^{({\check A}^{(o)}_{\perp})}
_{su}(\vec{\sigma}_{1},\vec{\sigma}_3;\tau )}{
\mid\vec{\sigma}_3-\vec{\sigma}_{1}\mid^{3}
\mid\vec{\sigma}_3-\vec{\sigma}_4\mid
\mid\vec{\sigma}_4-\vec{\sigma}_{2}\mid^{3}}\}.
\label{e9}
\end{eqnarray}

\noindent The first term, with the instantaneous Coulomb interaction, does not
contribute, because for every N we have $\sum_a{\check Q}_{1a}{\check Q}_{1a}
=0$, due to Eq.(\ref{a23}).

With regard to the other two terms, let us remark that in the limit ${\vec \eta}
_1={\vec \eta}_2$ [and thus also in the points of maximal divergence, i.e.
$\vec{\sigma}_{1}=\vec{\sigma}_{2}=\vec{\sigma}$ for the first two terms and 
 $\vec{\sigma}_{1}=\vec{\sigma}_{2}=\vec{\sigma}=\vec{\sigma}'$ for the third
one] the potential vanishes due to the multiplicative term containing the
Dirac functions.

With the substitutions $\vec{\sigma}=\vec{\sigma}_{1}+\vec{\xi}$ and
$\vec{\sigma}'=\vec{\sigma}_{2}+\vec{\xi}'$, the particle-particle potential
can be put in the form

\begin{eqnarray} 
V_{PP}&=& g^2_s \sum_{a,b}\check{Q}_{1a}\check{Q}_{1b}\{
c_{auv}\int\frac{d^{3}\xi}{(4\pi)^{2}\mid\vec{\xi}\mid^{2}}\cdot\nonumber\\
&\cdot&[\sum_{i=1}^{2}\frac{\vec{\xi}\cdot
\vec{A}^{(0)}_{v\perp}(\tau ,\vec{\eta}_{i}(\tau )+\vec{\xi})}{\mid\vec{\xi}
\mid^{2}}\zeta^{({\check A}^{(o)}_{\perp})}
_{ub}(\vec{\eta}_{i}(\tau ),\vec{\eta}_{i}(\tau )+\vec{\xi};\tau )-\nonumber\\
&-&            \frac{\vec{\xi}\cdot
\vec{A}^{(0)}_{v\perp}(\tau ,\vec{\eta}_{1}(\tau )+\vec{\xi})}
{\mid\vec{\xi}\mid\mid\vec{\eta}_{1}(\tau )-\vec{\eta}_{2}(\tau )+\vec{\xi}\mid}
\zeta^{({\check A}^{(o)}_{\perp})}
_{ub}(\vec{\eta}_{1}(\tau ),\vec{\eta}_{1}(\tau )+\vec{\xi};\tau )-\nonumber \\
&-&               \frac{\vec{\xi}\cdot
\vec{A}^{(0)}_{v\perp}(\tau ,\vec{\eta}_{2}(\tau )+\vec{\xi})}
{\mid\vec{\xi}\mid\mid\vec{\eta}_{1}(\tau )-\vec{\eta}_{2}(\tau )-\vec{\xi}\mid}
\zeta^{({\check A}^{(o)}_{\perp})}
_{ub}(\vec{\eta}_{2}(\tau ),\vec{\eta}_{2}(\tau )+\vec{\xi};\tau )]+\nonumber\\
&+&\frac{1}{2}c_{arv}c_{bst}\int\frac{d^{3}\xi d^{3}\xi'}{(4\pi)^{3}
\mid\vec{\xi}\mid^{3}\mid\vec{\xi}'\mid^{3}}\cdot\nonumber\\
&\cdot&[\sum_{i=1}^{2}\frac{
\vec{\xi}\cdot\vec{A}^{(0)}_{v\perp}(\tau ,\vec{\eta}_{i}(\tau )+\vec{\xi})
\vec{\xi'}\cdot\vec{A}^{(0)}_{t\perp}(\tau ,\vec{\eta}_{i}(\tau )+\vec{\xi}')}
{\mid\vec{\xi}-\vec{\xi}'\mid}\cdot \nonumber \\
&\cdot& \zeta^{({\check A}^{(o)}_{\perp})}
_{ru}(\vec{\eta}_{i}(\tau ),\vec{\eta}_{i}(\tau )+\vec{\xi};\tau )
\zeta^{({\check A}^{(o)}_{\perp})}
_{su}(\vec{\eta}_{i}(\tau ),\vec{\eta}_{i}(\tau )+\vec{\xi}';\tau )+\nonumber\\
&-&2\cdot\frac{
\vec{\xi}\cdot\vec{A}^{(0)}_{v\perp}(\tau ,\vec{\eta}_{1}(\tau )+\vec{\xi})
\vec{\xi'}\cdot\vec{A}^{(0)}_{t\perp}(\tau ,\vec{\eta}_{2}(\tau )+\vec{\xi}')}
{\mid\vec{\eta}_{1}(\tau )-\vec{\eta}_{2}(\tau )+\vec{\xi}-\vec{\xi}'\mid}
\cdot \nonumber \\
&\cdot& \zeta^{({\check A}^{(o)}_{\perp})}
_{ru}(\vec{\eta}_{1}(\tau ),\vec{\eta}_{1}(\tau )+\vec{\xi};\tau )
\zeta^{({\check A}^{(o)}_{\perp})}
_{su}(\vec{\eta}_{2}(\tau ),\vec{\eta}_{2}(\tau )+\vec{\xi};\tau )]=\nonumber \\
&=& {{g^2_s[{\vec \eta}_1-{\vec \eta}_2;{\vec \eta}_i;{\check {\vec A}}^{(o)}
_{\perp}](\tau )}\over {|{\vec \eta}_1(\tau )-{\vec \eta}_2(\tau )|}}\, 
{\rightarrow}_{{\vec \eta}_1\rightarrow {\vec \eta}_2}\,\, 0,\nonumber \\
&&{}\nonumber \\
\Rightarrow && g^2_s[{\vec \eta}_1-{\vec \eta}_2;{\vec \eta}_i;{\check {\vec 
A}}^{(o)}_{\perp}](\tau )\, {\rightarrow}_{{\vec
\eta}_1\rightarrow {\vec \eta}_2}\, 0.
\label{e10}
\end{eqnarray}

This is the pseudoclassical statement of asymptotic freedom for N=2.

By using the developments (\ref{e4}), the equation of motion for the color 
charge ${\check Q}_{1a}(\tau )=-{\check Q}_2(\tau )$ becomes

\begin{eqnarray}
{d\over {d\tau}}{\check Q}_{1a}(\tau )\, &{\buildrel \circ \over =}&\, c_{acd}
{\check Q}_{1d}(\tau ) [{\dot {\vec \eta}}_1(\tau )\cdot {\check {\vec A}}
^{(o)}_{c\perp}(\tau ,{\vec \eta}_1(\tau )-\nonumber \\
&&-g^2_s\int d^3\sigma K^{(o)}_{cb}({\vec \eta}_1(\tau ),\vec \sigma ;\tau )
c_{buv}{\check {\vec A}}^{(o)}_{u\perp}(\tau ,\vec \sigma )\cdot {\check {\vec
\pi}}^{(o)}_{v\perp}(\tau ,\vec \sigma )]+\nonumber \\
&&+c_{acd}{\check Q}_{1d}(\tau ){\check Q}_{1r}(\tau ) [{\dot {\vec \eta}}
_1(\tau )\cdot {\check {\vec A}}^{(1)}_{cr\perp})(\tau ,{\vec \eta}_1(\tau )+
\nonumber \\
&&+g^2_s\{ K^{(o)}_{cb}({\vec \eta}_1(\tau ),{\vec \eta}_1(\tau );\tau )-K^{(o)}
_{cb}({\vec \eta}_1(\tau ),{\vec \eta}_2(\tau );\tau ) \}-\nonumber \\
&&-g^2_s\int d^3\sigma \{ K^{(o)}_{cb}({\vec \eta}_1(\tau ),\vec \sigma ;\tau )
c_{buv}({\check {\vec A}}^{(o)}_{u\perp}\cdot {\check {\vec \pi}}^{(1)}
_{vr\perp}+{\check {\vec A}}^{(1)}_{ur\perp}\cdot {\check {\vec \pi}}^{(o)}
_{v\perp})(\tau ,\vec \sigma )+\nonumber \\
&&+K^{(1)}_{cbr}({\vec \eta}_1(\tau ),\vec \sigma ;\tau )c_{buv}{\check {\vec
A}}^{(o)}_{u\perp}(\tau ,\vec \sigma )\cdot {\check {|vec \pi}}^{(o0}_{v\perp}
(\tau ,\vec \sigma ) \},
\label{e12}
\end{eqnarray}

\noindent while the particle equations become

\begin{eqnarray}
{d\over {d\tau}}&&[m_1 {{{\dot \eta}^r_1(\tau )}\over
{\sqrt{1-{\dot {\vec \eta}}_1^2(\tau )}}}]\,
{\buildrel \circ \over =}\nonumber \\
&&{\buildrel \circ \over =}\, \sum_a{\check Q}_{1a}(\tau ) \{ {\check E}^{(o)r}
_{a\perp}(\tau ,{\vec \eta}_1(\tau ))+[{\dot {\vec \eta}}_1(\tau )
\times {\check {\vec B}}^{(o)}_a(\tau ,{\vec \eta}_1(\tau ))]^r \}-
\nonumber \\
&&-P^{rs}_{\perp}({\vec \eta}_1) c_{amd}{\check A}^{(o)s}_{m\perp}(\tau ,{\vec 
\eta}_1(\tau ))\int d^3\sigma K_{de}^{(o)}({\vec \eta}
_1(\tau ),\vec \sigma ;\tau ) c_{enb}{\check {\vec A}}^{(o)}_{n\perp}(\tau ,
\vec \sigma )\cdot {\check {\vec E}}^{(o)}_{b\perp}(\tau ,\vec \sigma )-
\nonumber \\
&&-g^2_s\int d^3\sigma [c_{adc}{\check A}^{(o)r}_{d\perp}(\tau ,{\vec \eta}
_1(\tau ))K^{(o)}_{cb}({\vec \eta}_1(\tau ),\vec \sigma;\tau )+\nonumber \\
&&+{{\partial K^{(o)}_{ab}({\vec \eta}_1(\tau ),\vec \sigma;\tau )}\over 
{\partial \eta^r_1}}]{\check \rho}^{(YM)(o)}_b(\tau ,\vec \sigma ) 
\} +\nonumber \\
&&{}\nonumber \\
&&+\sum_{au}{\check Q}_{1a}(\tau ){\check Q}_{1u}(\tau ) \{ {\check E}^{(1)r}
_{a\perp}(\tau ,{\vec \eta}_1(\tau ))+[{\dot {\vec \eta}}_1(\tau )
\times {\check {\vec B}}^{(1)}_a(\tau ,{\vec \eta}_1(\tau ))]^r-\nonumber \\
&&-P^{rs}_{\perp}({\vec \eta}_1) c_{amd}\int d^3\sigma [{\check A}^{(o)s}
_{m\perp}(\tau ,{\vec \eta}_1(\tau )) K_{de}^{(o)}({\vec \eta}_1(\tau ),\vec 
\sigma ;\tau ) \nonumber \\
&&c_{enb}({\check {\vec A}}^{(o)}_{n\perp}\cdot {\check {\vec E}}^{(1)}
_{bu\perp}+{\check {\vec A}}^{(1)}_{nu\perp}\cdot {\check {\vec E}}^{(o)}
_{b\perp})(\tau ,\vec \sigma )+({\check A}^{(o)s}
_{m\perp}(\tau ,{\vec \eta}_1(\tau )) K_{deu}^{(1)}({\vec \eta}_1(\tau ),\vec 
\sigma ;\tau )+\nonumber \\
&&+{\check A}^{(1)s}_{mu\perp}(\tau ,{\vec \eta}_1(\tau )) K_{de}^{(o)}({\vec 
\eta}_1(\tau ),\vec \sigma ;\tau ) )c_{enb}({\check {\vec A}}^{(o)}_{n\perp}
\cdot {\check {\vec E}}^{(o)}_{b\perp})(\tau ,\vec \sigma ) ]+\nonumber \\
&&+g^2_s[c_{adc}{\check A}^{(o)r}_{d\perp}(\tau ,{\vec \eta}_1(\tau )) 
(K^{(o)}_{cu}({\vec \eta}_1(\tau ),{\vec \eta}_1(\tau );\tau )-K^{(o)}_{cu}
({\vec \eta}_1(\tau ),{\vec \eta}_2(\tau );\tau ))+\nonumber \\
&&+{{\partial}\over {\partial \sigma^r}}{|}_{\vec \sigma ={\vec \eta}_1}\, 
(K^{(o)}_{au}(\vec \sigma ,{\vec \eta}_1(\tau );\tau )-K^{(o)}_{au}(\vec 
\sigma ,{\vec \eta}_2(\tau );\tau )\,)]-\nonumber \\
&&-g^2_s\int d^3\sigma [c_{adc}{\check A}^{(o)r}_{d\perp}(\tau ,{\vec \eta}
_1(\tau ))K^{(o)}_{cb}({\vec \eta}_1(\tau ),\vec \sigma;\tau ){\check \rho}
^{(YM)(1)}_{bu}(\tau ,\vec \sigma )+\nonumber \\
&&+(c_{adc}(\, {\check A}^{(o)r}_{d\perp}(\tau ,{\vec \eta}_1(\tau ))K^{(1)}
_{cbu}({\vec \eta}_1(\tau ),\vec \sigma;\tau )+{\check A}^{(1)r}_{du\perp}(\tau 
,{\vec \eta}_1(\tau ))K^{(o)}_{cb}({\vec \eta}_1(\tau ),\vec \sigma;\tau ) )+
\nonumber \\
&&+{{\partial K^{(1)}_{abu}({\vec \eta}_1(\tau ),\vec \sigma;\tau )}\over 
{\partial \eta^r_1}}]{\check \rho}^{(YM)(o)}_b(\tau ,\vec \sigma ) ]\, \} ,
\nonumber \\ 
&&{}\nonumber \\
{d\over {d\tau}}&&[m_2 {{{\dot \eta}^r_2(\tau )}\over
{\sqrt{1-{\dot {\vec \eta}}_2^2(\tau )}}}]\,
{\buildrel \circ \over =}\nonumber \\
&&{\buildrel \circ \over =}\, -\sum_a{\check Q}_{1a}(\tau ) \{ {\check E}^{(o)r}
_{a\perp}(\tau ,{\vec \eta}_2(\tau ))+[{\dot {\vec \eta}}_2(\tau )
\times {\check {\vec B}}^{(o)}_a(\tau ,{\vec \eta}_2(\tau ))]^r \}-
\nonumber \\
&&-P^{rs}_{\perp}({\vec \eta}_2) c_{amd}{\check A}^{(o)s}_{m\perp}(\tau ,{\vec 
\eta}_2(\tau ))\int d^3\sigma K_{de}^{(o)}({\vec \eta}
_2(\tau ),\vec \sigma ;\tau ) c_{enb}{\check {\vec A}}^{(o)}_{n\perp}(\tau ,
\vec \sigma )\cdot {\check {\vec E}}^{(o)}_{b\perp}(\tau ,\vec \sigma )-
\nonumber \\
&&-g^2_s\int d^3\sigma [c_{adc}{\check A}^{(o)r}_{d\perp}(\tau ,{\vec \eta}
_2(\tau ))K^{(o)}_{cb}({\vec \eta}_2(\tau ),\vec \sigma;\tau )+\nonumber \\
&&+{{\partial K^{(o)}_{ab}({\vec \eta}_2(\tau ),\vec \sigma;\tau )}\over 
{\partial \eta^r_2}}]{\check \rho}^{(YM)(o)}_b(\tau ,\vec \sigma ) 
\} -\nonumber \\
&&{}\nonumber \\
&&-\sum_{au}{\check Q}_{1a}(\tau ){\check Q}_{1u}(\tau ) \{ {\check E}^{(1)r}
_{a\perp}(\tau ,{\vec \eta}_2(\tau ))+[{\dot {\vec \eta}}_2(\tau )
\times {\check {\vec B}}^{(1)}_a(\tau ,{\vec \eta}_2(\tau ))]^r-\nonumber \\
&&-P^{rs}_{\perp}({\vec \eta}_2) c_{amd}\int d^3\sigma [{\check A}^{(o)s}
_{m\perp}(\tau ,{\vec \eta}_2(\tau )) K_{de}^{(o)}({\vec \eta}_2(\tau ),\vec 
\sigma ;\tau ) \nonumber \\
&&c_{enb}({\check {\vec A}}^{(o)}_{n\perp}\cdot {\check {\vec E}}^{(1)}
_{bu\perp}+{\check {\vec A}}^{(1)}_{nu\perp}\cdot {\check {\vec E}}^{(o)}
_{b\perp})(\tau ,\vec \sigma )+({\check A}^{(o)s}
_{m\perp}(\tau ,{\vec \eta}_2(\tau )) K_{deu}^{(1)}({\vec \eta}_2(\tau ),\vec 
\sigma ;\tau )+\nonumber \\
&&+{\check A}^{(1)s}_{mu\perp}(\tau ,{\vec \eta}_2(\tau )) K_{de}^{(o)}({\vec 
\eta}_2(\tau ),\vec \sigma ;\tau ) )c_{enb}({\check {\vec A}}^{(o)}_{n\perp}
\cdot {\check {\vec E}}^{(o)}_{b\perp})(\tau ,\vec \sigma ) ]+\nonumber \\
&&+g^2_s[c_{adc}{\check A}^{(o)r}_{d\perp}(\tau ,{\vec \eta}_2(\tau )) 
(K^{(o)}_{cu}({\vec \eta}_2(\tau ),{\vec \eta}_1(\tau );\tau )-K^{(o)}_{cu}
({\vec \eta}_2(\tau ),{\vec \eta}_2(\tau );\tau ))+\nonumber \\
&&+{{\partial}\over {\partial \sigma^r}}{|}_{\vec \sigma ={\vec \eta}_2}\, 
(K^{(o)}_{au}(\vec \sigma ,{\vec \eta}_1(\tau );\tau )-K^{(o)}_{au}(\vec 
\sigma ,{\vec \eta}_2(\tau );\tau )\,)]-\nonumber \\
&&-g^2_s\int d^3\sigma [c_{adc}{\check A}^{(o)r}_{d\perp}(\tau ,{\vec \eta}
_2(\tau ))K^{(o)}_{cb}({\vec \eta}_2(\tau ),\vec \sigma;\tau ){\check \rho}
^{(YM)(1)}_{bu}(\tau ,\vec \sigma )+\nonumber \\
&&+(c_{adc}(\, {\check A}^{(o)r}_{d\perp}(\tau ,{\vec \eta}_2(\tau ))K^{(1)}
_{cbu}({\vec \eta}_2(\tau ),\vec \sigma;\tau )+{\check A}^{(1)r}_{du\perp}(\tau 
,{\vec \eta}_2(\tau ))K^{(o)}_{cb}({\vec \eta}_2(\tau ),\vec \sigma;\tau ) )+
\nonumber \\
&&+{{\partial K^{(1)}_{abu}({\vec \eta}_2(\tau ),\vec \sigma;\tau )}\over 
{\partial \eta^r_2}}]{\check \rho}^{(YM)(o)}_b(\tau ,\vec \sigma ) ]\, \} ,
\label{e13}
\end{eqnarray}

\noindent and the equations defining the rest-frame are

\begin{eqnarray}
&&m_1{{ {\dot {\vec \eta}}_1(\tau )}\over
{\sqrt{1-{\dot {\vec \eta}}_1^2(\tau )} }}+
m_2{{ {\dot {\vec \eta}}_2(\tau )}\over
{\sqrt{1-{\dot {\vec \eta}}_2^2(\tau )} }}-
\nonumber \\
&&-\sum_a {\check Q}_{1a}(\tau ) [{\check {\vec A}}^{(o)}_{a\perp}(\tau ,{\vec
\eta}_1(\tau ))-{\check {\vec A}}^{(o)}_{a\perp}(\tau ,{\vec \eta}_2(\tau ))+
{\check Q}_{1u}(\tau ) ({\check {\vec A}}^{(1)}_{au\perp}(\tau ,{\vec \eta}_1
(\tau ))-{\check {\vec A}}^{(1)}_{au\perp}(\tau ,{\vec \eta}_2(\tau )) )]+
\nonumber \\
&&+g^{-2}_s\int d^3\sigma \{ \vec \partial {\check A}^{(o)s}_{a\perp}{\check
E}^{(o)s}_{a\perp} +{\check Q}_{1u}(\tau ) (\vec \partial {\check A}^{(o)s}
_{a\perp}{\check E}^{(1)s}_{au\perp}+\vec \partial {\check A}^{(1)s}_{au\perp}
{\check E}^{(o)s}_{a\perp})+\nonumber \\
&&+{\check Q}_{1u}(\tau ){\check Q}_{1v}(\tau )( \vec \partial {\check A}^{(o)s}
_{a\perp} {\check E}^{(2)s}_{auv\perp}+\vec \partial {\check A}^{(1)s}_{au\perp}
{\check E}^{(1)s}_{av\perp}+\vec \partial {\check A}^{(2)s}_{auv\perp}{\check
E}^{(o)s}_{a\perp} ) \} (\tau ,\vec \sigma )\, {\buildrel \circ \over =}\, 0.
\label{e14}
\end{eqnarray}

The invariant mass (the relative Hamiltonian) of the
pseudoclassical scalar quark model takes the form

\begin{eqnarray}
H_{rel}&=&\sqrt{
M_{1}^{2}[m_{1},{\vec \eta}_1,\check{\vec{\kappa}}_{1},\check {\vec{A}}^{(0)}
_{a\perp},{\check {\vec A}}^{(1)}_{\perp}](\tau )
+\check{\vec{\kappa}}_{1}^{2}(\tau )}+\sqrt{
M_{2}^{2}[m_{2},{\vec \eta}_2,\check{\vec{\kappa}}_{2},\check {\vec{A}}^{(0)}
_{a\perp},{\check {\vec A}}^{(1)}_{\perp}](\tau )
+\check{\vec{\kappa}}_{2}^{2}(\tau )}+\nonumber\\
&+&V_{PP}[\vec{\eta}_{1}-\vec{\eta}_{2};\vec{\eta}_{i};
\check{\vec{A}}^{(0)}_{a\perp}](\tau )
+\sum_{i=1}^{2}V_{(i)PF}[(\vec{\eta}_{i};\check{\vec{A}}^{(0)}
_{a\perp};{\check {\vec \pi}}^{(o)}_{a\perp};\check{\vec{A}}^{(1)}_{\perp}]
(\tau )+\nonumber\\
&+&V_{FF}[\check{\vec{A}}^{(k)}_{\perp},{\check {\vec \pi}}_{\perp}
^{(k)}](\tau )+
\frac{1}{2}\int d^{3}\sigma\:\sum_{a}[g^{2}_{s}\check{\vec{\pi}}^2_{a\perp}+
g^{-2}_{s}\check{\vec{B}}^2_{a}](\tau,\vec{\sigma})
\label{e15}
\end{eqnarray}

\noindent with

\begin{eqnarray}
M_{i}^{2}&=&m_{i}^{2}+(-)^{i+1}2\check{\vec{\kappa}}_{i}(\tau )\cdot\sum_{a}
\check{Q}_{1a}(\tau )[\check{\vec{A}}^{(0)}_{a\perp}+\sum_u{\check Q}_{1u}
(\tau ){\check {\vec A}}^{(1)}_{au\perp}](\tau,\vec{\eta}_{i}(\tau ))+
\nonumber \\
&+&\sum_{a,b}\check{Q}_{1a}(\tau ){\check Q}_{1b}(\tau )[\check{\vec{A}}^{(0)}
_{a\perp}\cdot {\check {\vec A}}^{(o)}_{b\perp}]
(\tau,\vec{\eta}_{i}(\tau )).
\label{e16}
\end{eqnarray}

\noindent The first three terms of $H_{rel}$, after a suitable average over the
field degrees of freedom, should give the effective rest-frame Hamiltonian 
for the pseudoclassical relativistic scalar quark model: while $m_i$ are the
quark current masses, the suitable average of $M_i$ should give the quark
constituent masses. The terms $V_{(i)PF}$ describe quark-field interactions. 
The last two terms should describe the pseudoclassical glueball degrees of 
freedom.

All the previous results can be reexpressed in terms of rescaled Yang-Mills
fields by putting
${\check {\vec A}}_{a\perp}=g_s {\check {\vec {\tilde A}}}_{a\perp}$,
${\check {\vec \pi}}_{a\perp}=g^{-1}_s {\check {\vec {\tilde \pi}}}_{a\perp}$,
${\check {\vec B}}_a=g^{-1}_s {\check {\vec {\tilde B}}}_a$.

Finally, the Berezin-Marinov distribution function is

\begin{eqnarray}
&&\rho =\rho_1\rho_2,\nonumber \\
&&\rho_i=-2 < {\check Q}_{1a}(\tau ) > {\check Q}_{1a}(\tau ) N_i + {1\over 6} 
N_i^3 \approx 0,
\label{e17}
\end{eqnarray}

\noindent and, by taking the mean value of equations with it, the ``classical 
equations" are [see Refs.\cite{lus2,wong,lus4}
for the case of an external Yang-Mills field]

\begin{eqnarray}
&&{d\over {d\tau}} < {\check Q}_{1a}(\tau ) > -c_{abc} < {\check Q}_{1c}(\tau 
) > [{\dot {\vec \eta}}_1(\tau )\cdot {\check {\vec A}}^{(o)}_{b\perp}(\tau ,
{\vec \eta}_1(\tau )) -\nonumber \\
&&-g^2_s \int d^3\sigma K^{(o)}_{ab}({\vec \eta}_1(\tau ),\vec \sigma 
;\tau ){\check \rho}^{(YM)(o)}_b(\tau ,\vec \sigma )]\, {\buildrel \circ \over
=}\, 0
\label{e18}
\end{eqnarray}

\begin{eqnarray}
{d\over {d\tau}}&&[m_1 {{{\dot \eta}^r_1(\tau )}\over
{\sqrt{1-{\dot {\vec \eta}}_1^2(\tau )}}}]\,
{\buildrel \circ \over =}\nonumber \\
&&{\buildrel \circ \over =}\, \sum_a
< {\check Q}_{1a}(\tau ) > \{ {\check E}^{(o)r}
_{a\perp}(\tau ,{\vec \eta}_1(\tau ))+[{\dot {\vec \eta}}_1(\tau )
\times {\check {\vec B}}^{(o)}_a(\tau ,{\vec \eta}_1(\tau ))]^r \}-
\nonumber \\
&&-P^{rs}_{\perp}({\vec \eta}_1) c_{amd}{\check A}^{(o)s}_{m\perp}(\tau ,{\vec 
\eta}_1(\tau ))\int d^3\sigma K_{de}^{(o)}({\vec \eta}
_1(\tau ),\vec \sigma ;\tau ) c_{enb}{\check {\vec A}}^{(o)}_{n\perp}(\tau ,
\vec \sigma )\cdot {\check {\vec E}}^{(o)}_{b\perp}(\tau ,\vec \sigma )-
\nonumber \\
&&-g^2_s\int d^3\sigma [c_{adc}{\check A}^{(o)r}_{d\perp}(\tau ,{\vec \eta}
_1(\tau ))K^{(o)}_{cb}({\vec \eta}_1(\tau ),\vec \sigma;\tau )+\nonumber \\
&&+{{\partial K^{(o)}_{ab}({\vec \eta}_1(\tau ),\vec \sigma;\tau )}\over 
{\partial \eta^r_1}}]{\check \rho}^{(YM)(o)}_b(\tau ,\vec \sigma ) 
\} ,\nonumber \\
&&{}\nonumber \\
{d\over {d\tau}}&&[m_2 {{{\dot \eta}^r_2(\tau )}\over
{\sqrt{1-{\dot {\vec \eta}}_2^2(\tau )}}}]\,
{\buildrel \circ \over =}\nonumber \\
&&{\buildrel \circ \over =}\, -\sum_a
< {\check Q}_{1a}(\tau ) > \{ {\check E}^{(o)r}
_{a\perp}(\tau ,{\vec \eta}_2(\tau ))+[{\dot {\vec \eta}}_2(\tau )
\times {\check {\vec B}}^{(o)}_a(\tau ,{\vec \eta}_2(\tau ))]^r \}-
\nonumber \\
&&-P^{rs}_{\perp}({\vec \eta}_2) c_{amd}{\check A}^{(o)s}_{m\perp}(\tau ,{\vec 
\eta}_2(\tau ))\int d^3\sigma K_{de}^{(o)}({\vec \eta}
_2(\tau ),\vec \sigma ;\tau ) c_{enb}{\check {\vec A}}^{(o)}_{n\perp}(\tau ,
\vec \sigma )\cdot {\check {\vec E}}^{(o)}_{b\perp}(\tau ,\vec \sigma )-
\nonumber \\
&&-g^2_s\int d^3\sigma [c_{adc}{\check A}^{(o)r}_{d\perp}(\tau ,{\vec \eta}
_2(\tau ))K^{(o)}_{cb}({\vec \eta}_2(\tau ),\vec \sigma;\tau )+\nonumber \\
&&+{{\partial K^{(o)}_{ab}({\vec \eta}_2(\tau ),\vec \sigma;\tau )}\over 
{\partial \eta^r_2}}]{\check \rho}^{(YM)(o)}_b(\tau ,\vec \sigma ) 
\} ,
\label{e20}
\end{eqnarray}

\begin{eqnarray}
&&m_1{{ {\dot {\vec \eta}}_1(\tau )}\over
{\sqrt{1-{\dot {\vec \eta}}_1^2(\tau )} }}+
m_2{{ {\dot {\vec \eta}}_2(\tau )}\over
{\sqrt{1-{\dot {\vec \eta}}_2^2(\tau )} }}+g^{-2}_s\int d^3\sigma (\vec \partial
{\check A}^{(o)s}_{a\perp}{\check E}^{(o)s}_{a\perp})(\tau ,\vec \sigma )-
\nonumber \\
&&-\sum_a < {\check Q}_{1a}(\tau ) > [{\check {\vec A}}^{(o)}_{a\perp}(\tau ,{\vec
\eta}_1(\tau ))-{\check {\vec A}}^{(o)}_{a\perp}(\tau ,{\vec \eta}_2(\tau ))]+
\nonumber \\
&&+g^{-2}_s\sum_{a,u}< {\check Q}_{1u}(\tau ) >\int d^3\sigma 
(\vec \partial {\check A}^{(o)s}
_{a\perp}{\check E}^{(1)s}_{au\perp}+\vec \partial {\check A}^{(1)s}_{au\perp}
{\check E}^{(o)s}_{a\perp})
(\tau ,\vec \sigma )\, {\buildrel \circ \over =}\, 0.
\label{e21}
\end{eqnarray}

The particle ``classical equations" are expected to have the same causal
pathologies (runaway solutions or preaccelerations) as the Abraham-Lorentz-
Dirac ones in the electromagnetic case. In that case, see Refs.\cite{lus2,alba}
, one can follow a different procedure: i) solve the field equations (the
solutions are incoming free linear waves plus the Lienard-Wiechert potential
if retarded solutions are selected); ii) put the solution in the particle
equations; iii) take the mean value of the new equations. This procedure gives
different ``classical equations" without pathologies, because one has $Q^2_i=0$ 
for the Grassmann-valued electric charges: in this way, at the pseudoclassical 
level, one regularizes the Coulomb self-energies\cite{lus1}, has no radiation
produced by the single Lienard-Wiechert potentials but has radiation from the
superposition of them \cite{alba} (terms in $Q_iQ_j$ with $i\not= j$).

In the non-Abelian case, we do not know solutions of Eqs.(\ref{ea3}) and the
superposition principle does not hold, so that we do not know how to apply the
second procedure. A first needed step would be to find the Green function of
the operator on the left side of Eq.(\ref{ea3}).

\vfill\eject

\section{Discussion and Conclusions.}

We have obtained the description of the isolated system of N scalar particles 
with Grassmann-valued color charges plus the color SU(3) Yang-Mills field on
spacelike hypersurfaces and then its canonical reduction to the Wigner 
hyperplane with only physical degrees of freedom (generalized Coulomb gauge in
the Wigner-covariant rest-frame instant form of dynamics) following the same 
steps of Refs.\cite{lus1,alba} for the case of N scalar particles with
Grassmann-valued electric charges plus the electromagnetic field.

In particular, we obtained the reduced Hamilton and Euler-Lagrange equations
for the reduced transverse Yang-Mills field and for the particles in this
rest-frame generalized Coulomb gauge, generalizing those of the electromagnetic
case. Four kinds of interactions between two color charge densities (associated
either with a particle or with the SU(3) field) were identified in the
resulting potential extracted from the gauge part of the Yang-mills field:
one of them is a Coulomb interaction, while the others involve Wilson lines
[$\zeta ({\vec \sigma}_1,{\vec \sigma}_2;\tau )=P\, e^{\int_{{\vec \sigma}_2}
^{{\vec \sigma}_1} d\vec \sigma \cdot {\check {\vec A}}_{a\perp}(\tau ,\vec
\sigma ){\hat T}^a_s}$] along straightlines between the two (simultaneous in the
rest frame) points where the two color densities are located. 
In the electromagnetic case\cite{lus1,alba}, where the field is not charged, 
there is only the interparticle Coulomb potential. Due to the high
nonlinearity of the field equations, it is extremely difficult to try to define
a non-Abelian analogue of the Lienard-Wiechert potentials of the Abelian
case\cite{alba}. Moreover, we have found the Berezin-Marinov distribution
function for the Grassmann sector\cite{lus2,casalb} and, then, the ``classical 
equations of motion" as the mean value of the previous equations. In absence
of a Lienard-Wiechert potential, solution of the field equations, to be put
in the particle equations of motion (which. otherwise, are going to have
causal pathologies
like in the electromagnetic Abraham-Lorentz-Dirac equations), we cannot take the
mean value of these new equations and check that the resulting equations do not
have causal pothologies as it happens in the electromagnetic case\cite{alba}.

We then studied the N=2 case, which is the pseudoclassical basis of the 
relativistic scalar-quark model but with the reduced transverse color field
present, which describes the pseudoclassical glueball degrees of freedom. With
suitable constraints on the Grassmann variables we obtain the description of
an isolated system corresponding to a meson: a quark-antiquark pair plus the 
transverse SU(3) field. The reduced Hamiltonian of the system is its invariant
mass expressed in the intrinsic rest frame. It contains the relativistic kinetic
terms of the two quarks minimally coupled to the transverse SU(3) field, an
interparticle field-dependent (but quark-mass-independent)
potential, a single particle - field potential and
the kinetic terms plus a self-interaction potential for the transverse SU(3)
field. If one would know the non-Abelian Lienard-Wiechert potentials of the
quarks (if this concept makes sense due to the nonlinear selfcoupling of the 
color field, i.e. due to the glueball degrees of freedom), one
could obtain a field-independent expression for the invariant mass of the
meson and therefore an explicit formulation of a field-independent relativistic
scalar-quark model starting from the pseudoclassical Lagrangian for QCD with
scalar quarks. This would fill the gap, at least at the pseudoclassical
level, between QCD and quark models [see Ref.\cite{q1} for the nonrelativistic
one]. Moreover, a field-independent (but quark-velocity-dependent) interparticle
potential would appear, whose static [$m_i\rightarrow \infty\, \Rightarrow \,
{\dot {\vec \eta}}_i(\tau )\rightarrow 0$] part should have connections with the
static potential in QCD \cite{peter}. Note that at the pseudoclassical level 
there is no notion of gluons (or gluon exchange): only SU(3) fields are present
and only a Lienard-Wiechert potential (if it exists in some sense) would
create a bridge towards the Wilson loop expectation value form of the static
potential between heavy quarks [in the quenched approximation in which quark 
loops from pair production in vacuum (absent in the pseudoclassical theory) are
neglected (sea-quarks of infinite mass) and glueballs (closed color loops) are
unambiguously defined] re-expressed in terms of perturbative QCD.

When the study of Dirac fields and spinning particles on spacelike hypersurfaces
will be finished \cite{dep}, one will get analogous results for spinning
particles (with Grassmann-valued spin), namely one will introduce the quark
spin structure in the pseudoclassical relativistic quark model.

In the N=2 (meson) case, we have explored the implications of the imposition of 
the condition that the isolated system (quarks+transverse SU(3) field) is a 
color singlet. The condition ${\check Q}_a={\check Q}^{(YM)}_a(\tau )+{\check
Q}_{1a}(\tau )+{\check Q}_{2a}(\tau ) =0$ is imposed by hand and, moreover, it
is asked to be fulfilled by asking separately the two conditions ${\check Q}
^{(YM)}_a(\tau )=0$ [no color flux of the transverse SU(3) field at space
infinity; it replaces the Abelian condition of no radiation field, i.e.
${\vec A}_{\perp}(\tau ,\vec \sigma )=0$ \cite{lus1,alba}] and ${\check
Q}_{1a}(\tau )+{\check Q}_{2a}(\tau ) =0$ [it is the color singlet condition
of field-independent quark models, which work very well phenomenologically
\cite{q1}]. Now, the condition ${\check Q}_a(\tau )=0$ can be imposed by
choosing suitable Hamiltonian boundary conditions at fixed $\tau$ on the
transverse SU(3) field. Therefore, if confinement exists, the condition
${\check Q}_a^{(YM)}(\tau )={\check Q}_{1a}(\tau )+{\check Q}_{2a}(\tau )\,
{\buildrel \circ \over =}\, 0$ should
emerge from the solution of the Hamilton equations [at least approximately
seen the phenomenological soundness of the quark model]. Moreover, the
condition ${\check Q}_{1a}(\tau )+{\check Q}_{2a}(\tau ) =0$, together with
the other constraints on Grassmann variables, implies the disappearance of the
Coulomb term in the interparticle potential [this is a statement
stronger than the
regularization of the Coulomb self-energies in the Abelian case \cite{lus1,alba}
] and that this potential tends to zero when the two quarks tend to the same
spatial position in the rest frame [$|{\vec \eta}_1(\tau )-{\vec \eta}_2(\tau )|
\rightarrow 0$]. This is the pseudoclassical statement of asymptotic freedom:
there is an antiscreening even stronger than in QCD, because the reducing
(screening) effect of pair production is here absent. This kind of asymptotic 
freedom is an algebraic consequence of the request of color singlets in the
quark model oriented form ${\check Q}_{1a}(\tau )+{\check Q}_{2a}(\tau ) =0$.

It would be interesting to study in this way [${\check Q}_{1a}(\tau )+{\check 
Q}_{2a}(\tau )+{\check Q}_{3a}(\tau ) =0$] the N=3 (baryon) case of 3 quarks 
or 3 antiquarks. What happens when either only two quarks or all three quarks tend to the same spatial position?

Coming back to the N=2 case, the main unsolved problem, connected with the
color singlet requirement, is confinement. Even if the boundary conditions for
the transverse SU(3) fields are chosen so to imply ${\check Q}_a(\tau )
=0$,
we do not know how to do either analytical or numerical calculations to check
whether the interparticle potential implies confinement and, if yes, whether 
confinement implies ${\check Q}_{1a}(\tau )+{\check Q}_{2a}(\tau )
=0$, not to speak of the possible glueball degrees of freedom.
The obstruction is the lack of control on the Wilson line
$\zeta ({\vec \sigma}_1,{\vec \sigma}_2;\tau )=P\, e^{\int_{{\vec \sigma}_2}
^{{\vec \sigma}_1} d\vec \sigma \cdot {\check {\vec A}}_{a\perp}(\tau ,\vec
\sigma ){\hat T}^a_s}$: given the function space for the transverse SU(3)
potential ${\check {\vec A}}_{a\perp}(\tau ,\vec \sigma )$, which is the
behaviour in ${\vec \sigma}_1$ and ${\vec \sigma}_2$ of $\zeta ({\vec \sigma}
_1,{\vec \sigma}_2;\tau )$ as an element of the group SU(3) in the adjoint 
representation? Does it belong to the same function space as the transverse 
SU(3) potential? How to simulate it on a lattice having eliminated all the
gauge degrees of freedom in favour of a transverse potential? Which is the
pseudoclassical analogue of the quantum Wilson criterion of confinement
\cite{wi,lat} in this Hamiltonian generalized Coulomb gauge? Moreover, an 
aspect of the confinement problem which has to be understood at the
pseudoclassical level is the role of the center $Z_3$ of SU(3), namely the
zero triality condition [fermions know SU(3), but the Yang-Mills field feels 
only the not simply connected group $SU(3)/Z_3$ \cite{lusa}],
relevant in the approaches of Refs.\cite{polo}. Following Ref.\cite{john},
this condition is probably hidden in the fact that the Wilson line
$\zeta ({\vec \sigma}_1,{\vec \sigma}_2;\tau )$ must take values in $SU(3)/Z_3$ 
and not in SU(3).

Let us also note that the results of Ref.\cite{lusa} on Yang-Mills fields plus
Grassmann-valued Dirac fields, once the description of Dirac fields in the
rest-frame instant form will be terminated\cite{dep}, suggest that the 
potential V will remain unchanged except for the replacement $\sum_i{\check 
\rho}_{ia}(\tau ,\vec \sigma ) \mapsto {\check \psi}^{\dagger}(\tau ,\vec
\sigma ) T^a {\check \psi}(\tau ,\vec \sigma )$.

Another yet unsolved problem (also in the electromagnetic case) is how to
eliminate the 3 constraints ${\vec {\cal H}}_p
(\tau )\approx 0$ defining the intrinsic rest frame. 
This requires the introduction of 3 gauge-fixings
identifying the Wigner 3-vector describing the intrinsic 3-center of mass
on the Wigner hyperplane. However, till now these gauge-fixings are known only
in the case of an isolated system containing only particles. When the center of
mass canonical decomposition of linear classical field theories will be
available (see Ref.\cite{lon} for the Klein-Gordon field), its reformulation
on spacelike hypersurfaces will allow the determination of these gauge-fixings
also when fields are present and a Hamiltonian description with only
Wigner-covariant relative variables with an explicit  control on the
action-reaction balance between fields and particles or between two types of
fields.

As said in Ref.\cite{alba}, the quantization of this relativistic scalar-quark
model has to overcome two problems. On the particle
side, the complication is the quantization of the square roots associated
with the relativistic kinetic energy terms. On the field side, the obstacle
is the absence (notwithstanding there is no  no-go theorem) of a complete
regularization and renormalization procedure of electrodynamics in the
Coulomb gauge: see Refs.\cite{cou,lav} for the existing results for QED.
However, as shown in Refs.\cite{lus1,lusa,re}, the rest-frame instant form of
dynamics automatically gives a physical ultraviolet cutoff: it is the
M$\o$ller radius $\rho =\sqrt{-W^2}c/P^2=|\vec S|c/\sqrt{P^2}$ ($W^2=-P^2{\vec 
S}^2$ is the Pauli-Lubanski Casimir), namely the classical intrinsic radius of 
the worldtube, around the covariant noncanonical Fokker-Price center of
inertia, inside which the noncovariance of the canonical center of mass ${\tilde
x}^{\mu}$ is concentrated. At the quantum level $\rho$ becomes the Compton 
wavelength of the isolated system multiplied its spin eigenvalue $\sqrt{s(s+1)}$
, $\rho \mapsto \hat \rho = \sqrt{s(s+1)} \hbar /M=\sqrt{s(s+1)} \lambda_M$ 
with $M=\sqrt{P^2}$ the invariant mass and $\lambda_M=\hbar /M$ its Compton
wavelength.

Let us remark that in the electromagnetic case all the dressings with Coulomb
clouds [of the scalar particles and of charged Klein-Gordon fields in
Ref.\cite{alba}
and of Grassmann-valued Dirac fields in Ref.\cite{lusa}] are done with the
Dirac phase $\eta_{em}=-{1\over {\triangle}} \vec \partial \cdot \vec A$ 
\cite{dira}. The same phase is used in Ref.\cite{lav} to dress fermions in QED.
In these papers there is a definition of dressing of fermion fields in the
non-Abelian quantum case, which is quite similar to the one of Ref.\cite{lusa}
(used in this paper) even if implemented only perturbatively. Essentially, one 
looks for a matrix $h \in SU(3)$ such that under a gauge transformation U one
has $h \mapsto h^U=U^{-1}h$; then one has $\psi = h \psi_{PHYS}$ and ${\vec
A}_a{\hat T}^a= h{\check {\vec A}}_{a\, PHYS}{\hat T}^a h^{-1}-\vec \partial
h\, h^{-1}$ with $\psi_{PHYS}$ and ${\check {\vec A}}_{a\, PHYS}$ gauge
invariant. Comparison with Eq.(\ref{c1}) and with Ref.\cite{lusa} shows that
at the classical level one has $h=P\, e^{\Omega^{(\hat \gamma )}_{s a}(\eta
^{(A)}){\hat T}^a}$.

Also in Ref.\cite{john} the solution of the quantum Gauss law constraint on
Schroedinger functional $\Psi [A]$ in the case of two static particles of
opposite charges, is able to reproduce the Coulomb potential and the Coulomb
self-energy with the same mechanism as in Refs.\cite{lusa,alba} [namely with 
the Abelian form of Eq.(\ref{c21})] only if $\Psi [A]=
e^{i\eta_{em}} \Phi [A]$ with $\Phi [A]$ gauge invariant and not with 
$\Psi^{'}[A]=e^{i\int_{x_o}^{x_1} d\vec x\cdot \vec A(x^o,\vec x)} \Phi^{'}[A]$
with a phase factor resembling the Wilson loop operator [one has $\Phi [A] 
=e^{i\int_{x_o}^{x_1}d\vec x\cdot {\vec A}_{\perp}(x^o,\vec x)} \Phi^{'}[A]$,
namely the Wilson line operator has been broken in the gauge part plus the
gauge invariant part using $\vec A=\vec \partial \eta_{em}+{\vec A}_{\perp}$].
However, it is difficult to see a connection between the phase of the
Schroedinger functional $\Psi [A]$ proposed in Ref.\cite{john} as a solution of 
the non-Abelian Gauss laws and our potential V, which is a consequence of Eqs.
(\ref{c1}), (\ref{c3})-(\ref{c5}), (\ref{c11}), (\ref{c21}) [now the Wilson
line $e^{i\int_{x_o}^{x_1}d\vec x\cdot {\vec A}_a(x^o,\vec x){\hat T}^a}$
cannot be broken in a gauge part and in a gauge invariant part due to
Eq.(\ref{c1}); in Ref.\cite{john} the gauge-dependent part is a product $U(x_1)
U^{\dagger}(x_o)$ of two gauge transformations].

Let us finish with some heuristic considerations about the M$\o$ller radius.
In QCD, due to asymptotic freedom and to the renormalization group equations,
the strong coupling constant $\alpha_s=g^2_s/4\pi$ is replaced by the
effective running coupling constant (see for instance REf.\cite{bbj})
$\alpha_s(Q^2)=12\pi /(33-2N_F)\, ln{{Q^2}\over {\Lambda^2_{QCD}}}$ at high
$Q^2$ ($N_F$ is the number of flavors, giving the screening contribution of the
fermions to the vacuum polarization). Therefore, the adimensional coupling 
constant $\alpha_s$ may be replaced with the fundamental QCD scale ($\hbar =c=
1$) $\Lambda_{QCD} \approx 300 Mev \approx 10^{-13} cm=1 fm$, which is now 
usually replaced by $\alpha_s(m_Z^2)\approx 0.116$ \cite{schme} 
[dimensional
transmutation, connected with the breaking of scale invariance at high energies 
and with the scale anomaly; the physical mechanism for generating the scale
at low energies is unclear (a candidate is the chiral symmetry breaking phase
transition of QCD which generates a constituent mass of order 300 Mev for the
light quarks)]. This implies \cite{q1} that in the nonrelativistic quark model
with confinement, one may choose (among the many possible phenomenological 
potentials) the simple potential $V(r)=-{4\over 3} {{\alpha_s(r)}\over r}+
\kappa r$ with the short distance behaviour $\alpha_s(r)=12\pi /(33-2N_F)\, ln 
{{r^2_o}\over {r^2}}$, i.e. for $r < r_o={1\over {\Lambda_{QCD}}} \approx 10
^{-13} cm= 1 fm$ [in Ref.\cite{peter} it is shown that QCD perturbative
results cannot be trusted for $r < 0.07 \Lambda^{-1}_{QCD}
=0.07 fm$ for $\Lambda_{QCD} \approx 210 Mev.$, i.e. $\alpha_s(m_Z^2)=0.118$]. 
One can consider $r_o$ as an effective radius of confinement for 
quarks and glueballs [the proton Compton wavelength is $\lambda_p=\hbar /m_pc
\approx 10^{-13} cm =1 fm < r_o$). For $r \approx r_o$ the expressions of
$\alpha_s(Q^2)$ and $\alpha_s(r)$ break down due to confinement, which is 
described by the linear term in V(r) [$\kappa \approx 0.2 Gev^2$ is the
``string tension", which turns out to be determined numerically as a function of
$\Lambda_{QCD}$ in lattice gauge theory\cite{lat} at least for heavy quarks, 
for which a string-like (chromoelectric flux tube) structure emerges; in the 
theoretical approach based on the analogy with type II superconductors (se the 
reviews in Refs.\cite{bbj,q1}), where the vacuum is a color-dielectric medium 
and a $\bar qq$ state is a confined color-electric flux tube (anti-Meissner 
effect), $\kappa$ is determined by the gluon condensate ${{\alpha_s}\over 
{\pi}} < \sum_aF_{a\mu\nu}F_a^{\mu\nu} > 0$ (as confirmed in strong coupling
expansion of lattice QCD\cite{wi}), which is present besides the $< \bar qq >$
condensate responsible for chiral symmetry breaking]. The crucial point for the
pseudoclassical relativistic quark model would be to see whether Eq.(\ref{e10})
implies $V_{PP}{\rightarrow}_{|{\vec \eta}_1-{\vec \eta}_2|\rightarrow \infty}
\,\, \kappa \, |{\vec \eta}_1-{\vec \eta}_2|+....$

Let us also note that in the MIT bag model (see Ref.\cite{bag} for a
review), the bag constant B is connected with the gluon condensate and the 
length of the chromoelectric flux tube for heavy quarks and the string tension
$\kappa$ are determined by B and $\alpha_s$.

The M$\o$ller radius $\rho =|\vec S|/M$ is going to play the
role of a ultraviolet cutoff $\hat \rho =\sqrt{s(s+1)} \lambda_M$ [$\lambda_M$ 
is the Compton wavelength of the isolated system with invariant mass
$M=\sqrt{P^2}=H_{rel}$] at the quantum level (like the lattice spacing in
lattice QCD). Since $\rho$ describes a nontestable classical short distance
region [impossibility of frame-independent determination of the location of the
relativistic canonical center of mass (also named Pryce center of mass and
having the same covariance of the Newton-Wigner position operator \cite{pnw}); 
its connection with the Mach's principle
according to which only relative motions are measurable], it sounds reasonable 
that for a confined system of effective radius $r_o=1/\Lambda_{QCD}$ one has 
$\hat \rho \approx r_o$. However, this is not correct because it implies a
mass-spin relation $|\vec S| \approx r_o M$, while the phenomenological Regge
trajectories are $|\vec S| = \alpha_s^{'} M^2+\alpha_o$ 
[$\alpha_s^{'}=1 Gev^{-2}$],
implying, at least for heavy quarks, an effective string theory inside QCD.
Now, in string theory\cite{ven} the relevant dimensional quantity is the
tension $T_s=1/2\pi \alpha^{'}_s$ (the energy per unit length), which, at the 
quantum level, determines a minimal length $L_s=\sqrt{\hbar /T_s} =\sqrt{2\pi
\hbar \alpha^{'}_s}{\buildrel {\hbar =1} \over =}\, \sqrt{2\pi \alpha^{'}_s}$. 
For a classical string one has
$|\vec S| \leq \alpha_s^{'} M^2$, so that its M$\o$ller radius is $\rho \leq
\alpha_s^{'} M$: at the quantum level one has $\hat \rho =\sqrt{s(s+1)} 
\lambda_M \leq \alpha_s^{'} M= L^2_s/\lambda_M$. Therefore, if the QCD string
has $L_s=\sqrt{2\pi \alpha_s^{'}}\approx 10^{-13}cm =1 fm \leq
 r_o=1/\Lambda_{QCD}\approx 1 fm$, one gets that the
M$\o$ller radius of a confined system must be of the order $\rho \leq
r^2_oM =M/\Lambda^2_{QCD}$ [$\hat \rho \leq r^2_o/\lambda_M$]. This corresponds
to a QCD string with $2\pi \alpha^{'}_s \leq r^2_o=\Lambda^{-2}_{QCD}$: see
Ref.\cite{sun} for a long glue string giving rise to an effective Nambu-Goto 
string with this $\alpha_s^{'}$ and more generally see Ref.\cite{pol}.

This effective QCD string theory (whose final formulation has still to be 
found) must not be confused with string cosmology\cite{ven,ven1}, in which, at 
the quantum level, the string tension $T_{cs}=1/2\pi \alpha^{'}_{cs}=
L^2_{cs}/\hbar$ gives
rise to a minimal length $L_{cs}{\buildrel {\hbar =1} \over =}\, \sqrt{2\pi 
\alpha^{'}_{cs}} \geq L_P$ [$L_P=1.6\, 10^{-33} cm$ is 
the Planck length] and is determined by the vacuum 
expectation value of the background metric of the vacuum (if the ground state is
flat Minkowski spacetime), while the grand unified coupling constant $\alpha
_{GUT}$ (replacing $\alpha_s$ of QCD) is determined by the vacuum expectation 
value of the background dilaton field. This minimal length $L_{cs}\geq
L_P$ (suppressing the gravitational corrections) could be a lower bound 
for the M$\o$ller radius of an asymptotically flat universe, built with the
Poincar\'e Casimirs of the asymptotic ADM Poincar\'e charges (see Ref.
\cite{rus} for the canonical reduction of tetrad gravity). The upper bound on 
$\rho$ (namely a physical infrared cutoff) could be the Hubble distance 
$cH_o^{-1}\approx 10^{28} cm$ considered as an effective radius of the universe.
Therefore, it seems reasonable that our physical ultraviolet cutoff $\rho$
is meaningful in the range $L_P\leq L_{cs} < \rho < cH_o^{-1}$.

Let us remember that $\rho$ is also a remnant in flat Minkowski spacetime of 
the energy conditions of general relativity\cite{lus1}: since the M$\o$ller
noncanonical, noncovariant center of energy has its noncovariance localized
inside the same worldtube with radius $\rho$ (it was discovered in this way)
\cite{mol}, it turns out that an extended relativistic system with the
material radius smaller of its intrinsic radius $\rho$ has: i) the peripheral
rotation velocity can exceed the velocity of light; ii) its classical energy
density cannot be positive definite everywhere in every frame.

Moreover, the extended Heisenberg relations  of string theory\cite{ven}, i.e.
$\triangle x ={{\hbar}\over {\triangle p}}+{{\triangle p}\over {T_{cs}}}=
{{\hbar}\over {\triangle p}}+{{\hbar \triangle p}\over {L^2_{cs}}}$ (see Ref.
\cite{ven} for the meaning of $\triangle p$ in string theory) implying the
lower bound $\triangle x > L_{cs}=\sqrt{\hbar /T_{cs}}$,
have a counterpart in the quantization of the M$\o$ller radius\cite{lus1}:
if we ask that, also at the quantum level, one cannot test the inside of the 
worldtube, we must ask $\triangle x > \hat \rho$ which is the lower bound
implied by the modified uncertainty relation $\triangle x ={{\hbar}\over 
{\triangle p}}+{{\hbar \triangle p}\over {{\hat \rho}^2}}$. This would imply 
that the center-of-mass canonical noncovariant (Pryce) 3-coordinate 
$\vec z=\sqrt{P^2}({\vec {\tilde x}}-{{\vec P}\over {P^o}}{\tilde x}^o)$ 
\cite{lus1} cannot become a
self-adjoint operator. See Hegerfeldt's theorems \cite{heg}, his interpretation 
pointing at the impossibility of a good localization of relativistic particles
(experimentally one determines only a worldtube in spacetime emerging from the 
interaction region) and also the comments of Ref.\cite{kal} against this
interpretation. Since the eigenfunctions of the canonical center-of-mass
operator are playing the role of the wave function of the universe, one could 
also say that the center-of-mass variable has not to be quantized, because it
lies on the classical macroscopic side of Copenhagen's interpretation and,
moreover, because, in the spirit of Mach's principle that only relative 
motions can be observed, no one can observe it. On the other hand, if one 
rejects the canonical noncovariant center of mass in favor of the covariant
noncanonical Fokker-Pryce center of inertia \cite{fp,lus1,pau}
$Y^{\mu}$, $\{ Y^{\mu},Y^{\nu} \}
\not= 0$, one could invoke the philosophy of quantum groups to quantize 
$Y^{\mu}$ to get some kind of quantum plane for the center-of-mass 
description.

\vfill\eject

\end{document}